%% file: tesi.tex
\documentclass[12pt,dvips,oneside]{book}
\usepackage{cite}
\usepackage[]{graphicx}
\usepackage{makeidx}
\newcommand{\mathsym}[1]{{}}

\usepackage[]{caption}
\makeindex
\def\gsim{\mathrel{\raise.3ex\hbox{$>$\kern-.75em\lower1ex\hbox{$\sim$}}}}
\def\lsim{\mathrel{\raise.3ex\hbox{$<$\kern-.75em\lower1ex\hbox{$\sim$}}}}

\begin{document}
\setlength{\baselineskip}{16.5pt}

\input{files/frontespizio.tex}

\pagenumbering{roman}

\input{files/Preface.tex}
\newpage

\tableofcontents
\label{contents}
\markboth{CONTENTS}{CONTENTS}

\newpage

\pagenumbering{arabic}
\setcounter{page}{1}

\input{files/Introduction.tex}
\input{files/SM.tex}
\input{files/std_cosmology.tex}
\input{files/electric_ch_quant.tex}
\input{files/MCP.tex}
\input{files/mirror_cosmology.tex}

\input{files/mirror_kin_mix.tex}
\newpage
\input{files/Conclusions.tex}

\newpage
\input{files/Ringraziamenti.tex}
\appendix
\input{files/appendix_cosmology.tex}
\input{files/appendix_cross_sections.tex}

\input{bibliografy.tex}


%
%


\end{document}

%% file: files/frontespizio.tex


\begin{titlepage}
\topmargin=-10pt
        \begin{figure} [htbp]
        \begin{center}
        \includegraphics[scale=0.4]{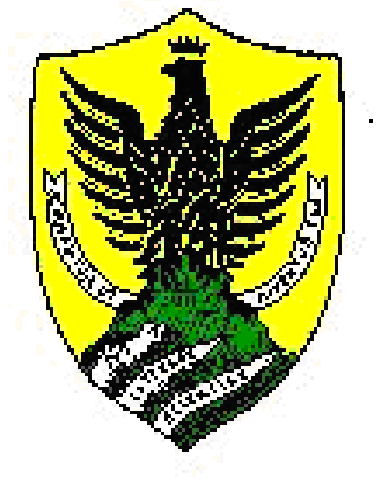}
        \end{center}
        \end{figure} 
\begin{center}
\vspace{-.5cm}
{\Large UNIVERSIT\`A DEGLI STUDI DELL'AQUILA}\\
\vspace{.3cm}
{\normalsize FACOLT\`A DI SCIENZE MATEMATICHE, FISICHE E NATURALI}\\
\vspace{2.cm}
{\Large TESI DI LAUREA SPECIALISTICA IN FISICA}\\
\vspace{0.3cm}
{\large A.A. 2006/2007}\\
\vspace{2.cm}
{\Large \bf Phenomenology and cosmology \\ \vspace{.2cm}
of millicharged particles \\ \vspace{.35cm}
 and experimental prospects for their search
}\\
\vspace{2.5cm}
        \begin{tabular}{p{3in} p{3in} p{0in}}
        {\center \large Candidato:} &  {\center \large Relatore:} \\
        {\center \large \textbf{Angela Lepidi}} & 
        {\center \large \textbf{Prof. Zurab Berezhiani}} \\
         \end{tabular}
\end{center}
\end{titlepage}


%% file: files/Preface.tex

\noindent {\huge \noindent  {\bf Preface}}
\noindent  \addcontentsline{toc}{chapter}{{Preface}}
\vspace{1.5cm}
\label{chap-pref}
\markboth{Preface}{}
\noindent

This is my Diploma Thesis performed under the supervision of Prof. Z. Berezhiani, defended at L'Aquila University in 10 October 2007 and deposited 
 in the official acta of the University of L'Aquila. Only some minor misprints are corrected with respect to the deposited copy. So it does not contain references on the works that have been done later, in particular regarding the data released by DAMA/Libra in April 2008 and related papers. 



The main results of my Thesis and in particular the ones regarding the cosmological limits of the photon-mirror photon kinetic mixing were presented by me at the IMPRS Workshop, Max Planck Institute, Munich, 26 Oct. 2007 and by Z. Berezhiani at the Int. Workshop "Hot Topics in Cosmology", Cargese, 12-16 May 2008.




\vspace{1cm}
\hspace{9cm}
Angela Lepidi

\vspace{0.5cm}
\hspace{8cm}
Ferrara, 28th September 2008


%% file: files/Introduction.tex


\noindent {\huge \noindent  {\bf Introduction}}
\noindent  \addcontentsline{toc}{chapter}{{Introduction}}
\vspace{1.5cm}
\label{chap-intro}
\markboth{Introduction}{}


\noindent
The problem of whether the electric charge is quantized or not has been longly investigated, both from experimental and theoretical points of view. 
Indeed, there are several experimental observations and theoretical arguments suggesting that the electric charges of the elementary particles we know are integer multiples of a fundamental unity: the charge of d-quark ($-1/3 \, e$). 

Nevertheless, the existence of other particles carrying smaller electric charges, which electromagnetically interact with the ordinary photon, has not been ruled out so far; since these particles were supposed to have charge of order $10^{-3} \, e$, thay are usually referred to as millicharged particles, or MCPs.


In 1966 Kobzarev, Okun and Pomeranchuk proposed that there may be a mirror sector, that is a second particle sector identical to our one, but where weak interactions are right-handed; such a sector would provide to restore the strange lack of the left-right symmetry observed in nature.
$20$ years later, in 1986, Holdom proposed a mechanism to induce MCPs via the kinetic mixing of two photons; this mechanism naturally applies to theories containing a mirror sector. 

If existing, mirror MCPs would have effects on the cosmological evolution of the universe (big bang nucleosynthesis, structure formation, cooling of stars and so on) as well on laboratory experiments, for instance the laser ones or those searching for dark matter.
An other phenomenon where mirror MCPs may lead to observable effects is the positronium decay, as pointed out by Glashow in 1986.

The aim of these thesis is studying in detail the big bang nucleosynthesis (BBN) in models containing the mirror sector, which can be or not millicharged.

The first part of this work (chapters \ref{chap-The_standard_model} and \ref{chap-exp-univ}) consists in a review of basical concepts and formalisms used in particle physics and cosmology.
Chapter \ref{chap-The_standard_model} contains an introduction to the standard model of elementary particles, while chapter \ref{chap-exp-univ} is devoted to the standard model of cosmology.

The problem concerning the quantization of the electric charge in theoretical physics is dealt in chapter \ref{chap-Electric_charge_quantization}; three of the most challenging physics models (quantum standard model, grand unified theories and magnetic monopoles) are analyzed together with their predictions about the elementary particle electric charges; these charges end up to be quantized in all the three models.

In chapter \ref{The_mirror_universe} the photon kinetic mixing, which leads the mirror matter to be millicharged, is introduced; then a review of  the mirror theory and its cosmological features is given. We will argue that mirror MCPs are compatible with the theoretical bounds analyzed in the previous chapter.

The last two chapters are devoted to the original work which has been done for this thesis.
In chapter \ref{chap-mir-BBN} big bang nucleosynthesis in the mirror scenario is analyzed in detail and the primordial abundances of light elements are worked out, but without considering at the moment the possible existence of millicharged interactions.

Finally, in chapter \ref{chap-mir-MCP}, BBN in presence of the millicharged mirror sector is analyzed. After a review of the main processes leading to particle exchanges between the two sectors, we work out two approximated bounds on the photon mixing parameter $\epsilon$, which is related to the magnitude of the millicharges. Finally, we propose a scenario worth to be analyzed in detail in future researches.




%% file: files/SM.tex


\def \chap-The_standard_model{The standard model of elementary particles}
\chapter{\chap-The_standard_model}
\label{chap-The_standard_model}
\markboth{Chapter \ref{chap-The_standard_model}. ~ \chap-The_standard_model}
                    {Chapter \ref{chap-The_standard_model}. ~ \chap-The_standard_model}

The standard model of elementary particles (SM) is a field theory which describes electromagnetic, weak and strong forces by mean of the group $\mathcal{SU}(3)_c \times \mathcal{SU}(2)_L \times \mathcal{U}(1)_Y$ and contains the fundamental particles which make up all matter.   

In the literature the term "standard model" sometimes refers to the only electroweak sector, gauged by the group $\mathcal{SU}(2)_L \times \mathcal{U}(1)_Y$; nevertheless, if not differently specified, by SM we will mean the whole $\mathcal{SU}(3)_c \times \mathcal{SU}(2)_L \times \mathcal{U}(1)_Y$.
Its building required great efforts and contributes by various authors \cite{Glashow:1961tr,Weinberg:1967tq,Salam:1968}, who received the Nobel Prize in Physics in 1979. 

The SM had several experimental successes; among them, it predicted the existence of several particles unknown at that time which were afterwards experimentally observed, like the 3 massive bosons which mediate weak interactions - $W^{\pm}$ and $Z^0$ - gluons, charm and top quarks.

Nevertheless, the SM can not be considered a complete theory of fundamental interactions because it leaves many unanswered questions. For example, it does not include gravity, which is the fourth known fundamental force, and it contains a wide number of arbitrary parameters%
\footnote{There are at least 19: 3 gauge couplings - one possible choice is $\alpha_{em}$, $\alpha_s$ and $sin^2\theta_W$ -, 9 coupling constants of the Higgs boson with leptons and quarks, from which the fermion masses arise - see \S \ref{Coupling_of_the_leptons_to_the_Higgs}, 4 - three mixing angles and one phase - from the CKM matrix, 2 from the Higgs potential - one possible choice are the VEV $v$ and the quartic coupling $\lambda$ 
- and finally the QCD $\theta$ parameter. This number is still higher if neutrinos are massive particles, as indicated by several recent experimental and theoretical hints.}
%
%
%
, which are chosen to fit the experimental data but can not be derived from first principles.

In this section we briefly review the SM main features and its structure as a gauge theory. Wider treatments of these topics can be found in books, such as \cite{Leader:1996hk}. 

\def \sec-symmetries{Symmetries and conservation laws}
\section{\sec-symmetries}
\label{sec-symmetries}
\markboth{Chapter \ref{chap-The_standard_model}. ~ \chap-The_standard_model}
                    {\S \ref{sec-symmetries} ~ \sec-symmetries}

Let us consider a system specified by mean of several fields $\phi_j(x)$, $j=1,...,N$ and their gradients $\partial_{\mu}\phi_j(x)$; the theories we are considering can be derived applying a variational principle to the lagrangian density $\mathcal{L}$, which is a function of $\phi_j(x)$ and $\partial_{\mu}\phi_j(x)$:
	\begin{eqnarray}
	\label{General_lagrangian_density_in_FT}
	\mathcal{L} (\mathbf{x},t)= \mathcal{L} \left(\phi_j(x), \partial_{\mu}\phi_j(x) \right)
	\end{eqnarray}
The action integral $S (\Omega)$, where $\Omega$ is an arbitrary region of the four-dimensional space-time, is defined as:
	\begin{eqnarray}
	\label{}
	S (\Omega) = \int_{-\infty}^{+\infty} L(t) \;dt
	= \int_{\Omega} d^4x \mathcal{L} (\phi_j(x), \partial_{\mu}\phi_j(x))
	\end{eqnarray}
The variational principle we will use is analogous to Hamilton's principle in mechanics and states that for any variations of the fields vanishing on the region surface, that is for any
	\begin{eqnarray}
	\label{}
	\phi_j (x) \rightarrow \phi_j(x) + \partial \phi_j (x),
	\hspace{1.5cm}
	\partial \phi_j(x \in \hspace{.1cm} \Omega \hspace{.1cm} \mathrm{surface}) = 0 
	\end{eqnarray}
the action has a stationary value:
	\begin{eqnarray}
	\label{}
	\partial S (\Omega) = 0
	\end{eqnarray}
From this principle we can derive the Euler-Lagrange equations of motion:
	\begin{eqnarray}
	\label{}
	\frac{\partial \mathcal{L}}{\partial \phi_j} - 
	\frac{\partial}{\partial x^{\mu}}
	 \frac{\partial \mathcal{L}}{\partial (\delta_{\mu} \phi_j)} = 0
	\end{eqnarray}
In classic theories 
\footnote{Generalization to quantum theories is not trivial, see \S \ref{sec-Anomalies}.}
\bf Noether's theorem \rm holds \cite{Noether:1918}. 
It states that a conserved current defined as follows:
	\begin{eqnarray}
	\label{Noether_current}
	J^{\mu} = \frac{\delta \mathcal{L}}{\delta \partial_{\mu} \phi_a} \delta \phi_a
	\end{eqnarray}
is associated with each generator of a continuous symmetry in the lagrangian. Moreover, the charge defined as:
	\begin{eqnarray}
	\label{}
	Q (t) = \int d^3x J_0 (x)
	\end{eqnarray}
is a constant of the motion:
	\begin{eqnarray}
	\label{}
	\frac{dQ}{dt} = 0
	\end{eqnarray}
Derivations of the Noether theorem can be found in any book about quantum field theory - see for instance \cite{Cheng:1985bj,Kaku:1993ym,Mandl:1985bg}.

\def \sec-Gauge_symmetries{Gauge symmetries}
\section{\sec-Gauge_symmetries}
\label{sec-Gauge_symmetries}
\markboth{Chapter \ref{chap-The_standard_model}. ~ \chap-The_standard_model}
                    {\S \ref{sec-Gauge_symmetries} ~ \sec-Gauge_symmetries}


In the previous section we stressed the link of symmetries and conservation laws. We will show now that if the symmetry transformations are space-time dependent, they can be used to generate dynamics, that is, interactions between particles.
These local symmetries are also called \it gauge symmetries\rm .

\subsection{Gauge theories for abelian groups}
\label{Gauge_theories_for_abelian_groups}

A group is called \it abelian \rm if all the generators commute with each other. The simplest abelian group is the one-dimensional $\mathcal{U}(1)$, having the unity matrix as generator. Because of the important physical applications of 
this group - it enters both QED and SM - we will refer to the only $\mathcal{U}(1)$ in this section.

If a system is invariant under $\mathcal{U}(1)$, the fields can be changed by an arbitrary phase $\theta$:
	\begin{eqnarray}
	\label{global_abelian_field_tranformation}
	\phi_j(x) \rightarrow \phi'_j(x) = \exp^{-i q_j \theta} \phi_j (x)
	\end{eqnarray}
without any effects on the physically measurable quantities. In the equation above $q_j$ is the charge of the particle represented by the field $\phi_j$ in units of the electron's charge $e$. 

If $\theta$ does not depend on the position in the four-dimensional space-time, the theory is globally symmetric; if $\theta$ depends on $x$ instead, the theory is locally invariant and is called \it gauge theory\rm. In this case the transformation
%
	\begin{eqnarray}
	\label{local_abelian_field_tranformation}
	\phi_j(x) \rightarrow \phi'_j(x) = \exp^{-i q_j \theta(x)} \phi_j (x)
	\end{eqnarray}
leaves $\mathcal{L}$ unchanged.

Let us consider a lagrangian density $\mathcal{L} (\phi,\partial_{\mu} \phi)$ invariant under the global tranformation (\ref{global_abelian_field_tranformation}). 
The terms in $\mathcal{L}$ containing only the fields $\phi$ are also invariant under the local transformation (\ref{local_abelian_field_tranformation}), but the ones involving gradients are not because
	\begin{eqnarray}
	\label{gradient_transformation}
	\partial_{\mu} \phi_j (x) 
	\rightarrow \exp^{-i q_j \theta(x)} \partial_{\mu}\phi_j (x) + \ldots 
	\neq \exp^{-i q_j \theta(x)} \partial_{\mu}\phi_j (x)
	\end{eqnarray}
To generalize a global invariance to a local one therefore, the derivatives must be replaced by special combinations $D_{\mu}$ called \it covariant derivatives \rm which are chosen such that
	\begin{eqnarray}
	\label{}
	D_{\mu} \phi_j (x) \rightarrow \exp^{-i q_j \theta(x)} D_{\mu}\phi_j (x)
	\end{eqnarray}
The derivative $D_{\mu}$ depends on a field which is called \it gauge boson \rm and is used to reabsorbe the extra terms in (\ref{gradient_transformation}).

\subsubsection{QED}

In QED, which is gauged by the group $\mathcal{U}(1)$, the covariant derivative has the form:
	\begin{eqnarray}
	\label{}
	D_{\mu}^j = \partial_{\mu} + i e q_j A_{\mu}
	\end{eqnarray}
and the field $A_{\mu}$, which is the vector potential, transforms as
	\begin{eqnarray}
	\label{}
	A_{\mu} \rightarrow A_{\mu}' = A_{\mu} + \frac{1}{e} \partial_{\mu} \theta
	\end{eqnarray}
If we want the theory to be gauge invariant, we have to replace the simple derivative with the covariant one in the Dirac lagrangian for free spinors
	\begin{eqnarray}
	\label{Dirac_lagrangian_SM}
	\mathcal{L}_{Dirac} &=& \overline{\psi} (i \partial \!\!\! / - m) \psi \cr\cr
	&\rightarrow& \overline{\psi} (i D \!\!\!\! / - m) \psi =
	\overline{\psi} (i \partial \!\!\! / - e q_j A \!\!\!/ - m) \psi
	\end{eqnarray}
This way we get interactions between fermions and photons through the extra term $- e q_j \overline{\psi} A \!\!\!/  \psi$. This phenomenon is also present in theories gauged by more complicated groups - see \S \ref{Gauge_theories_for_non-abelian_groups}: in general transforming a global symmetry to a local one in a lagrangian for free fields , we get interactions between the fields and the gauge bosons.

\subsection{Gauge theories for non-abelian groups}
\label{Gauge_theories_for_non-abelian_groups}

Let us now consider a non-abelian  gauge group $G$ 
having a set of generators $T_j$ with $ j = 1,...,N_G$; the $T_i$s obey the commutation relations:
	\begin{eqnarray}
	\label{}
	[T_i,T_j] = i f_{ijk} T_k
	\end{eqnarray}
where $f_{ijk}$ are called the structure constants of the group and are antisymmetric under interchange of any pair of indices.

If the fields tranform according to some representations of $G$, the generators $T_j$ will be represented by $n \times n$ matrices $L_j$; the field transformations are specified by $N_G$ parameters which we will call $\theta_j$ and can be written as:
	\begin{eqnarray}
	\label{}
	\phi \rightarrow \phi' = \exp^{-i \, \mathbf{L \cdot \theta}} \phi
	\end{eqnarray}
where $\phi$ is a multiplet of fields
	\begin{eqnarray}
	\label{}
	\phi = 
	\left( \begin{array}{c}
	\phi_1 \\ \phi_2 \\ \vdots \\ \phi_n
	\end{array} \right)
	\end{eqnarray}
Similarly to the abelian case, the lagrangian can be invariant under a global transformation  $\mathcal{U} (\theta_j)$. But, if the parameters $\theta_j$ depend on the position $x$, the gradients contained in the lagrangian will break the invariance at local level. This problem is avoided defining a covariant derivative with the property
	\begin{eqnarray}
	\label{Non_abelian_covariant_derivative_property}
	D_{\mu} \phi (x) \rightarrow 
	D_{\mu}' \phi' (x) =  \mathcal{U} (\theta) D_{\mu}\phi (x)
	\end{eqnarray}
Assuming that the lagrangian contains gradients only through the covariant derivative, the invariance under local non-abelian gauge transformations is ensured.
Generalizing the procedure we adopted for QED, we can introduce one vector field $W^j_{\mu}(x)$ for each generator of the group and then write the covariant derivative as
	\begin{eqnarray}
	\label{Non_abelian_covariant_derivative}
	D_{\mu} \phi (x) = [\partial_{\mu} + i g \, \mathbf{L \cdot W}_{\mu} (x)] \phi (x)
	\end{eqnarray}
where $g$ plays the role of a coupling constant. Finally, if we impose that the property (\ref{Non_abelian_covariant_derivative_property}) must be valid also for $D_{\mu}$ defined in (\ref{Non_abelian_covariant_derivative}), we will derive how the gauge fields $W^j_{\mu}(x)$ must transform:
	\begin{eqnarray}
	\label{}
	\mathbf{L \cdot W}_{\mu}' = 
	\mathcal{U} (\theta) 
	\left[ \mathbf{L \cdot W}_{\mu} + \frac{i}{g} \, \mathcal{U}^{-1} (\theta)
	\partial_{\mu} (\theta) \mathcal{U} (\theta) 
	\right ] \mathcal{U}^{-1} (\theta)
	\end{eqnarray}
As disclosed at the end of \S \ref{Gauge_theories_for_abelian_groups}, passing from a global to a local symmetry induces interactions in free particle lagrangians for both abelian and non-abelian gauge theories. 
The SM is based on this formalism, as we will see in the following sections.

\def \sec-SSB{Spontaneous symmetry breaking}
\section{\sec-SSB}
\label{sec-SSB}
\markboth{Chapter \ref{chap-The_standard_model}. ~ \chap-The_standard_model}
                    {\S \ref{sec-SSB} ~ \sec-SSB}


Spontaneous symmetry breaking (SSB) occours when a lagrangian has some exact symmetry but the solutions of the problem do not. 
The \bf Goldstone theorem \rm states that the spontaneous breakdown of a continuous symmetry in field theory implies the existence of massless spinless particles. These particles are referred to as \it Nambu-Goldstone bosons \rm or simply \it Goldstone bosons\rm. Their number is equal to the number of the broken generators. The Goldstone theorem has been first studied by Nambu \cite{Nambu:1960xd, Nambu:1961fr,Nambu:1961tp}, and later proved by Goldstone and others \cite{Goldstone:1961eq,Goldstone:1962es}.

Dynamical systems in which the ground state does not possess the same symmetry properties of the lagrangian are very interesting for particle physics. Indeed it comes out that when SSB occours in a gauge theory involving massless vector fields and scalar fields, the Goldstone bosons disappear and re-emerge as the longitudinal mode of the vector fields, which therefore behave like massive particles (\it Higgs phenomenon\rm). This way we get the three massive bosons - $W^{\pm}$ are $Z^0$ - which mediate weak interactions.

\subsection{SSB of a global symmetry and Goldstone theorem}
\label{SSB_of_a_global_symmetry_and_Goldstone_theorem}

In this section an example of SSB in classical field theory is briefly reported. Let us consider a complex scalar field with the lagrangian:
	\begin{eqnarray}
	\label{Lagrangian_Goldstone}
	\mathcal{L}_{SSB} = (\partial_{\mu} \phi) (\partial^{\mu} \phi^*)
	-\mu^2 \phi \phi^* - \lambda (\phi \phi^*)^2
	\end{eqnarray}
which is invariant under the global $\mathcal{U} (1)$ transformation 
	\begin{eqnarray}
	\label{}
	\phi (x) \rightarrow \phi' (x) = \exp^{-i \theta} \phi (x)
	\end{eqnarray}
$\mathcal{L}$ can be thought as the sum of a kinetic term and a "potential" one:
	\begin{eqnarray}
	\label{}
	V (\phi) = \mu^2 \phi \phi^* + \lambda (\phi \phi^*)^2
	= \mu^2 \rho + \lambda \rho^2,
	\hspace{1.2cm}
	\rho \equiv \phi \phi^*
	\end{eqnarray}
Since the kinetic term in $\mathcal{L}$ vanishes for $\phi = const$, minimum points for $V$ are also minima of the total energy and are therefore called vacua or ground states. 
Clearly $V$ has a minimum only if $\lambda > 0$. Assuming $\lambda > 0$ and $\mu^2 > 0$, $V$ has one minimum in $\rho = 0$. 
If $\mu^2 <0$ instead, that is $\mu$ does not represent a physical mass for the field $\phi$, there are infinitely many minimum points for $V$ in
	\begin{eqnarray}
	\label{}
	\phi_{vac} = \frac{v}{\sqrt{2}} \exp^{i \Lambda}
	\hspace{1.cm} \mathrm{where} \hspace{1.cm}
	v = \sqrt{\frac{-\mu^2}{\lambda}}
	\end{eqnarray}
and $\Lambda$ is a real arbitrary number. 
All the minima are equivalent with each other, therefore anyone of them can be arbitratly chosen. In particular we can choose the one laying on the real axis, having $\Lambda = 0$, and therefore get:
	\begin{eqnarray}
	\label{phi_Goldstone}
	\phi (x) = \frac{1}{\sqrt{2}} [v + \xi (x) + i \chi (x) ]
	\end{eqnarray}
Sobstituting the field in (\ref{phi_Goldstone}) in the lagrangian in (\ref{Lagrangian_Goldstone}) we get, ignoring some constant terms:
	\begin{eqnarray}
	\label{}
	\mathcal{L} = \frac{1}{2} (\partial_{\mu} \xi)^2 + \frac{1}{2} (\partial_{\mu} \chi)^2 
	- \lambda v^2 \xi^2 -  \lambda v \xi (\xi^2 + \chi^2) - 
	\frac{1}{4} \lambda (\xi^2 + \chi^2)^2
	\end{eqnarray}
This way we rearranged $\mathcal{L}$ as a functions of two fields $\chi$ and $\xi$, where $\xi$ is massive:
	\begin{eqnarray}
	\label{}
	m_{\xi}^2 = 2 \lambda v^2
	\end{eqnarray}
The massless field $\chi$ is the Goldstone boson.

\subsection{SSB of a local symmetry and Higgs phenomenon}

We saw in \S \ref{Gauge_theories_for_abelian_groups} and \ref{Gauge_theories_for_non-abelian_groups} that the imposition of a local symmetry implies the existence of a certain number of massless gauge bosons. If we want these bosons to be massive and introduce this feature explicitly by a symmetry-breaking mass term, we will spoil the renormalizability of the theory. 
The high momentum limit of the propagators indeed is dominated by
	\begin{eqnarray}
	\label{}
	\frac{k_{\mu} k_{\nu} / m^2}{k^2 - m^2} \rightarrow \mathrm{const}
	\end{eqnarray}
which leads to divergent terms. 

A solution to this problem is found spontaneously breaking the involved symmetry. 
%
This way we get a certain number of Goldstone bosons, which we would like to eliminate. Remarkably the Goldstone theorem is evaded in gauge theories because its derivations requires axioms which are not all compatible with the gauge-fixing condition. 

What finally comes out is the disappearance of the Goldstone bosons, which re-emerge as longitudinal DOFs of the gauge bosons, which therefore becomes massive.

This phenomenon was first studied by Anderson \cite{Anderson:1963pc} for condensed-matter physics and later generalized to relativistic field theory by Englert and Brout \cite{Englert:1964et} and by Guralnik et al. \cite{Guralnik:1964eu}. But it is known as \it Higgs phenomenon \rm because this author gave the most complete treatment \cite{Higgs:1964ia,Higgs:1966ev}.
The first demonstration that field theories are renormalizable also in the presence of SSB has been provided by 't Hooft \cite{Hooft:1971rn}.

\subsubsection{A short example}

Let us consider again the lagrangian density (\ref{Lagrangian_Goldstone}); local invariance is achieved substituting ordinary derivatives with covariant ones; moreover, a kinetic term for the gauge bosons must be added. This way, the $\mathcal{U}(1)$ gauge invariant lagrangian becomes:
	\begin{eqnarray}
	\label{L_U(1)_Higgs}
	\mathcal{L} = &-& \frac{1}{4} F_{\mu\nu} F^{\mu\nu} + 
	\left[ \left(\partial_{\mu} - ie A_{\mu} \right) \phi^* \right]
	\left[ \left(\partial_{\mu} + ie A_{\mu} \right) \phi \right] \cr\cr
	&-& \mu^2 \phi \phi^* - \lambda ( \phi \phi^* )^2
	\end{eqnarray}
This lagrangian has in total four DOFs, two from the massless vector boson and two from the scalar complex field.
As for the globally invariant case, if we assume $\lambda > 0$ and $\mu^2 < 0$ there is a ring of degenerate ground states.
If we re-defyne the scalar field as:
	\begin{eqnarray}
	\label{}
	\phi (x) = \frac{1}{\sqrt{2}} \left[ v + \xi (x) + i \chi (x) \right]
	\hspace{1.5cm} v = \sqrt{\frac{-\mu^2}{\lambda}}
	\end{eqnarray}
and substitute it in the lagrangian (\ref{L_U(1)_Higgs}) we get:
	\begin{eqnarray}
	\label{}
	\mathcal{L} = - \frac{1}{4} F_{\mu\nu} F^{\mu\nu} + 
	\frac{e^2 v^2}{2} A_{\mu} A^{\mu} +
	\frac{1}{2} (\partial_{\mu} \xi)^2 + \frac{1}{2} (\partial_{\mu} \chi)^2 - \lambda v^2 \xi^2 + \ldots
	\end{eqnarray}
which seems to describe the interactions of a massive vector field - having three DOFs - and two scalars - having a total of two DOFs. Therefore we have in total five DOFs and not four as in the original lagrangian in eq.(\ref{L_U(1)_Higgs}). 
This one more degree of freedom is only apparent and can be eliminated choosing a particular gauge called the \it U gauge \rm in which the unphysical DOF is absorbed in the arbitrary field phase.

Let us see how this mechanism works in a $\mathcal{U}(1)$ gauge theory. Since the parameter $\theta (x)$ can be arbitrarly chosen, we can set it equal to the phase of $\phi (x)$ at each space-time point, so that:
	\begin{eqnarray}
	\label{}
	\phi ' (x) = \exp^{-i \theta (x)} \phi (x) = \frac{1}{\sqrt{2}} \left[ v + \eta (x) \right] 
	\end{eqnarray}
This way both $\phi'$ and $\eta$ are real and the lagrangian becomes
	\begin{eqnarray}
	\label{}
	\mathcal{L} = - \frac{1}{4} F'_{\mu\nu} F'^{\mu\nu} + 
	\frac{e^2 v^2}{2} A'_{\mu} A'^{\mu} +
	\frac{1}{2} (\partial_{\mu} \eta)^2 
	- \lambda v^2 \eta^2 + \ldots
	\end{eqnarray}
where
	\begin{eqnarray}
	\label{}
	A'_{\mu} (x) = A_{\mu} (x) + \frac{1}{e} \partial_{\mu} \theta (x) \cr\cr
	F'_{\mu\nu} = \partial_{\mu} A'_{\nu} - \partial_{\nu} A'_{\mu} 
	\end{eqnarray}
This way $\mathcal{L}$ describes the interactions of a massive vector boson $A'_{\mu}$ and a real scalar field $\eta$, called the \it Higgs boson \rm with mass
	\begin{eqnarray}
	\label{}
	m_{\eta}^2 = 2 \lambda v^2 = -2 \mu^2
	\end{eqnarray}
In conclusion, when a symmetry is spontaneously broken the gauge boson acquires a mass while the Goldstone boson disappears, leaving  the Higgs boson as the only present scalar field. The non-abelian extensions of this mechanism are called \it Yang-Mills theories \rm and have a great importance in the SM construction.

\def \sec-Parity_violation_weak_interactions{Parity violation in weak interactions}
\section{\sec-Parity_violation_weak_interactions}
\label{sec-Parity_violation_weak_interactions}
\markboth{Chapter \ref{chap-The_standard_model}. ~ \chap-The_standard_model}
                    {\S \ref{sec-Parity_violation_weak_interactions} ~ \sec-Parity_violation_weak_interactions}


Let us introduce the \it chirality projection operators: \rm $P_{R/L}$:
	\begin{eqnarray}
	\label{}
	P_R &=& \frac{1}{2} \left(1 + \gamma^5 \right) \hspace{1cm} \Longrightarrow \hspace{1cm}
	P_R \, \psi \equiv \psi_R \cr\cr
	P_L &=& \frac{1}{2} \left(1 - \gamma^5 \right) \hspace{1cm} \Longrightarrow \hspace{1cm}
	P_L \, \psi \equiv \psi_L
	\end{eqnarray}
where $ \psi_{R/L} $ are called respectively \it right- \rm and \it left-handed \rm components of a spinor. It can be shown that for a massive fermion handedness does not commute with the Dirac hamiltonian:
	\begin{eqnarray}
	\label{H_Dirac}
	H_{\mathrm{Dirac}} = \left( \mathbf{\alpha \cdot P} + \beta m \right) 
	\end{eqnarray}
	\begin{eqnarray}
	\label{Dirac_alpha_beta}
	\mathbf{\alpha} = \left( \begin{array}{cc} 0 & \mathbf{\sigma} \\ \mathbf{\sigma} & 0
	\\ \end{array} \right) \hspace{1.5cm}
	\beta = \left( \begin{array}{cc} 1_{2\times2} & 0 \\ 0 & - 1_{2\times2}
	\\ \end{array} \right) \hspace{1.5cm}
	\end{eqnarray}
and therefore it is not a good quantum number.

In 1956 Lee and Yang proposed that weak interactions may be not invariant under parity transformations \cite{Lee:1956qn}. Actually, the problem of wether parity is conserved or not  had been investigated in several contexts at that time  but Lee and Yang where the first who observed that there were no experimental tests of parity conservation in weak interactions. Moreover, there was evidence for the $K^+$ decay in two modes ($\tau - \theta$ paradox) in which the final states have opposite parities.

In their paper they also proposed several possible experimental tests; the first one was carried out by Wu et al. \cite{Wu:1957my}, who studied $\beta$ decay of $^{60}$Co. They found a clear violation of parity conservation: \it weak interactions are left-handed\rm.

Let us now introduce the quantum number \it helicity\rm , which is defined as:
	\begin{eqnarray}
	\label{}
	\mathbf{\Sigma \cdot \widehat{p}} =
	\left( \begin{array}{cc} \frac{\mathbf{\sigma \cdot p}}{|\mathbf{p}|}  & 0 \\
	0 & \frac{\mathbf{\sigma \cdot p} }{|\mathbf{p}|} \\ \end{array} \right)
	\end{eqnarray}
Helicity is therefore a measure of the allignement of the spin and momentum vectors; the corresponding helicity projection operators are:
	\begin{eqnarray}
	\label{}
	\Pi^{\pm} (\mathbf{p}) = \frac{1}{2} \left(1 \pm \mathbf{\Sigma \cdot \widehat{p}} \right)
	\end{eqnarray}
Helicity commutes with the Dirac hamiltonian (\ref{H_Dirac}) and is therefore a good quantum number for spinors.
It can be shown that in the high energy limit, that is for $m=0$, the helicity projection operators become:
	\begin{eqnarray}
	\label{}
	\lim_{m \rightarrow 0}\Pi^{\pm} (\mathbf{p}) = \frac{1}{2} \left(1 \pm \gamma^5 \right) = P_{R/L}
	\end{eqnarray}
that is, they coincide with the chirality projection operators.

\def \sec-Construction_of_the_SM{Construction of the SM} 
\section{\sec-Construction_of_the_SM}
\label{sec-Construction_of_the_SM}
\markboth{Chapter \ref{chap-The_standard_model}. ~ \chap-The_standard_model}
                    {\S \ref{sec-Construction_of_the_SM} ~ \sec-Construction_of_the_SM}


This section's aim is to introduce some methods and ideas which will be useful later on in this thesis. It does not give an exhaustive nor complete introduction to the standard model, which can be found in textbooks.

The \it electro-weak interactions \rm are gauged by the group $\mathcal{SU} (2) \times \mathcal{U} (1)$, having a total of $4$ generators: $Y$, that is the unit matrix $1$ divided by $2$, and $\tau_i, \; i=1,2,3$, that is the Pauli matrices $\sigma_i$ also divided by $2$. 

The color group $\mathcal{SU} (3)$, when gauged, gives rise to \it quantum chromodynamics - or QCD\rm , which describes the strong interactions of colored quarks and gluons. This group has $8$ generators $T_a$, $a=1, \ldots , 8$ which corresponds to $8$ gauge fields $G_i$, $i = 1,\ldots , 8$ which are called gluons. Since the color symmetry is unbroken, gluons do not acquire any masses. They do not take part in electro-weak interactions, that is they are singulets of the symmetry $\mathcal{SU} (2) \times \mathcal{U} (1)$, but they carry color. They can therefore interact with each other as well as with quarks, which carry color too, but not with leptons, which are color singulets. 

QCD has two peculiar properties; the first one, called \it Asymptotic freedom\rm , was discovered in the early 1970s by David Politzer \cite{Politzer:1973fx} and by Frank Wilczek and David Gross \cite{Gross:1973ju}. Asymptotic freedom means that in very high-energy reactions, quarks and gluons interact very weakly; this implies that in high-energy scattering the quarks move within nucleons, such as the neutron and proton, essentially as free, non-interacting particles.

The second property, called \it Confinement\rm , implies that the force between colored particles (such as quarks) does not diminish as they are separated. Because of this, it would take an infinite amount of energy to separate two of these particles, which thus can not be isolated and are confined in bound states of neutral net color. Although analytically unproven, confinement is widely believed to be true because it explains the consistent failure of free quark searches, and it is easy to demonstrate in lattice QCD.

Let us consider now the only electro-weak part of the theory; in a model with four generators we have four gauge bosons, which we call:
	\begin{eqnarray}
	\label{}
	B_{\mu} \hspace{5cm} 
	\mathrm{for} \; \mathcal{U} (1) 
	\cr\cr
	W_{\mu}^i \hspace{1.5cm} i=1,2,3  \hspace{1.5cm}  \mathrm{for} \; \mathcal{SU} (2) 
	\end{eqnarray}
If we want three of them to be massive, we need at least $4$ independent scalar fields. The simplest choice of them (\it minimal standard model\rm ) is a douplet of complex scalar fields:
	\begin{eqnarray}
	\label{}
	\phi = \left( \begin{array}{c} \phi^+ \\ \phi^0 \\ \end{array} \right)
	\end{eqnarray}
We want the model to include also leptons, which are electron, muon and tauon with the corresponding neutrinos and anti-particles, and quarks; neutrinos are massless
\footnote{There is now convincing evidence that neutrinos change from one flavor to another, what implies that neutrinos have non-zero masses. There are several possible ways to add neutrino masses to the model, for instance via a Dirac or a Majorana term. Nevertheless, theories containing such masses are considered beyond the standard model, which contains instead no mass terms for neutrinos.}%
, that is they are helicity eigenstates. Moreover they experience only weak interactions, which are - see \S \ref{sec-Parity_violation_weak_interactions} - left-handed; therefore the only left-handed part of the neutrino has to enter the model and leptons are represented as
\footnote{There are three lepton families in the SM; nevertheless, since they have the same properties with respect to the $\mathcal{SU} (3) \times  \mathcal{SU} (2) \times \mathcal{U} (1)$ symmetries, we consider for sake of semplicity the only electronic family with the corresponding light quarks $u$ and $d$.}%
:
	\begin{eqnarray}
	\label{SM-leptons}
	l_L = \left( \begin{array}{c} \nu_e \\ e \\ \end{array} \right)_L \sim (1,2,-1)
	\hspace{1.5cm}
	e_R = (e)_R \sim (1,1,-2)
	\end{eqnarray}
The first is a $\mathcal{SU} (2)$ douplet containing left-handed neutrino and electron 
which are at the present both massless. The second is a singulet representing the right-handed electron, also massless
. The brackets indicate the $\mathcal{SU} (3)$ and $\mathcal{SU} (2)$ content of the multiplets and their $\mathcal{U} (1)$ hypercharges.

Similarly $u$ and $d$ quarks enter the model as:
	\begin{eqnarray}
	\label{SM-left-q}
	q_L = \left( \begin{array}{c} u \\ d \\ \end{array} \right)_L \sim \left(3,2,\frac{1}{3}\right)
	\end{eqnarray}
	\begin{eqnarray}
	\label{SM-right-q}
	u_R = (u)_R \sim \left(3,1,\frac{4}{3}\right) \hspace{1.5cm}
	d_R = (d)_R \sim \left(3,1,- \frac{2}{3}\right)
	\end{eqnarray}
Moreover, leptons have a global lepton charge $L=1$ and a family lepton charge $L_e = 1$, while quarks have a baryon charge $B=\frac{1}{3}$, so that baryons consisting of three quarks have $B=1$.

The charge conjugation matrix $C$ makes the transformation
	\begin{eqnarray}
	\label{}
	f_R^c = C \gamma_0 f_L^* \hspace{1.5cm}
	f_L^c = C \gamma_0 f_R^*
	\end{eqnarray}
where $f$ stands for a generic fermion field - lepton or quark. The fields marked with $^c$ have opposite gauge charges as well as opposite chiralities with respect to fermions. Also barion and lepton numbers are assigned with opposite signs.

The tranformations of any fields $f$ under the gauge groups are:
	\begin{eqnarray}
	\label{Higgs_field_transformations}
	f \rightarrow f' &=& \exp^{-i \frac{1 Y_W \theta (x)}{2}} f
	\simeq \left(1 -i \frac{1 Y_W \theta (x)}{2} \right) f
	\hspace{1.cm} \Big(\mathrm{under} \: \mathcal{U} (1) \Big)
	\cr\cr\cr
	f \rightarrow f' &=& \exp^{-i I_W \frac{\mathbf{\sigma \cdot \theta} (x)}{2}} f
	\simeq \left(1 -i I_W \frac{\mathbf{\sigma \cdot \theta} (x)}{2} \right) f
	\hspace{.6cm} \Big(\mathrm{under} \: \mathcal{SU} (2) \Big)
	\hspace{0.6cm}
	\end{eqnarray}
where $Y_W$ and $I_W$ are the field eigenvalues of the hypercharge $Y$ and the weak isospin $I$. These numbers, together with the transformation properties under $\mathcal{SU}(2)_L$ and $\mathcal{SU}(3)_c$ and the electric charge are reported in Table \ref{tab-SM_quant_num}. 
	\begin{table}[htdp]
	\caption{The SM particle content.}
	\begin{center}
	\begin{tabular}{|c|c|c|c|c|c|c|c|}
	\hline 
	& $I_W$ & $Y_W = \mathcal{U}(1)_Y$ & $\mathcal{SU}(2)_L$ & $\mathcal{SU}(3)_c$ 
	& $\tau_{3} = I_{3W}$ & $Q = \tau_{3} + \frac{Y_W}{2}$ \\
	\hline
	$l_L$ & $\frac{1}{2}$ & $ - 1$ & 2 & 1 & 
	$\Biggl\{ \begin{array}{c}   \nu_L \hspace{0.9cm} \frac{1}{2} 
	\\ e_L \hspace{0.5cm} - \frac{1}{2} \end{array}$ 
	& $\begin{array}{c} 0 \\ - 1 \end{array}$ \\
	\hline 
	$e_R$ & 0 & -2 & 1 & 1 & 0 & -1 \\
	\hline
	$q_L$ & $\frac{1}{2}$ & $ \frac{1}{3} $ & 2 & 3 & 
	$\Biggl\{ \begin{array}{c}   u_L \hspace{0.9cm} \frac{1}{2} 
	\\ d_L \hspace{0.5cm} - \frac{1}{2} \end{array}$ 
	& $\begin{array}{c} \frac{2}{3} \\ - \frac{1}{3} \end{array}$ \\
	\hline 
	$u_R$ & 0 & $\frac{4}{3}$ & 1 & 3 & 0 & $\frac{2}{3}$ \\
	\hline
	$d_R$ & 0 & $-\frac{2}{3}$ & 1 & 3 & 0 & $- \frac{1}{3}$ \\
	\hline
	$\phi$ & $\frac{1}{2}$ & 1 & 2 & 1 &
	$\Biggl\{ \begin{array}{c} \phi^+ \hspace{0.9cm} \frac{1}{2} 
	\\ \phi^0 \hspace{0.5cm} - \frac{1}{2} \end{array}$ 
	& $\begin{array}{c} 1 \\ 0\end{array}$ \\
	\hline 
	\end{tabular}\\ 
	\end{center} 
	\label{tab-SM_quant_num}
	\end{table}
We analyze below the main interactions occourring in the model;
most of them naturally arise when we impose gauge invariance - and therefore replace ordinary derivatives with covariant ones - in the lagrangian:
	\begin{eqnarray}
	\label{}
	\mathcal{L} = \mathcal{L}_{Dirac} (l_L, e_R) 
	+  \mathcal{L}_{Maxwell} (B_{\mu} , W_{\mu}^i ) 
	+  \mathcal{L}_{SSB} (\phi) 
	\end{eqnarray}

\subsubsection{Coupling of the gauge bosons to the Higgs scalars: the electric charge}

Coupling  of the gauge bosons (GB) to the Higgs scalars (S) are found imposing gauge invariance to the SSB lagrangian, which therefore becomes:
%
	\begin{eqnarray}
	\label{L_GB_S}
	\mathcal{L}_{GB-S} = 
	\left\{ \left[ \partial_{\mu} + i \, \frac{g}{2} \mathbf{W_{\mu} \cdot \tau} +
	i \, \frac{g'}{2} 1 B_{\mu} \right] \phi \right\} ^{\dag}
	\times \Bigg\{ \ldots \Bigg\}
	- V (\phi^{\dag} \phi) 
	\end{eqnarray}
The potential involving the Higgs fields is the same than in \S \ref{SSB_of_a_global_symmetry_and_Goldstone_theorem} and its parameters $\lambda$ and $\mu^2$ are chosen so that there are infinitely many minima:
	\begin{eqnarray}
	V (\phi^{\dag} \phi) = 
	\mu^2 \phi^{\dag} \phi + \lambda (\phi^{\dag} \phi)^2
	\hspace{1.5cm}
	\lambda > 0 ,
	\hspace{.5cm}
	\mu^2 < 0 
	\end{eqnarray}
The most common choice for the ground state is:
	\begin{eqnarray}
	\label{}
	\phi_{vac} = \left( \begin{array}{c} 0 \\ \frac{v}{\sqrt{2}} \\ \end{array} \right)
	\end{eqnarray}
None of the infinitesimal transformations in (\ref{Higgs_field_transformations}) leaves it unchanged and thus it seems that all the guage bosons acquire a mass via the Higgs mechanism. But we know that the photon is massless! Paying more attention we can see that the combination of generators
	\begin{eqnarray}
	\label{}
	Q =  \tau_{3L} + \frac{Y}{2}
	\end{eqnarray}
has no effects on $\phi_{vac}$. To identify the massless vector boson - that is the photon - it is therefore useful to re-define the fields as:
	\begin{eqnarray}
	\label{B_W_mixing}
	B_{\mu} &=& \cos \theta_W A_{\mu} + \sin \theta_W Z_{\mu} \cr\cr
	W^3_{\mu} &=& \sin \theta_W A_{\mu} - \cos \theta_W Z_{\mu}
	\end{eqnarray}
where $\theta_W$ is the Weinberg angle, which is a free parameter with value $\sin^2 \theta_W \simeq 0.223$ chosen to fit the experimental data. Making the calculations we can see that $A_{\mu}$ couples to the only $Q$ if
	\begin{eqnarray}
	\label{}
	e = g \sin \theta_W = g' \cos \theta_W
	\end{eqnarray}
With these substitutions in the lagrangian in (\ref{L_GB_S}) we finally get a theory in which:
	\begin{itemize}
	\item $A_{\mu}$ remains massless because it couples to $\phi$ only through the unbroken generator $Q$.
	\item $Q$ measures the electric charge in units of $e$.
	\item The two charged generators defined as
		\begin{eqnarray}
		\label{}
		W_{\mu}^{\pm} = \frac{1}{\sqrt{2}} \left( W^1_{\mu} \mp W^2_{\mu}
		\right)
		\end{eqnarray}
	and the neutral one $Z_{\mu}^0$ becomes massive via the Higgs mechanism with
		\begin{eqnarray}
		\label{}
		M_W = \frac{gv}{2} \hspace{1.5cm} M_Z = \frac{gv}{2\cos \theta_W} 
		\end{eqnarray}
	\item The Higgs scalar which survives becomes massive with $m_H = \sqrt{-2\mu^2}$. 
	\end{itemize}

\subsubsection{Coupling of the gauge bosons to the leptons}

Interactions of the gauge bosons (GB) and the leptons (l) arise when imposing gauge invariance to the Dirac lagrangian for massless spinors:
	\begin{eqnarray}
	\label{}
	\mathcal{L}_{GB-l} = \bar l_L i \gamma^{\mu} 
	\left( \partial_{\mu} - \frac{1}{2} i g' B_{\mu} + \frac{1}{2} i g \mathbf{\tau \cdot W_{\mu}}
	\right) l_L
	+ \bar e_R i \gamma^{\mu} (\partial_{\mu} - i g' B_{\mu}) e_R
	\end{eqnarray}
Left-handed components of the leptons couple to charged bosons $W^{\pm}$:
	\begin{eqnarray}
	\label{}
	\mathcal{L}_{L-W^{\pm}}
	- \frac{g}{\sqrt{2}} \left(\bar\nu_L \gamma^{\mu} e_L W_{\mu}^+ +
	\bar e_L  \gamma^{\mu} \nu_L W_{\mu}^- \right)
	\end{eqnarray}
while both left- and right-handed components couple to the neutral bosons $A_{\mu}$ and $Z^0_{\mu}$:
	\begin{eqnarray}
	\label{}
	\mathcal{L}_{l-A} =
	- g \sin \theta_W \left( \bar e_L \gamma^{\mu} e_L + e_R \gamma^{\mu} e_R \right) A_{\mu} =
	-e J^{\mu}_{QED} A_{\mu}
	\end{eqnarray}
	\begin{eqnarray}
	\label{}
	\mathcal{L}_{l-Z} &=& 
	e \tan \theta_W \left[ \frac{1}{2} \csc^2 \theta_W 
	\left( \bar\nu_L \gamma^{\mu} \nu_L - \bar e_L \gamma^{\mu} e_L \right)
	+ \bar e \gamma^{\mu} e \right] Z_{\mu} \cr\cr
	&=& \frac{e}{2 \cos \theta_W \sin \theta_W} 
	\left[ l_3^{\mu} - 2 \sin^2 \theta_W J^{\mu}_{QED}
	\right] Z_{\mu}
	\end{eqnarray}
where we used the electro-magnetic current
	\begin{eqnarray}
	\label{}
	J^{\mu}_{QED} = \bar e \gamma^{\mu} e 
	\end{eqnarray}
and the third component of a triplet of weak-isospin currents:
	\begin{eqnarray}
	\label{}
	l_3^{\mu} = l_L \gamma^{\mu} \sigma_3  l_L
	\end{eqnarray}

\subsubsection{Coupling of the fermions to the Higgs: fermion masses and CP violation}
\label{Coupling_of_the_leptons_to_the_Higgs}

The last interaction we analyze arises from the \it Yukawa term\rm, which must be added by hand. Let us now take in account of all the three fermion generations; the Yukawa term has the form:
	\begin{eqnarray}
	\label{L_Yukawa_SM}
	\mathcal{L}_{Yukawa} &=& 
	- Y_e^{ij} \left( \bar l_{L,i} \phi e_{R,j} \right)
	- Y_d^{kl} \left( \bar q_{L,k} \phi d_{R,l} \right)
	- Y_u^{mn} \left( \bar q_{L,m} \phi u_{R,n} \right) + h.c. \cr\cr
	i,j &=& e,\mu,\tau \hspace{1.5cm} 
	k,l = d,s,b  \hspace{1.5cm} 
	m,n = u,c,t
	\end{eqnarray}
Diagonalizing the three matrices in (\ref{L_Yukawa_SM}) we get $9$ of the SM free parameters:
	\begin{eqnarray}
	\label{}
	Y_{e,d,u}^D =
	\left( \begin{array}{ccc}
	 y_{e,d,u} & 0  & 0  \\
	 0 & y_{\mu,s,c}  & 0  \\
	 0 & 0  &  y_{\tau,b,t} 
	\end{array} \right) 
	\end{eqnarray}
which are proportional to the fermion masses: assuming indeed $\phi = \phi_{vac}$ we get for instance from the leptonic term in (\ref{L_Yukawa_SM}) a mass for the electron ($\mu$, $\tau$):
	\begin{eqnarray}
	\label{}
	\mathcal{L}_{Yukawa-el}^{vac} =
	- \frac{Y_e v}{\sqrt{2}} \bar e e  
	\hspace{0.5cm} \Longrightarrow \hspace{0.5cm}
	m_e = \frac{Y_e v}{\sqrt{2}}
	\end{eqnarray}
without explicitely break the symmetry in the lagrangian by mean of a Dirac mass term. The same procedure applies to quarks; note that, in this model, the neutrino remains massless.
%
%


But we can not perform the diagonalization of the matrices $Y$ for free; indeed we must use four ausiliary matrices (per generation) $U_{L/R}$ and $V_{L/R}$ such that:
	\begin{eqnarray}
	\label{}
	Y_d^D= U_L^{\dag} Y_d U_R
	\hspace{1.5cm}
	Y_u^D= V_L^{\dag} Y_d V_R
	\end{eqnarray}
but the three quark fields (per generation) we have are not enough to re-absorbe all them; one of them survives and is responsible for the flavor violating charged currents (\it Cabibbo-Kobayashi-Maskawa or CKM matrix\rm ). 

This matrix leads the SM to have $4$ more physical parameters, one of which is a phase which makes the CKM matrix not real.
It can be shown that the CP invariance, which is valid in the leptonic sector of the SM, does not hold  in the hadronic sector because the presence of this phase of the CKM matrix.
A small CP violating effect was first observed in 1964 studying the $K^0$ and $\bar K^0$ decay 
- see \cite{Perkins:1982xb} for a more detailed review.




%% file: files/std_cosmology.tex

\def \chap-exp-univ{Standard cosmology}
\chapter{\chap-exp-univ}
\label{chap-exp-univ}
\markboth{Chapter \ref{chap-exp-univ}. ~ \chap-exp-univ}
                    {Chapter \ref{chap-exp-univ}. ~ \chap-exp-univ}

In this chapter some topics of standard cosmology which will be used in chapters \ref{chap-mir-BBN} and \ref{chap-mir-MCP} are summarised.
These topics can be also found in many textbooks and reviews - see for instance \cite{Kolb:1990vq,paddybook,Ciarcelluti:2003wm}.

\def \sec-universe{The expanding universe}
\section{\sec-universe}
\label{sec-universe}
\markboth{Chapter \ref{chap-exp-univ}. ~ \chap-exp-univ}
                    {\S \ref{sec-universe} ~ \sec-universe}

\def \sec-RW-metric{The Friedmann-Robertson-Walker metric}
\subsubsection{\sec-RW-metric}
\label{sec-RW-metric}

The {\sl cosmological principle} states that on large scales the Universe is to a good approximation homogeneous and isotropic. By homogeneity we mean that the geometrical properties are the same at all spatial locations, while by isotropy we mean that the geometrical properties do not sigle out any special direction in space.

Assumed that the cosmological principle holds, it is convenient to describe the universe using a particular coordinate system, appropried to the special class of observers for whom the universe appears isotropic.

The most general space-time metric describing a Universe compatible with the cosmological principle can be determined entirely by symmetry considerations without any references to the energy sources or Einstein's equations. It is called \it Friedmann-Robertson-Walker metric \rm (FRW) and can be written as:
	\begin{eqnarray}
	\label{frw}
	ds^2 = dt^2 - a^2(t) 
	\left[ {dr^2\over 1 - kr^2} + r^2 d\theta^2 + r^2 \sin^2\theta ~ d\phi^2 \right] \,,
	\end{eqnarray}
where ($t$, $r$, $\theta$, $\phi$) are {\sl comoving} coordinates
\footnote{ A ``comoving observer' follows the expansion of the Universe including the effects of any inhomogeneities that may be present.  }%
, $a(t)$ is the {\sl cosmic scale factor}, and, with an appropriate rescaling of the coordinates, {\sl k} can be chosen to be +1, $-$1, or 0 for spaces of constant positive, negative, or zero spatial curvature, respectively. The coordinate $r$ is dimensionless, while $a(t)$ has dimensions of length.
The conventions for signs we used are reported in appendix \ref{gen_rel}.

The value of $k$ and the form of the function $a(t)$ can not be determined by geometrical consideration; they have to be determined with Einstein's equations once the matter distribution is specified.

%
%

\def \sec-z-H{Redshift}
\subsubsection{\sec-z-H}

%
%

Let us consider an expanding universe. 
The light emitted by a distant object is described in the quantum mechanics scenario in terms of freely-propagating photons. Since the wavelength of a photon is 
	\begin{eqnarray}
	\label{}
	\lambda=\frac{h}{p}
	\end{eqnarray}
and the momentum $p$ changes in proportion to $a^{-1}$, the wavelength at time $t_0$, denoted as $\lambda_0$, will differ from that at time $t$, denoted as $\lambda$, by
	\begin{eqnarray}
	\label{l_redsh}
	{\lambda \over \lambda_0} = {a(t) \over a(t_0)}
	\end{eqnarray}
As the Universe expands, the wavelength of a freely-propagating photon increases, just as all physical distances increase with the expansion.

Hence, we introduce a new variable related to the scale factor $a$ which is directly observable, the {\bf redshift} of an object, $z$: 
%
	\begin{eqnarray}
	\label{redshift}
	z = {{\lambda_0 -\lambda} \over \lambda} \hspace{2cm} 1+z = {a(t_0) \over a(t)}
	\end{eqnarray}
where $\lambda _0$ is the detected wavelength and $\lambda$ is emitted wavelength.
Any increase (decrease) in $a(t)$ leads to a red shift (blue shift) of the light from distant sources. Since today astronomers observe distant galaxies to have red shifted spectra, we can conclude that the Universe is expanding.

\def \sec-H-law{The Hubble's law}
\subsubsection{\sec-H-law}

The Hubble's law is a linear relationship between the distance of an object and its observed red shift; it can be expressed as 
	\begin{eqnarray}
	\label{}
	z \approx Hd \approx 10^{-28} {\rm cm}^{-1} \times d
	\end{eqnarray}
where $d$ is the {\sl proper distance} \footnote{The proper distance of a point $P$ from another point $P_0$ is the distance measured by a chain of rulers held by observers which connect $P$ to $P_0$.} of a source, and $H$ is the {\sl Hubble constant} or, more accurately, the {\sl Hubble parameter} (because it is not constant in time, and in general varies as $t^{-1}$), defined by
	\begin{eqnarray}
	\label{}
	H = {\dot a(t) \over a(t)}
	\end{eqnarray}
%
%
%
%
%
%
At present time the Hubble parameter value $H_0$ is not known with great accuracy, so it is indicated by
	\begin{eqnarray}
	\label{consthubble}
	H_0 = 100 ~h ~ {\rm km ~ s^{-1}Mpc^{-1}} ~~~~~~~~~~~~~  h = 0.73 \pm 0.02
	\end{eqnarray}
where the dimensionless parameter $h$ contains the uncertainty on $H_0$ \cite{Tegmark:2006az}.

The present age and the local spatial scale for the Universe are set by the {\sl 
Hubble time}\footnote{Note that earlier than some time, say $t_X$, or better for $a$ less than some $a_X$, our knowledge of the Universe is uncertain, so that the time elapsed from $a = 0$ to $a = a_X$ cannot be reliably calculated. However, this contribution to the age of the Universe is very small, and most of the time elapsed since $a = 0$ accumulated during the most recent few Hubble times.} and {\sl radius}
	\begin{eqnarray}
	\label{}
	H_0^{-1} \simeq 9.778 \times 10^9 h ^{-1} ~{\rm yr} \simeq 3000 ~h ^{-1} ~{\rm Mpc} 
	\end{eqnarray}

\def \sec-Friedmann{The Friedmann equations and the equation of state}
\subsubsection{\sec-Friedmann}
\label{sec-Friedmann}

The expansion of the Universe is determined by the Einstein equations
	\begin{eqnarray}
	\label{}
	G_{\alpha \beta } \equiv  R_{\alpha \beta } - {1\over 2} R g_{\alpha \beta } =
	 {8\pi G \over c^4} T_{\alpha \beta } + \Lambda g_{\alpha \beta }
	\end{eqnarray}
where $R_{\alpha \beta}$ is the {\sl Ricci tensor}, $R$ is the {\sl Ricci scalar}, $g_{\alpha \beta}$ is the {\sl metric tensor}, $T_{\alpha \beta}$ is the {\sl energy-momentum tensor}, and $\Lambda$ is the {\sl cosmological constant} \footnote{The cosmological constant $\Lambda $ was introduced by Einstein for the need to have a static Universe. 
However, now we know for sure that the Universe is expanding. In fact, one of the biggest theoretical problems of the modern physics is to explain why the cosmological term is small and not order $ M_P^2 $.}.
For the FRW metric (\ref{frw}), they are reduced to the form (see appendix \ref{gen_rel})
	\begin{eqnarray}
	\label{friedmann1}
	{{\ddot a}\over a} = -{{4 \pi}\over 3}G {\left (\rho + 3p \right)}
	\end{eqnarray}
for the time-time component, and
	\begin{eqnarray}
	\label{friedmann2}
	{{\ddot a}\over a} + 2{\left( {\dot a} \over a \right)}^2 + 2{k \over a^2} = 4 \pi G {\left( \rho - p \right)}
	\end{eqnarray}
for the space-space components, where, if the cosmological constant is present, $ p $ and $ \rho $ are modified according to (\ref{p_rho_wl}). From equations (\ref{friedmann1}) and (\ref{friedmann2}) we obtain also
	\begin{eqnarray}
	\label{friedmann3}
	{\left( {\dot a} \over a \right)}^2 + {k \over {a^2}} = {{8 \pi}\over 3}G \rho
	\end{eqnarray}
Equations (\ref{friedmann1}) and (\ref{friedmann3}) are called the {\bf Friedmann equations}; they are not independent since the second can be recovered from the first if the adiabatic expansion of the Universe is taken into account.
The Friedmann equation (\ref{friedmann3}) can be recast as
	\begin{eqnarray}
	\label{friedmann4}
	{k \over {H^2 a^2}} = {\rho \over {3 H^2} / 8 \pi G} -1 \equiv  \Omega -1
	\end{eqnarray}
where $\Omega $ is the ratio of the density to the {\sl critical density} $\rho _c$ necessary for closing the Universe 
\footnote{The present value of the critical density is $\rho_{0c} = 1.88 ~h^2 \times 10^{-29}$ g cm$^{-3}$; taking into account the range of permitted values for $h$, we get $\rho_{0c} \sim (3 - 12) \times 10^{-27}$ kg m$^{-3}$, which corresponds to a few H atoms/m$^3$. Just to compare, a `really good' vacuum (in the laboratory) of $10^{-9}$ N$/$m$^{2}$ at 300 K contains $\sim 2\times 10^{11}$ molecules/m$^3$. The Universe seems to be empty indeed!} 
	\begin{eqnarray}
	\label{omega}
	\Omega \equiv  {\rho \over \rho _c} ~~~~~~,~~~~~~~~~~ \rho _c \equiv {{3 H^2} \over {8 \pi G}} 
	\end{eqnarray}
Since $ H^2 a^2 \geq 0 $, there is a correspondence between the sign of $ k $, and the sign of ($ \Omega - 1 $)
	\begin{center}
	CLOSED $\;\;\;\;\;\;\;\;\;\;\;\; k = +1\;\;\;\;\;\;\;\; \Longrightarrow \;\;\;\;\;\; \Omega > 1 $ \\
	FLAT $\;\;\;\;\;\;\;\;\;\;\;\;\;\;\;\;\; k = 0 \;\;\;\;\;\;\;\;\;\;\; \Longrightarrow \;\;\;\;\;\; \Omega = 1 $ \\ 
	OPEN $\;\;\;\;\;\;\;\;\;\;\;\;\;\;\;\; k = -1 \;\;\;\;\;\;\;\;\, \Longrightarrow \;\;\;\;\;\; \Omega < 1 $ 
	\end{center}
From equation (\ref{friedmann4}) we find for the scale factor today
	\begin{eqnarray}
	\label{a0}
	a_0 \equiv H_0^{-1} \left( k \over {\Omega _0 -1} \right)^{1/2} 
	\approx {{3000 ~h^{-1}~{\rm Mpc}}\over |\:{\Omega _0 -1}|^{1/2}}
	\end{eqnarray}
which can be interpreted as the current radius of curvature of the Universe. If $\Omega _0 = 1$, then $a_0$ has no physical meaning and can be chosen arbitrarily (it will always cancel out when some physical quantity is computed).

In order to derive the dynamical evolution of the scale factor $a(t)$, it is necessary to specify the {\bf equation of state} for the fluid, $p = p (\rho)$. It is standard to assume the form
	\begin{eqnarray}
	\label{eos}
	p = w \rho 
	\end{eqnarray}
and consider different types of components by choosing different values for $w$.

From equation (\ref{friedmann1}), models of the Universe made from fluids with $-1/3 < w < 1$ have $\ddot a$ always negative; then, because today $\dot a \geq  0$, they possess a point in time where $a$ vanishes and the density diverges; this instant is called the {\sl Big Bang singularity}. Note that the expansion of the Universe described in the Big Bang model is not due in any way to the effect of pressure, which always acts to decelerate the expansion, but is a result of the initial conditions describing a homogeneous and isotropic Universe. 

If the Universe is filled with pressureless {\sl non--relativistic matter} (dust), we have $p \ll \rho$ and thus $w = 0$.
Instead, for radiation, the {\sl relativistic ideal gas} equation of state $p = 1/3 \rho$ will be used, from which follows $w = 1/3$.
Another interesting equation of state is $p = - \rho$, corresponding to $w = -1$.
This is the case of {\sl vacuum energy} 
, which will be the relevant form of energy during the so--called inflationary epoch.

The $\alpha = 0$ component of the conservation equation for the energy--momentum tensor, $T^{\alpha\beta}_{~~;\beta} = 0$, gives the 1st law of thermodynamics:
	\begin{eqnarray}
	\label{FirstLaw2}
	d(\rho a^3) = - p d(a^3)
	\end{eqnarray}
where the change of energy in a comoving volume element is equal to minus the pressure times the change in the volume. 

From equations (\ref{eos}) and (\ref{FirstLaw2}), assuming $w$ independent of time, we can obtain the relation
	\begin{eqnarray}
	\label{conserv}
	\rho \propto a^{-3(1+w)} 
	\end{eqnarray}
Some special cases are:
	\begin{eqnarray}
	\label{}
	\mathrm{Radiation:} \hspace{1cm} p_R &=& \frac{1}{3} \rho_R \hspace{0.5cm}
	\Rightarrow \hspace{0.5cm} \rho_R \propto a^{-4} \cr\cr 
	\mathrm{Matter:} \hspace{1cm} p_M &=& 0 \hspace{1cm}
	\Rightarrow \hspace{0.5cm} \rho_M \propto a^{-3} \cr\cr 
	\mathrm{Vacuum \: energy:} \hspace{1cm} p_V &=& -\rho_V \hspace{0.5cm}
	\Rightarrow \hspace{0.5cm} \rho_V \propto \mathrm{const}
	\end{eqnarray}
The overall density in the Friedmann equation is:
	\begin{eqnarray}
	\label{}
	\rho = \rho_R + \rho_M + \rho_V
	\end{eqnarray}
At the very beginning, the universe was radiation dominated; then it became matter dominated and, in absence of vacuum energy, it will continue to be matter dominated.

\def \sec-flatmod{Flat models}
\section{\sec-flatmod}
\label{sec-flatmod}
\markboth{Chapter \ref{chap-exp-univ}. ~ \chap-exp-univ}
                    {\S \ref{sec-flatmod} ~ \sec-flatmod}

From equations (\ref{friedmann3}) and (\ref{conserv}) we get for $ k = 0 $ 
	\begin{eqnarray}
	\label{friedmann5}
	{\left( \dot a \over a_0 \right)}^2 = 
	H_0^2{\left[ \Omega_0 {\left( a_0 \over a \right)}^{1+3w} + {\left( 1 - \Omega_0 \right) } \right]} 
	\end{eqnarray}
Now we shall find the solution to this equation appropriate to a {\sl flat Universe}. For $\Omega_0 = 1$ integrating equation (\ref{friedmann5}) gives
	\begin{eqnarray}
	\label{aflat}
	a(t) = a_0 {\left( t \over t_0 \right)}^{2/3 \left( 1+w \right)}
	\end{eqnarray}
which shows that the expansion of a flat Universe is indefinetely long in time; equation (\ref{aflat}) is equivalent to the relation between the cosmic time $t$ and the redshift 
	\begin{eqnarray}
	\label{tflat}
	t = t_0 {\left( 1+z \right)}^{-3 \left( 1+w \right)/2}
	\end{eqnarray}
From equations (\ref{aflat}), (\ref{tflat}) and (\ref{conserv}), we can derive
	\begin{eqnarray}
	\label{Hflat}
	H & \equiv & {\dot a \over a} = 
	{2 \over 3 \left( 1+w \right)t} = H_0{t_0 \over t} = H_0 {\left( 1+z \right)}^{3 \left( 1+w \right)/2} ~  \;,\\
	\label{qflat}
	q & \equiv & -{{a \ddot a} \over \dot a^2} = {{1+3w} \over 2} = {\rm const.} = q_0 ~  \;,\\
	\label{t0flat}
	t_0 & = & {2 \over {3(1+w) H_0}} ~  \;,\\
	\label{rhoflat}
	\rho & = & \rho_0 { \left( t \over t_0 \right) }^{-2} = { 1 \over {6 (1+w)^2 \pi G t^2}} ~
	\end{eqnarray}
In appendix \ref{app-flat} we report the above relations for the special cases of a Universe dominated by matter or radiation. 
A general property of flat Universe models is that the scale factor $a$ grows indefinitely with time, with constant deceleration parameter $q_0$. 
The pressure role can be stressed again by observing that increasing $w$, and therefore the pressure, 
leads also to an increase of the deceleration parameter.

A cosmological model in which the Universe is empty of matter and has a positive cosmological constant is called the {\sl de Sitter Universe}. From equations (\ref{p_rho_wl}) 
and (\ref{friedmann3}) we obtain
	\begin{eqnarray}
	\label{}
	\left( {\dot a} \over a \right)^2 = {8 \pi G \over 3} \rho_\Lambda
	\end{eqnarray}
which for positive $\Lambda $ implies the exponentially fast expansion
	\begin{eqnarray}
	\label{a_exp}
	a(t) = a_0 \, e^{H (t - t_0)} ~,~~~~~~~~~~~ H = {8 \pi G \over 3} \rho_\Lambda = {\Lambda \over 3}
	\end{eqnarray}
corresponding to a Hubble parameter constant in time. In the de Sitter vacuum Universe test particles move away from each other because of the repulsive gravitational effect of the positive cosmological constant. 

Finally, the age of a flat Universe containing both matter and positive vacuum energy ($\Omega = \Omega_{\rm m} + \Omega_\Lambda = 1$) is
	\begin{eqnarray}
	\label{t0_flat_vacuum}
	t_0 = {2 \over 3 H_0} {1 \over \Omega _{\Lambda}^{1/2}} 
	\ln \left[{1 + \Omega_{\Lambda}^{1/2}} \over { \left( 1 -\Omega_{\Lambda} \right)^{1/2}} \right] 
	~,~~~~~~ \Omega_\Lambda = {\rho_\Lambda \over \rho_{\rm c}}
	\end{eqnarray}
It is interesting to note that, unlike previous models, a Universe with $\Omega _{\Lambda} \geq 0.74$ is older than $H_0^{-1}$ because the expansion rate accelerates. Also, when $\Omega _{\Lambda} \rightarrow 1$ the time $t_0 \rightarrow \infty $. 
For this reason the problem of reconciling a young expansion age with other independent age determinations (like for example the globular clusters) has often led cosmologists to invoke a cosmological constant.

\def \sec-eq-therm{Equilibrium thermodynamics}
\section{\sec-eq-therm}
\label{sec-eq-therm}
\markboth{Chapter \ref{chap-exp-univ}. ~ \chap-exp-univ} 
	{\S \ref{sec-eq-therm} ~ \sec-eq-therm}

In this section we will study the properties of the Universe considered as a thermodynamic system composed by different species (electrons, photons, neutrinos, nucleons, etc.) which, in the early phases, were to a good approximation in thermodynamic equilibrium, established through rapid interactions. Of course, going back to the past, the cosmic scale factor decreases while the temperature becomes higher.

To work out the physical processes at some time $t$, we need the distribution function $f_A$(\bf x\rm,\bf p\rm ,t)  of the present particle species. 
We assume, coherently with the hypothesis a homogeneous universe, that $f_A$ does not actually depend on  the coordinates \bf x\rm, therefore $f_A$(\bf x\rm,\bf p\rm ,t) $=f_A$(\bf p\rm ,t).

All interactions that involve elementary particles have a short range
\footnote{The only exception is the Coulomb force, that is anyway shielded in plasma by the Debye effect.}.
We therefore assume that the role of these interactions is providing a mechanism for thermalization without affecting the form of the distribution function (ideal gas approximation)
\footnote{The ideal gas approximation is valid at low densities, that is $\rho < \frac{3}{4\pi r^3}$ where $r$ is the average distance between two particles.}.
We can therefore use the equilibrium Bose-Einstein or Fermi-Dirac distribution functions:
	\begin{eqnarray}
	\label{eq_distr_funct}
	f_A(\mathbf{p},t)d^3\mathbf{p}
	=\frac{g_{spin_A}}{(2\pi)^3} 
	\cdot \frac{1}{\exp^{\frac{E_\mathbf{p}-\mu_A}{T(t)}} \pm 1} 
	d^3\mathbf{p}
	\end{eqnarray}
where $g_{spin_A}$ is the spin-degeneracy factor of the species $A$, $\mu_A$ is the chemical potential, the signs + and - correspond respectively to fermions and bosons and $E_p$ is given by the mass-shell equation:
	\begin{eqnarray}
	\label{mass_shell}
	E_p^2 = \mathbf{p}^2 + m^2
	\end{eqnarray}
From the distribution function in (\ref{eq_distr_funct}) it is straightforward to calculate the number density $n$, the energy density $\rho$ and the pressure $p$ for every particle species in thermal equilibrium:
	\begin{eqnarray}
	\label{number_density}
	n_A = \int f(\mathbf{k}) d^3\mathbf{k} = \frac{g_{spin_A}}{2\pi^2} 
	\int_{m_A}^{\infty}  \frac{\sqrt{E^2-m_A^2} \:E}{\exp^{\frac{E-\mu_A}{T}} \pm 1} \:dE
	\end{eqnarray}
	\begin{eqnarray}
	\label{energy_density}
	\rho_A = \int E f(\mathbf{k}) d^3\mathbf{k} = \frac{g_{spin_A}}{2\pi^2} 
	\int_{m_A}^{\infty}  \frac{\sqrt{E^2-m_A^2} \:E^2}{\exp^{\frac{E-\mu_A}{T}} \pm 1} \:dE
	\end{eqnarray}
	\begin{eqnarray}
	\label{pressure}
	p_A = \frac{1}{3}\int \mathbf{k}v(\mathbf{k})f(\mathbf{k}) d^3\mathbf{k} = \frac{g_{spin_A}}{6\pi^2} 
	\int_{m_A}^{\infty}  \frac{(E^2-m_A^2)^{\frac{3}{2}} }{\exp^{\frac{E-\mu_A}{T}} \pm 1} \:dE
	\end{eqnarray}

\def \sec-cosm-E-S{Entropy and energy}
\subsection{\sec-cosm-E-S}
\label{sec-cosm-E-S}
\markboth{Chapter \ref{chap-exp-univ}. ~ \chap-exp-univ} 
	{\S \ref{sec-cosm-E-S} ~ \sec-cosm-E-S}

Entropy per comoving volume can be defined, up to an additive constant, as (see \S \ref{entropy_derivation}):
	\begin{eqnarray}
	\label{entropy_generic}
	S_A(T) = V \cdot \frac{p_A(T) + \rho_A(T) - \mu_A n_A(T)}{T} 
	\end{eqnarray}
This quantity is conserved, that is
	\begin{eqnarray}
	\label{}
	dS=0
	\end{eqnarray}
We will assume below that $\mu \ll T$ and use the entropy density $s$, defined as:
	\begin{eqnarray}
	\label{entropy}
	s (T) &\equiv& \frac{S(T)}{V}
	= \frac{2\pi^2}{45}\: q_{tot}(T)\: T^3 
	\end{eqnarray}
	\begin{eqnarray}
	\label{q_tot__def}
	q_{tot}(T)&\equiv& \sum_{\tiny{bosons}} q_b (T) \left(\frac{T_b}{T}\right)^3 +
        \frac{7}{8}\sum_{\tiny{fermions}} q_f (T)
        \left(\frac{T_f}{T}\right)^3
	\end{eqnarray}
This way, the functions $q_{b/f}(T)$, called \it entropy DOFs\rm, contain all the integrals in (\ref{entropy_generic}). This formalism is convenient because, as we can see from equations (\ref{td_fcts_non_rev}),
the energy density and the pressure of a non relativistic species is exponentially smaller than that of a relativistic species and thus we can often ignore its contribution.

An analogous formalism can be used for the total energy density $\rho$, which can be defined as:
	\begin{eqnarray}
	\label{energy}
	\rho (T)&=& \frac{\pi^2}{30}\: g_{tot}(T)\: T^4 
	\end{eqnarray}
	\begin{eqnarray}
	\label{g_tot__def}
	g_{tot} (T)&\equiv& \sum_{\tiny{bosons}} g_b (T) \left(\frac{T_b}{T}\right)^4 +
        	\frac{7}{8}\sum_{\tiny{fermions}} g_f (T)
        	\left(\frac{T_f}{T}\right)^4
        	\end{eqnarray}
The functions $g_{b/f}(T)$ are analogous to the functions $q_{b/f}(T)$ and are called \it energy DOFs\rm.
For ultrarelativistic species $q(T)$ and $g(T)$ are constants of the temperature and we have:
	\begin{eqnarray}
	\label{ultrarel_DOFs}
	g_{spin_A}&=&g_{A}=q_{A} \cr\cr
	g_{\nu}= 2 \cdot 3 = 6 \hspace{1.5cm} g_{\gamma}&=&2  
	\hspace{1.5cm} g_{e^{\pm}} = 2 \cdot 2 = 4
	\end{eqnarray}
When photons, neutrinos and electron-positrons are in thermal equilibrium, that is $T \gg 2\div 3 MeV$, the number of DOFs is easy to work out:
	\begin{eqnarray}
	\label{std_DOF_tot_eq}
	g_{tot}=q_{tot}= 2+\frac{7}{8} \cdot (4+6)=10.75
	\end{eqnarray}
Also, when the annihilation process is over, we get:
	\begin{eqnarray}
	\label{std_DOF_tot_after_ann}
	\frac{T_{\nu}}{T}= \left( \frac{4}{11}\right)^{\frac{1}{3}} 
	\Rightarrow g_{tot} &=& 2 + \frac{7}{8} \cdot 6 \cdot \left( \frac{4}{11}\right)^{\frac{4}{3}} 
	\simeq 3.36 \cr\cr
	q_{tot} &=& 2 + \frac{7}{8} \cdot 6 \cdot \left( \frac{4}{11}\right) \simeq 3.91
	\end{eqnarray}
The temperature factor $\left( \frac{4}{11}\right)^{\frac{1}{3}}$ for neutrinos will be calculated in \S \ref{sec-nu-dec}.

In the following we shall omit the index $tot$ for semplicity, while we will always explicitely write down the index when considering a specific species (for instance $g_{e}$ stands for electron energy DOFs).

\subsubsection{Time - temperature relationship}

%
From the definition $H=\frac{\dot{a}}{a}$ we get:
	\begin{equation}
	t \; = \; \int_0^{a(t)} {1 \over H} \, {da \over a} \;.
	\end{equation}
By using the equations (\ref{friedmann3}), neglecting the curvature term in the radiation dominated era and recalling equation (\ref{energy}) we get:
	\begin{equation}
	\label{H_Mp}
	H \; \simeq \; \sqrt{{4 \pi^3} \over {45 M_P^2}} \, g^{1/2} \, T^2 ~ \simeq ~ 1.66 \; 
	g^{1/2} {T^2 \over M_P} \;,
	\end{equation}
where we used $G \sim M_P^{-2}$. 
From (\ref{H_Mp}) we can derive the relationship between the time $t$ and the background (photon) temperature $T$, using the entropy conservation and assuming both $g$ and $q$ approximately constant. 
This way we find
	\begin{equation}
	\label{t_vs_T}
	t \; \simeq \; 0.301 \; g^{- 1/2} \; {M_P \over T^2} \; \; \Longrightarrow 
	\; \; t \sim \left(\frac{T}{MeV}\right)^{-2} ({\rm sec})
	\end{equation}

\def \sec-th-eq{Thermal equilibrium criterion}
\subsection{\sec-th-eq}
\label{sec-th-eq}
\markboth{Chapter \ref{chap-exp-univ}. ~ \chap-exp-univ} 
	{\S \ref{sec-th-eq} ~ \sec-th-eq}


The correct way to evolve particle distributions is to integrate the Boltzmann equation; nevetheless a rough criterion to evaluate whether a particle species is in thermal equilibrium or not can be found considering that a species which is not in equilibrium at time $t$ will never thermalize if every particle has in average less than one interaction from $t$ to $\infty$. 

Let us consider two interacting particles species, one of which will be treated as a target. The interaction rate is defined as:
	\begin{eqnarray}
	\label{interaction_rate}
	\Gamma \equiv n\langle v \sigma \rangle 
	\end{eqnarray}
where $n$ is the number density of terget particles, $v$ is the relative velocity and $\sigma$ is the interaction cross section. If the particles are in thermal equilibrium with the photon bath, we can assume that $\Gamma$ is some power of the photon temperature $T$; therefore, using equation (\ref{t_vs_T}) we have:
	\begin{eqnarray}
	\label{}
	\Gamma \propto T^n \propto t^{-n/2}
	\end{eqnarray}
From equation (\ref{H_Mp}) we can see that the Hubble constant $H\propto t^{-1}$. Therefore the number of interactions per particle for time going from $t$ to $\infty$ is:
	\begin{eqnarray}
	N_{int} &=& \int_t^{\infty} \Gamma(t') ~dt' 
	\propto \int_t^{\infty} T(t')^n ~dt' \cr\cr
	&\propto& \int_t^{\infty} t'^{-n/2} ~dt' \propto t^{-n/2+1} \propto 
	\frac{\Gamma}{H}
	\end{eqnarray}
Therefore the commonly used rule:
	\begin{eqnarray}
	\label{eq_criterion}
	\Gamma(T)> H(T) \Longrightarrow \rm{implies \; thermal \; equilibrium}
	\end{eqnarray}
from which we can define the \it decoupling temperature \rm $T_D$ as:
	\begin{eqnarray}
	\label{T_dec_def}
	\frac{\Gamma(T_D)}{H(T_D)} =1
	\end{eqnarray}
Note that $\Gamma(T) < H(T)$ does not imply departure from thermal equilibrium because a non-interacting species once in equilibrium will ever keep an equilibrium distribution.

\def \sec-nu-dec{Neutrino decoupling and $e^{\pm}$ annihilation}
\section{\sec-nu-dec}
\label{sec-nu-dec}
\markboth{Chapter \ref{chap-exp-univ}. ~ \chap-exp-univ} 
	{\S \ref{sec-nu-dec} ~ \sec-nu-dec}


In the early universe neutrinos are kept in equilibrium via the reactions $\bar{\nu} \nu \leftrightarrow e^+ e^-$, $\nu e \leftrightarrow \nu e$ etc. 
Let us assume that all the involved particles are ultra-relativistic. The cross sections - see \S \ref{sec-nu-scat} - are of order:
	\begin{eqnarray}
	\label{}
	\sigma \simeq G_F^2 s \simeq G_F^2 T^2
	\end{eqnarray}
where $G_F$ is the Fermi constant $G_F \simeq 1.1664 \cdot 10^{-5} GeV^{-2}$ and we used $s  \propto T^2$ as in \S \ref{s_average}. Since $n \simeq T^3$ and $c=1$ the interaction rate defined in (\ref{interaction_rate}) becomes:
	\begin{eqnarray}
	\label{Gama_nu_dec}
	\Gamma \simeq G_F^2 T^5
	\end{eqnarray}
Using (\ref{H_Mp}) and (\ref{Gama_nu_dec})  in the definition of the decoupling temperature given in (\ref{T_dec_def}) we can see that the neutrinos decouple at the temperature $T_D$ given by:
	\begin{eqnarray}
	\label{}
	\frac{\Gamma (T_D)}{H (T_D)} \simeq \frac{G_F^2 T_D^5}{T_D^2/M_P} = 1 
	\hspace{0.8cm}  \Longrightarrow \hspace{0.8cm} 
	T_D \simeq 1MeV
	\end{eqnarray}
Therefore at $T \gg 1MeV$ neutrinos are in thermal equilibrium with the plasma and their temperature is $T_{\nu} = T$, while after decoupling their temperature scales as $a^{-1}$. 

Shortly after the $\nu$ decoupling, the $e^{\pm}$ annihilate because $T$ becomes smaller than $2 m_e$, which is the threshold for the reaction $\gamma \leftrightarrow e^+e^-$. Thus electrons and positrons transfer their entropy to photons, which become hotter than neutrinos.

A more quantitative understandment of these temperature relationships will be necessary in chapters \ref{chap-mir-BBN} and \ref{chap-mir-MCP} and can be found below.

\subsubsection{Neutrino temperature after decoupling}

After the $\nu$s decoupling
, the $\nu$s' and the remenant ultrarelativistic
particles' entropies $S$ are separately conserved:
	\begin{eqnarray}
	\label{entr_cons_e_annich}
	S_{\nu}(T) &=& s_{\nu}(T) \cdot a^3 = const = c_1 \cr\cr
	S_{\gamma\: e^{\pm}} (T) &=& s_{\gamma\: e^{\pm}} (T) \cdot a^3 = const = c_2
	\end{eqnarray}
The ratio of these entropies is also a constant, which we will call $c_3$; in the ratio the acceleration factor $a^3$ simplyfies, giving
	\begin{eqnarray}
	\label{ratio_s_nu_egamma}
	\frac{s_{\nu}  (T_{\nu})}{s_{\gamma\: e^{\pm}}(T)} = const = c_3
	\end{eqnarray}
When the $e^{\pm}$ annihilation takes place, photons and neutrinos are ultrarelativistic; therefore their entopy densities can be worked out from equations (\ref{entropy}) and (\ref{ultrarel_DOFs}):
	\begin{eqnarray}
	\label{s_nu_gamma}
	s_{\nu} (T_{\nu}) &=& \frac{2\pi^2}{45}\: \frac{7}{8}6\: T_{\nu} ^3 =  \frac{7}{2}a_B\: T_{\nu} ^3\cr\cr
	s_{\gamma}(T) &=& \frac{2\pi^2}{45}\: 2\: T ^3=\frac{4}{3} a_B T^3
	\end{eqnarray}
where $T$ is the photon temperature and $a_B$ is the 
radiation constant:
	\begin{eqnarray}
	a_B &=& \frac{\pi^2}{15} \;\;\;\; (h\!\!\! /=c=1 \; \rm{units}) \cr\cr
	&=& \frac{\pi^2 \,k_B^4}{15 (h\!\!\! / c)^3}  \;\;\;\; \rm{(SI \: units)}
	\end{eqnarray}
Electrons and positrons are non relativistic but in equilibrium with the photons; their entropy density can be therefore worked out using equations (\ref{energy_density}), (\ref{pressure}) and (\ref{entropy_generic}):
	\begin{eqnarray}
	\label{electron_entropy_density}
	s_{ e^{\pm}} (T) 
	=  \frac{ 2}{\pi^2 T} \left(
	\frac{1}{3}
	\int_{m_e}^{\infty} \frac{\sqrt{(E^2-m^2)^3}}{\exp^{\frac{E- \mu}{T}} + 1}\: dE\: +
	\int_{m_e}^{\infty} \frac{\sqrt{(E^2-m^2)}E^2}{\exp^{\frac{E- \mu}{T}} + 1}\: dE\: 
	\right)
	\end{eqnarray}
where we assumed $\mu_{ e^{\pm}}  = 0$. 
Equation (\ref{ratio_s_nu_egamma}) finally becomes, using the entropy densities in (\ref{s_nu_gamma}) and (\ref{electron_entropy_density}):
	\begin{eqnarray}
	\label{entropy_equality}
	 \frac{7}{2}a_B\: T_{\nu} ^3 = 
	\frac{1}{c_3} \left[ \frac{2\pi^2}{45}\: 2\: T ^3 + s _{ e^{\pm}} (T)  \right] 
	= \frac{1}{c_3}  \frac{4}{3} a_B T^3 \left[f \left(\frac{m_e}{T}\right) \right]^3
	\end{eqnarray}
where the function $f$ is 
	\begin{eqnarray}
	\label{f_function}
	[f(x)]^3 = 1 + \frac{15}{2 \pi^4} \int_0^{\infty} \frac{y^2}{\sqrt{x^2+y^2}} \frac{3x^2 + 4 y^2}{e^{\sqrt		{x^2+y^2}}+1} \; dy
	\end{eqnarray}
and the new variables $x$ and $y$ are defined as
	\begin{eqnarray}
	\label{change_var_x_y}
	x \equiv \frac{m_e}{T} \hspace{2cm}
	y \equiv \frac{p_e}{T} 
	\end{eqnarray}
Using the same formalism it is straightforward to write the electron-positron energy density as:
	\begin{eqnarray}
	\rho_{e^{\pm}} = \frac{2}{\pi^2} 
	 \int_{m_e}^{\infty} \frac{\sqrt{(E^2-m^2)}E^2}{\exp^{\frac{E- \mu}{T}} + 1}\: dE 
	 = a_B \:T^4 \:\frac{30}{\pi^4}  \int_{0}^{\infty} \frac{\sqrt{(x^2+y^2)}y^2}{\exp^{\sqrt{x^2		+y^2}} + 1}\: dy
\end{eqnarray}
and therefore the total energy density as a function of $x$ and $T$:
	\begin{eqnarray}
	\rho (T, x) = \frac{a_B T^4}{2} \left[ 2 + \frac{42}{8} \left(\frac{4}{11}\right)^\frac{4}{3} [f(x)] ^4 + \frac		{60}{\pi^4}  \int_{0}^{\infty} \frac{\sqrt{(x^2+y^2)}y^2}{\exp^{\sqrt{x^2+y^2}} + 1}\: dy
 	\right]
	\end{eqnarray}
The first term is the photon's energy, the second one is the neutrinos' and the third is the electron-positron's. The neutrino term is weighted by the number of degrees of freedom $\frac{7\cdot 6}{8}$ and the temperature factor. \\

Equation (\ref{entropy_equality}) totally determines the neutrino temperature $T_{\nu}$ as a function of the photon temperature $T$ once the initial conditions are specified by mean of the costant $c_3$. \\
It is worth to stress that the status of the annihilation process is totally contained in the $f$ function as the ratio $x = \frac{m_e}{T}$.  Its extreme cases are:
	\begin{itemize}
	\item $x \ll 1 \Longrightarrow T \gg m_e$, that is ultra-relativistic electron. 
	In this limit the reactions that convert  $e^{\pm}	$ in photons and viceversa are in equilibrium and
	the annihilation process is not begun yet. In 
	this limit $f$ becomes:
		\begin{eqnarray}
		[f(x<<1)]^3 \simeq 1 + \frac{15}{2 \pi^4} \int_0^{\infty} \frac{y^3}{e^{y}} \; dy \simeq \frac{11}{4} 
		\end{eqnarray}
	%
%
	\item $x \gg 1 \Longrightarrow T \ll m_e$. The $e^{-x}$ leads the integral to $0$ giving
		\begin{eqnarray}
		[f(x>>1)]^3 \simeq 1
		\end{eqnarray}
	This limit describes the situation we have after the end of the annihilation process.
	\end{itemize}

\subsubsection{Working out the constant $c_3$: the instantaneous decoupling approximation}

Let us assume that the neutrino decoupling is an instantaneous process taking place when photons have the temperature $T=T_{D\nu}$; before the decoupling, neutrinos and photons have the same temperature and therefore
	\begin{eqnarray}
	\label{c_3_instantaneous}
	c_3=\frac{\frac{7}{8}q_{e}(T)+q_{\gamma}}{\frac{7}{8}q_{\nu}
	\left(\frac{T_{D\nu}}{T
	}\right)^3} 
	=\frac{\frac{7}{8}q_{el}(T_{D\nu})+2}{\frac{7}{8}\cdot 6}
	\end{eqnarray}
%
The value of the constant $c_3$ is affected by the difference of $T_{D\nu}$ and the $e^{\pm}$ annihilation temperature $T_{ann}$. 
If these temperatures are close indeed and the $e^{\pm}$ annihilation begins before the $\nu$ decoupling is completed, the $e^{\pm}$ transfer a fraction of their entropy to neutrinos, raising their temperature. 

The asymptotic values of $c_3$ are:
	\begin{eqnarray}
	\label{c_3_asympt}
	c_3=\frac{\frac{7}{8}\cdot 4+2}{\frac{7}{8}\cdot 6} =  \frac{22}{21} = 1.048 \hspace{1.5cm}
	T_{D\nu} \gg T_{ann}\cr\cr
	c_3=\frac{2}{\frac{7}{8}\cdot 6} =  \frac{8}{21} = 0.38 \hspace{1.5cm}
	T_{D\nu} \ll T_{ann}
	\end{eqnarray}
Tipical values for $T_{D\nu}$ are $1\div3 \;MeV$; numerical calculations show that in this range $c_3=\frac{22}{21}$ within 2\% of error. Therefore in chapters \ref{chap-mir-BBN} and \ref{chap-mir-MCP} we will just use $c_3=\frac{22}{21}$.

Using this value for $c_3$ we also get the standard asymptotic ratio of the neutrino and the photon temperature:
%
	\begin{eqnarray}
	\label{}
	T_{\nu} = \left( \frac{4}{11} \right)^{\frac{1}{3}} T
	\end{eqnarray}
which is valid when $T \ll T_{ann}$.

\def \sec-std-BBN{Primordial nucleosynthesis (BBN)}
\section{\sec-std-BBN}
\label{sec-std-BBN}
\markboth{Chapter \ref{chap-exp-univ}. ~ \chap-exp-univ} 
	{\S \ref{sec-std-BBN} ~ \sec-std-BBN}

The Big Bang Nucleosynthesis (BBN) is the earliest test of the cosmological and of the particle interaction models. 
Energy considerations suggest that light nuclei could be formed when the temperature of the universe is in the range $1-30MeV$; nevertheless, we will demonstrate below that, because of the high entropy of the universe, the actual synthesis takes place at a much lower temperature, $T_N \sim 0.1MeV$. 
%
%

Let us start introducing the \it mass fraction \rm of a certain nuclear species, which is defined as
	\begin{eqnarray}
	\label{}
	X_A \equiv \frac{n_A A}{n_N} 
	\hspace{0.5cm} , \hspace{0.5cm} \sum_i X_i = 1
	\end{eqnarray}
where $A$ is the atomic number, $N$ stands for nucleons and $n_A$ is the number density
- see \S \ref{sec-eq-therm} and \S \ref{Td_ultrarev_ptc}. $X_A$ can be recasted as
%
%
%
	\begin{equation}
	\label{X_A}
	X_A = g_A A^{\frac{5}{2}} 2^{\frac{3A-5}{2}} \zeta_3^{A-1}
	\pi^{\frac{1-A}{2}} 
	\left(\frac{T}{m_N}\right) ^{3\frac{A-1}{2}}
	\eta^{A-1} X_p^Z X_n^{A-Z}
	\exp^{\frac{B_A}{T}}
	\end{equation}
where we neglected the difference in mass of proton ($p$) and neutron ($n$). In equation (\ref{X_A}) $B_A$ is the binding energy of the nuclear species $A$, defined as:
	\begin{eqnarray}
	\label{}
	B_A \equiv Z m_p + (A-Z) m_n - m_A
	\end{eqnarray}
This quantity has values which vary from $2.22 MeV$ (for $^2H$) to $92.2 MeV$ (for $^{12}C$), corresponding to a binding energy per nucleon of order of $1$ to $8 MeV$; finally $\eta$ is the barion to photon ratio in the universe and is proportional to $s^{-1}$: 
	\begin{eqnarray}
	\label{eta}
	\eta &=& \frac{n_B}{n_{\gamma}} =
	\frac{\pi^4}{45\zeta_3} \: \frac{n_B q(T)}{s} \simeq 1.8 \frac{n_B q(T)}{s}
	\cr\cr 
	n_B &=& n_b - n_{\overline{b}} 
	\simeq n_{N_{today}} 
	\end{eqnarray}
The main elements produced during BBN are $D$, $^3He$, $^4He$, $^7Li$ with a predominance of $^4He$, having a mass fraction of about $24\%$ at the end of BBN.
Among the light elements, $D$ and $^4 He$ play a crucial role because there are apparently no astrophysical processes that can account for their observed abundances;
it is therefore nowadays possible to measure to high precision their primordial abundances and, by comparing the measured and the theoretical values, it is possible to test the cosmological model and also to work out bounds on MCP - see \S \ref{sec-MCP-bounds} and chapter \ref{chap-mir-MCP}.

We can infer from equation (\ref{X_A}) that, although the binding energies per nucleon are $1 \div 8 MeV$, the equilibrium abundance of nuclear species do not become of order unity until a temperature of order $0.3 MeV$ because the high entropy of the universe leads to low values of $\eta$ - see equation (\ref{eta}). A rough estimate of when a species becomes thermodinamically favored can be worked out assuming $X_n \sim X_p \sim 1$ and solving for $X_A \sim 1$, from which follows
	\begin{equation} 
	\label{T_A} 
	T_{A} \simeq {{B_{A}}/(A-1) 
	\over {-\ln(\eta)+1.5\ln(m_{N}/T)}}
	\end{equation}
This temperature is much lower than the binding energy per nucleon $B_A/A$: for instance
	\begin{eqnarray}
	\label{T_N}
	T_N = T_{^2D} &=& 0.07 MeV \hspace{1.5cm} \frac{B_D}{2} = 1.11 MeV \cr\cr
	T_{^3He} &=& 0.11 MeV \hspace{1.5cm} \frac{B_{^3He}}{3} = 2.57 MeV \cr\cr
	T_{^4He} &=& 0.28 MeV \hspace{1.5cm} \frac{B_{^4He}}{4} = 7.07 MeV \cr\cr
	T_{^{12}C} &=& 0.25 MeV \hspace{1.5cm} \frac{B_{^{12}C}}{4} = 7.68 MeV
	\end{eqnarray}
Therefore the high universe entropy favour free nucleons for $T<T_A$; since these temperatures are much lower than the nuclear statistic equilibrium ones, BBN can not begin when the species go out of equilibrium, but has to wait till $T \simeq T_A$. Moreover, the reactions producing helium and heavier elements are all based on $D$:
	\begin{eqnarray}
	\label{}
	D + D &\rightarrow& ^3He + n \cr
	^3He + D &\rightarrow& ^4 He + p 
	\end{eqnarray}
	\begin{eqnarray}
	\label{}
	D + D &\rightarrow& ^3H + p \cr
	^3H + D &\rightarrow& ^4 He + n
	\end{eqnarray}
	\begin{eqnarray}
	\label{}
	D + D &\rightarrow& ^4He + \gamma
	\end{eqnarray}
and thus the reactions can not be fast enough to produce an equilibrium abundance of $^3He$ for $T<0.1MeV$.

Protons and neutrons are kept in chemical equilibrium by mean of the weak interactions:
	\begin{eqnarray}
	\label{reaz_eq_p_n}
	n & \leftrightarrow & p + e^- + \bar \nu_e \cr
	\nu + n & \leftrightarrow & p + e^- \cr
	n +e^+ & \leftrightarrow & p + \bar \nu_e
	\end{eqnarray}
These processes have approximately the rate $\Gamma_W \simeq G_F^2 T^5$ and go out of equilibrium at the 
``freeze-out'' temperature of weak interactions, when $\Gamma (T_W) \simeq H(T_W)$, that is:
	\begin{eqnarray}
	\label{T_W}
	T_{W}\simeq (0.7 - 0.8) MeV \hspace{1cm} (t_W\sim 1 s) 
	\end{eqnarray}
%
%
%
%
%
The neutron to proton density ratio, that is $\frac{n_n}{n_p} = \frac{X_n}{X_p}$, is given, when chemical equilibrium holds, by 
	\begin{eqnarray}
	\label{}
	\frac{n}{p} \equiv \frac{n_n}{n_p} \simeq \exp \left[ \frac{  -Q 
	}{T}\right] 
	\end{eqnarray}
where $Q$ is the neutron-proton mass difference:
	\begin{eqnarray}
	\label{}
	Q \equiv m_n - m_p \simeq 1.293 \; MeV
	\end{eqnarray}
For $T < T_W$ the weak reaction rate $\Gamma_W
$ drops below the Hubble expansion rate $H(T) \simeq 5.5 \, T^2/M_{Pl}$, the neutron abundance freezes out at the equilibrium value $X_{n}(T_{W})$ and it then evolves following the neutron decay exponential law: 
	\begin{eqnarray}
	\label{}
	X_{n}(t)=X_{n}(T_{W})\exp(-t/\tau)
	\end{eqnarray}
where $\tau= (886.7 \pm 1.9) s$ is the neutron lifetime. At this time the neutron to proton density ratio is
	\begin{eqnarray}
	\label{}
	\frac{n}{p}(T_W) \simeq \frac{1}{6}
	\end{eqnarray}
This quantity further decreases to 
	\begin{eqnarray}
	\label{}
	\frac{n}{p}(T_N) \simeq \frac{1}{7}
	\end{eqnarray}
due to the occasional weak interactions, mainly neutron decays.
At temperatures $T < T_{N} \sim 0.1 MeV$, the process $p+n \leftrightarrow D+\gamma$ is faster than the Universe expansion, and free nucleons and deuterium nuclei are in chemical equilibrium.  
Also, reactions are fast enough to produce an equilibrium abundance of $^4He$, which can be worked out assuming that all neutrons end up in $^4He$.
%
%
In conclusion, the primordial $^{4}$He mass fraction is
	\begin{equation} 
	\label{helium}
	X_{4} = \frac{4 (n_n / 2)}{n_n + n_p} = 
	\frac{2 \frac{n}{p}(T_N)}{1 + \frac{n}{p} (T_N)} =
	{{2\exp(-t_N/\tau)} \over {1+\exp(Q/T_{W})}}
	\simeq 0.25
	\end{equation}
The $^4He$ production is influenced by the number of degrees of freedom $g$, which enters $H$, by the neutron half life $\tau$ and by $\eta$.
Comparing the theoretical calculation of the primordial $^4He$ abundance, today accurate to within $\pm 0.4\%$ \cite{Lopez:1998vk}
with the experimental observations - concordant within $\pm 4\%$ \cite{Izotov:1994tg} - it is possible to bound these three parameters. In particular, the number of degrees of freedom can not be higher than a certain value, otherwise too much $^4He$ is produced. 
It is common to re-parametrize the DOFs number in terms of extra-neutrinos, defined as
	\begin{eqnarray}
	\label{}
	\Delta N_{\nu} = N_{\nu} - 3 = 
	\frac{8}{7} \sum_{b=bosons} \frac{g_i}{2} \left( \frac{T_b}{T_{\nu}}\right)^4 + 
	 \sum_{f=fermions} \frac{g_i}{2} \left( \frac{T_f}{T_{\nu}}\right)^4
	\end{eqnarray}
Bounds on $\Delta N_{\nu}$ have been calculated by several authors - for instance Lisi et al. \cite{Lisi:1999ng}, who obtained
	\begin{eqnarray}
	\label{extra-nu}
	\Delta N_{\nu} = 0 \pm 1 \hspace{1.5cm} (95\% \; \mathrm{C. L.})
	\end{eqnarray}

%% file: files/electric_ch_quant.tex
\def \chap-Electric_charge_quantization{Electric charge quantization in theoretical physics}
\chapter{\chap-Electric_charge_quantization}
\label{chap-Electric_charge_quantization}
\markboth{Chapter \ref{chap-Electric_charge_quantization}. ~ \chap-Electric_charge_quantization}
                    {Chapter \ref{chap-Electric_charge_quantization}. ~ \chap-Electric_charge_quantization}

%
%
%
%
%
The puzzle concerning the electric charge quantization has been studied since the beginning of the $20^{th}$ century and is nowadays still challenging. 

There are several experimental and theoretical hints suggesting that the electric charges of the particles entering the SM are quantized. For instance measures of the neutrality of matter from binary pulsars \cite{Sengupta:2000be} and magnetic effects \cite{Marinelli:1983nd} set respectively
	\begin{eqnarray}
	\label{}
	\frac{|q_p + q_e|}{e} < 
	\Biggl\{ 
	\begin{array}{c} 3.2 \times 10^{-20}
	\\ 1.0 \times 10^{-21}
	\end{array}\  
	\end{eqnarray}
where $q_p$ and $q_e$ are the proton and electron electric charges, while the neutron one $q_n$ is \cite{Baumann:1988ue}
	\begin{eqnarray}
	\label{}
	q_n = (-0.4 \pm 1.1) \times 10^{-21} e
	\end{eqnarray}
Other interesting measures concern the muon-electron charge ratio anomaly 
%
	$\frac{q_{\mu^+}}{q_{e^-}} +1 = (1.1 \pm 2.1) \times 10^{-9}$ \cite{Meyer:1999cx}
%
and the neutrino charge $q_{\nu} < 10^{-14}$ \cite{Raffelt:1999gv}.

We defined in \S \ref{sec-Construction_of_the_SM} the SM electric charge in the SM scenario as
        \begin{eqnarray}
        \label{electric_ch_sm}
        Q_{em} = \tau_{3} + \frac{Y}{2}
        \end{eqnarray}
This quantity is not quantized in the classical field theory since the hypercharge is not. Nevertheless, we will show in \S \ref{sec-Anomalies} that imposing the SM renormalizability at quantum level we will get for each generation four relations on the five present hypercharges. 
This topic can be found in many books about quantum field theory, such as \cite{Peskin:1995ev,Cheng:1985bj,Kaku:1993ym}.

Also several models beyond the standard one provide mechanisms to quantize the electric charge. Some of them achieve quantization adding new particles or imposing features to the existing ones (such as the neutrino being a Majorana particle in \cite{Babu:1989ex,Babu:1989tq}) or using gauge groups different from the SM one (for instance $[\mathcal{SU} (3) \times \mathcal{SU} (2) \times \mathcal{U} (1)]_L \times [\mathcal{SU} (3) \times \mathcal{SU} (2) \times \mathcal{U} (1)]_R$ in \cite{Berezhiani:1982ww}).

Electric charge quantization also arises as a natural consequence in Grand Unification Theories and in Dirac quantistic theory of magnetic monopoles, analyzed in \S \ref{sec-SU5} and \ref{sec-Dirac_monopoles}. These topics can be also found in textbooks, see for instance \cite{Cheng:1985bj,Kaku:1993ym}. 

We conclude this brief review stressing that, in spite of these constraints, there are mechanisms, like the photon kinetic mixing introduced in \S \ref{sec-Holdom_s_mechanism}, which are able to add particles with unquantized charge to SM extensions gauged by two $\mathcal{U}(1)$ and to some gauge theories without spoil their renormalizability or its other features.

\def \sec-Anomalies{Anomalies and their cancellation}
\section{\sec-Anomalies}
\label{sec-Anomalies}
\markboth{Chapter \ref{chap-Electric_charge_quantization}. ~ \chap-Electric_charge_quantization}
                    {\S \ref{sec-Anomalies} ~ \sec-Anomalies}

An anomaly is the failure of a classical symmetry of the lagrangian $\mathcal{L}$ to survive the process of quantization and regularization. Indeed, if we have a classic symmetry, the transformation $\phi \rightarrow \phi + \delta \phi$ will leave the action $S(\phi)$ invariant, while, if we have a quantum symmetry, the same transformation will leave the path integral $\int D\phi e^{iS(\phi)}$ invariant, where $D\phi$ is the measure. Therefore it looks reasonable that some classic symmetries may be not valid in quantum theories. 

An important example of anomaly concerns the chiral symmetry of theories with massless fermions. Before entering this topic, let us define the vector, vector axial and pseudoscalar currents, which are respectively:
	\begin{eqnarray}
	\label{vector_current}
	V^{\mu} (x) = \overline{\psi} (x) \gamma^{\mu} \psi (x)
	\end{eqnarray}
	\begin{eqnarray}
	\label{axial_vector_current}
	A^{\mu} (x) = \overline{\psi} (x) \gamma^{\mu} \gamma^5 \psi (x)
	\end{eqnarray}
	\begin{eqnarray}
	\label{pseudo_scalar_current}
	P (x) = \overline{\psi} (x) \gamma^5 \psi (x)
	\end{eqnarray}
We will see that the currents (\ref{vector_current}) and (\ref{axial_vector_current}) are conserved in the classic theory, but it is impossible to preserve both these conservations in the quantum one.

\subsection{Chiral symmetry for free Dirac spinors}

Dirac spinors $\psi$ and their adjoints $\overline{\psi}= \psi^{\dag} \gamma^0$ obey the Dirac equations:
	\begin{eqnarray}
	\label{Dirac_equation}
	(i\partial \!\!\! / -m)\psi=0 \hspace{1.2cm} \overline{\psi} (i\partial \!\!\! / + m)= 0
	\end{eqnarray}
which can be derived from the Dirac lagrangian
	\begin{eqnarray}
	\label{Dirac_lagrangian}
	\mathcal{L}_{Dirac}=\overline{\psi} (i \partial \!\!\! / - m) \psi
	\end{eqnarray}
%
%
%
The Dirac lagrangian (\ref{Dirac_lagrangian}) is manifestly invariant under the phase transformation:
	\begin{eqnarray}
	\label{Noether_vector_transformation}
	\psi(x) \rightarrow e^{i\alpha}\psi(x)
	\end{eqnarray}
Since this continuous symmetry of $\mathcal{L}$, there must exist for the Noether theorem - see \S \ref{sec-symmetries} - the conserved current defined in equation (\ref{Noether_current}). For the symmetry we are analyzing the Noether current is the same $V^{\mu}$ than in equation (\ref{vector_current}) and thus
%
	\begin{eqnarray}
	\label{vector_current_cons}
	\partial_{\mu} V^{\mu} (x) = 0
	\end{eqnarray}
The conservation of this current can be also derived using directly the Dirac equations (\ref{Dirac_equation}). 
Let us consider now the axial vector current (\ref{axial_vector_current}); using the Dirac equations it is straightforward that: 
	\begin{eqnarray}
	\label{axial_vector_current_cons_QED}
	\partial_{\mu} A^{\mu} (x) = 2im P(x)
	\hspace{1.2cm} \lim_{m\longrightarrow 0} \partial_{\mu} j^{\mu 5} (x) = 0
	\end{eqnarray}
Thus, when fermions are massless, the theory has a second Noether current corresponding to the chiral transformation:
	\begin{eqnarray}
	\label{Noether_axial_transformation}
	\psi(x) \rightarrow e^{i\alpha \gamma^5}\psi(x)
	\end{eqnarray}
The chiral symmetry pertains to the only derivative term in the lagrangian (\ref{Dirac_lagrangian}); this is why the axial vector current is not conserved for massive fermions. 

When $m=0$ the currents in (\ref{vector_current}) and (\ref{axial_vector_current}) can be used to define the electric current densities of left- and right-handed particles, which are both conserved:
	\begin{eqnarray}
	\label{chiral_curent_left_ptcs}
	j^{\mu}_L(x)=\overline{\psi} (x)\gamma^{\mu} \left( 
	\frac{1-\gamma^5}{2} \right) \psi (x)
	= \frac{1}{2} \left[ V^{\mu} (x) - A^{\mu} (x) 
	\right] 
	\equiv \overline{\psi}_L \gamma^{\mu} \psi_L
	\end{eqnarray}
	\begin{eqnarray}
	\label{chiral_curent_right_ptcs}
	j^{\mu}_R  (x) =\overline{\psi}  (x) \gamma^{\mu} \left( 
	\frac{1+ \gamma^5}{2} \right) \psi  (x)
	= \frac{1}{2} \left[ V^{\mu} (x) + A^{\mu} (x)
	\right]
	\equiv \overline{\psi}_R \gamma^{\mu} \psi_R
	\end{eqnarray}
In conclusion, the massless Dirac lagrangian has a symmetry associated with the separate number conservation of left- and right-handed fermions, generated by the axial vector current $j^{\mu 5}$. 

The conservation laws presented above are also valid in the quantum theory of free Dirac spinors.

\subsection{Chiral anomaly in abelian gauge theories}
\label{Chiral_anomaly_in_abelian_gauge_theories}

In gauge theories interactions are present - see \S \ref{sec-Gauge_symmetries}. 
When quantizing the theory, the functions we have in the classic model are replaced by quantum operators; if we assume these operators as a simple generalization of the classic functions, they may be not well defined and thus behave in a different way with respect to the classic ones. 
In this section we will see that one of the arising effects is the spoiling of the axial vector current conservation law. 

Let us consider now QED, which is an abelian gauge theory - see \S \ref{Gauge_theories_for_abelian_groups} -; its lagrangian is
\footnote{In this section we use the same conventions of \cite{Peskin:1995ev}; several books, such as \cite{Mandl:1985bg} define $F^{\mu\nu}$ and other quantities with opposite sign.}:
	\begin{eqnarray}
	\label{QED_lagrangian}
	\mathcal{L}_{QED} &=& \overline{\psi} (i D \!\!\!\!/ - m ) \psi - \frac{1}{4} F_{\mu\nu}F^{\mu\nu}
	\end{eqnarray}
	\begin{eqnarray}
	\label{}
	D^{\mu} = \partial^{\mu} - ieA^{\mu} \hspace{1.2cm} 
	F^{\mu\nu} = \partial^{\mu} A^{\nu} - \partial^{\nu} A^{\mu}
	\end{eqnarray}
The lagrangian in (\ref{QED_lagrangian}) does not change under the local gauge transformations:
	\begin{eqnarray}
	\label{}
	\psi \rightarrow \exp^{ie\alpha(x)} \psi \hspace{1.5cm}
	A_{\mu} \rightarrow A_{\mu} - ie \partial_{\mu} \alpha(x)
	\end{eqnarray}
As it happens for free spinors, if $m=0$ the lagrangian is also invariant under the local axial symmetry:
	\begin{eqnarray}
	\label{axial_lagrangian_invariance}
	\psi \rightarrow \exp^{ie\alpha(x) \gamma^5 } \psi \hspace{1.5cm}
	A_{\mu} \rightarrow A_{\mu} - ie \partial_{\mu} \alpha(x) \gamma^5
	\end{eqnarray}
which is broken by the mass term. 
Therefore we would naively expect $\partial_{\mu} j^{\mu 5} = 0$ in massless gauge theories. 

Nevetheless, performing the calculations more carefully, we can see that the actual picture is more complicated in the quantum scenario. 
To get an intuitive idea, recall the equal-time anticommutation relationship for Dirac particles in second quantization:
	\begin{eqnarray}
	\label{}
	\{\psi_a(\mathbf{x}), \psi_b^{\dag}(\mathbf{y})\}
	= \delta_{ab} \: \delta^{(3)}(\mathbf{x} - \mathbf{y})
	\end{eqnarray}
When $\mathbf{x} = \mathbf{y}$ this anticommutator diverges; thus we may wonder if the quantity $\overline{\psi}(\mathbf{x})\psi(\mathbf{x})$ is physically meaningless or not. 
Because this divergence, also the classic currents defined in (\ref{vector_current}), (\ref{axial_vector_current}) and (\ref{pseudo_scalar_current}), which contain the product $\overline{\psi} (\mathbf{x})\psi (\mathbf{x})$, may get  problems when generalized to quantum operators. 

What comes out indeed is the impossibility to preserve both vector and axial vector current conservation. 
Both these currents finally depend on an arbitrary parameter, which is chosen such that the gauge vector invariance is preserved and the theory of electromagnetic interactions with massless photon is safe. But this also implies that the derivative of the axial vector current gets non-zero value in the massless limit%
\footnote{$\epsilon^{\alpha\beta\mu\nu}$ is the totally antisymmetric tensor defined such that $\epsilon^{0123}=-\epsilon_{0123}=+1$.}:
	\begin{eqnarray}
	\label{axial_vector_current_anomaly_QED}
	\partial_{\mu} A^{\mu} &=& 2im P
	- \frac{e^2}{16\pi^2} \epsilon^{\alpha\beta\mu\nu}
	F_{\alpha\beta} F_{\mu\nu} \cr\cr
	\lim_{m\rightarrow 0} \partial_{\mu} A^{\mu} &=& - \frac{e^2}{16\pi^2} \epsilon^{\alpha\beta\mu\nu}
	F_{\alpha\beta} F_{\mu\nu}
	\end{eqnarray}
Using the current densities of left- and right-handed particles introduced in eq.(\ref{chiral_curent_left_ptcs}) and (\ref{chiral_curent_right_ptcs}) we can rewrite the anomaly for massless particles as:
	\begin{eqnarray}
	\label{}
	\partial_{\mu} j_L^{\mu} &=& + \frac{1}{2} \, \frac{e^2}{16\pi^2} \epsilon^{\alpha\beta\mu\nu}
	F_{\alpha\beta} F_{\mu\nu} \cr\cr
	\partial_{\mu} j_R^{\mu} &=& - \frac{1}{2} \, \frac{e^2}{16\pi^2} \epsilon^{\alpha\beta\mu\nu}
	F_{\alpha\beta} F_{\mu\nu}
	\end{eqnarray}
In conclusion, the massless lagrangian $\mathcal{L}$ is invariant under the chiral transformation in eqs.(\ref{axial_lagrangian_invariance}), but the associated current is non conserved.
This property of the quantum gauge theories is called Adler-Bell-Jackiw (or ABJ) anomaly \cite{Adler:1969gk,Bell:1969ts}. Its validity to all orders in QED perturbation theory has been proved by Adler and Bardeen \cite{Adler:1969er}. \\

\subsection{Chiral anomaly in non-abelian gauge theories}
\label{Chiral_anomaly_in_non-abelian_gauge_theories}

Non-abelian gauge theories are also anomalous as we can easily demonstrate once that equation (\ref{axial_vector_current_anomaly_QED}) is assumed. 
Let us consider a gauge theory with a group having a set of generators $t^a$ and the commutation relation between them in the form:
	\begin{eqnarray}
	\label{}
	[t^a,t^b] = i f^{abc} t^c
	\end{eqnarray}
where $f^{abc}$ is an antisymmetric set of numbers called structure constants.
In this picture a theory containing massless Dirac fermions has the lagrangian:
	\begin{eqnarray}
	\label{}
	\mathcal{L} \owns \overline{\psi} \gamma^{\mu } i \left[ \partial_{\mu} - i gA^a_{\mu} t^a
	\left( \frac{1-\gamma^5}{2} \right) 
	\right] \psi
	\end{eqnarray}
The same mathematical procedure leading to equation (\ref{axial_vector_current_anomaly_QED}) can be applied to the gauge symmetry current:
	\begin{eqnarray}
	\label{left_chiral_current_non_abelian}
	j^{\mu a} = \overline{\psi} \, \gamma^{\mu} \left( \frac{1-\gamma^5}{2} \right) t^a \psi
	\end{eqnarray}
Assuming massless particles we finally get \cite{Bardeen:1969md}:
	\begin{eqnarray}
	\label{axial_current_anomalous_divergence}
	\langle p,\nu,b;k,\lambda,c| \partial_{\mu} j^{\mu a} |0\rangle = 
	\frac{g^2}{4\pi^2} \epsilon^{\alpha\nu\beta\lambda}
	p_{\alpha} k_{\beta} \left(
	\frac{1}{2} \rm{tr} [t^a\{ t^b, t^c \}] \right)
	\end{eqnarray}
Of course, if right-handed fermions enter the triangle loop there will be analogous terms but with opposite sign.
	\begin{figure}[htbp]
	\begin{center}
	 \includegraphics[scale=1.]{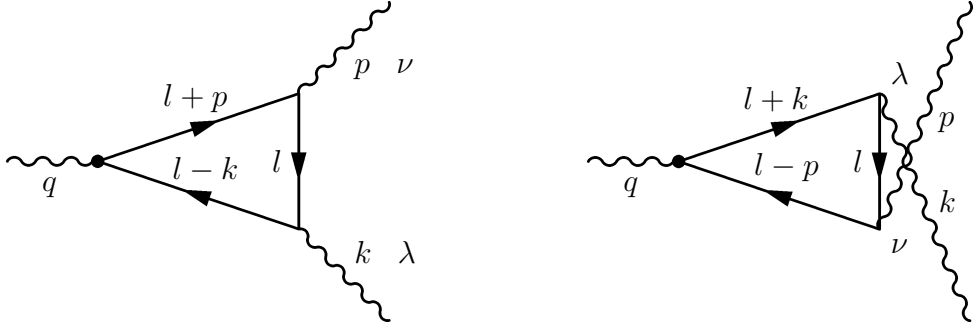}	
	\caption{{\it 	Triangle diagrams contributing to the axial vector anomaly.}}
	\label{Non_abelian_ABJ_anomaly}
	\end{center}
	\end{figure}
\noindent
It can be shown that the total anomaly is proportional to the one arising at triangle level, where the anticommutator in the trace is easily explained if we look at the diagrams in Figure \ref{Non_abelian_ABJ_anomaly}: the total triangular anomaly is worked out summing the two diagrams having $(p,\nu)$ and $(k,\lambda)$ interchanged and this leads to the anticommutator in the trace. 

The factor $\frac{1}{2}$ in front of the trace is a kind of weight which can be understood requiring that if we take $t^a=t^b=t^c=1$, we should get QED and therefore the term between round brackets should be $1$. Indeed:
%
	\begin{eqnarray}
	\label{}
	\frac{1}{2} \rm{tr} [1\{ 1, 1 \}] =1
	\end{eqnarray}
%
Physically consistent gauge theories must be anomaly free. From equation (\ref{axial_current_anomalous_divergence}) we can see that this condition is achieved if the generators satisfy
	\begin{eqnarray}
	\label{anom_cond_tr}
	\rm{tr} [t^a\{ t^b, t^c \}] = 0
	\end{eqnarray}
But there are also anomaly-free theories containing anomalous diagrams which cancel each other. 
In \S \ref{Anomaly_cancellation_in_the_standard_model} we will see that one of these theories is the standard model of elementary particles.

\subsection{Chiral symmetries in QCD and the $\pi^0$ decay}

Let us consider QCD with the only $u$ and $d$ quarks; its lagrangian is:
	\begin{eqnarray}
	\label{}
	\mathcal{L} = 
	\overline{u} i D\!\!\!\!/ u  +
	\overline{d} i D\!\!\!\!/ d
	-m_u \overline{u}u 
	-m_d \overline{d}d 
	\end{eqnarray}
In the following we will neglet the quark masses. Besides the $\mathcal{U}(1)_V \times \mathcal{U}(1)_A$ symmetry, where $V$ and $A$ stands respectively for vector and axial, this lagrangian is symmetric under the unitary transformations:
	\begin{eqnarray}
	\label{}
	\left( \begin{array}{c} u \\ d \\ \end{array} \right)_L
	\rightarrow U_L \left( \begin{array}{c} u \\ d \\ \end{array} \right)_L 
	\hspace{1.2cm}
	\left( \begin{array}{c} u \\ d \\ \end{array} \right)_R
	\rightarrow U_R \left( \begin{array}{c} u \\ d \\ \end{array} \right)_R
	\end{eqnarray}
because there are no couplings between left- and right-handed quarks. Let $Q = Q_L + Q_R$ denote the quark doublet, with chiral components:
	\begin{eqnarray}
	\label{}
	Q_L = \left( \frac{1-\gamma^5}{2} \right) \left( \begin{array}{c} u \\ d \\ \end{array} \right)
	\hspace{1.2cm}
	Q_R = \left( \frac{1+\gamma^5}{2} \right) \left( \begin{array}{c} u \\ d \\ \end{array} \right)
	\end{eqnarray}
The theory is symmetric under the group $\mathcal{SU}(2)_L \times \mathcal{SU}(2)_R \times \mathcal{U}(1)_L \times \mathcal{U}(1)_R$, which corresponds to the conserved currents:
	\begin{eqnarray}
	\label{QCD_left_right_currents}
	j^{\mu}_L &=& \overline{Q}_L \gamma^{\mu} Q_L
	\hspace{1.7cm}
	j^{\mu}_R = \overline{Q}_R \gamma^{\mu} Q_R \cr\cr
	j^{\mu a}_L &=& \overline{Q}_L \gamma^{\mu} \tau^a Q_L 
	\hspace{1.2cm}
	j^{\mu a}_R = \overline{Q}_R \gamma^{\mu} \tau^a Q_R 
	\end{eqnarray}
with $\tau^a=\frac{\sigma^a}{2}$ generators of $\mathcal{SU}(2)$. 
Equivalently we can say that the theory is symmetric under the group $\mathcal{SU}(2)_V \times \mathcal{SU}(2)_A \times \mathcal{U}(1)_V \times \mathcal{U}(1)_A$; sums and differences of the currents in (\ref{QCD_left_right_currents}) indeed lead to: 
	\begin{eqnarray}
	\label{QCD_vector_raxial_currents}
	V^{\mu} &=& \overline{Q} \gamma^{\mu} Q
	\hspace{1.7cm}
	V^{\mu a} = \overline{Q} \gamma^{\mu} \tau^a Q \cr\cr
	A^{\mu} &=& \overline{Q} \gamma^{\mu} \gamma^5 Q
	\hspace{1.2cm}
	A^{\mu a} = \overline{Q} \gamma^{\mu} \gamma^5 \tau^a Q 
	\end{eqnarray}
Classically the four currents in (\ref{QCD_vector_raxial_currents}) are conserved; nevertheless, when passing to the quantum theory, we expect anomalies to arise for axial currents - see \S \ref{Chiral_anomaly_in_abelian_gauge_theories}. \\
	\begin{figure}[htbp]
	\begin{center}
	 \includegraphics[scale=1.]{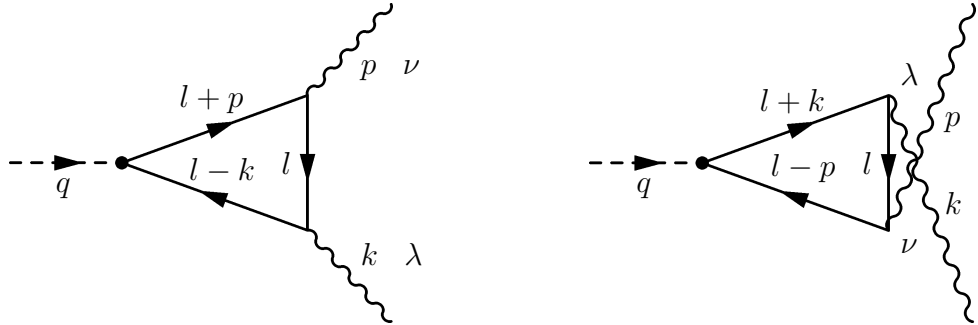}	
	\caption{{\it 	Triangle diagrams contributing to the $\pi^0$
	decay in two photons.}}
	\label{Neutral_pion_decay}
	\end{center}
	\end{figure}
An important application of the ABJ anomaly in QCD is in the derivation of the theorem for the $\pi^0 \rightarrow 2\gamma$ decay in two photons - see figure \ref{Neutral_pion_decay}. 
Applying the formalism we developed in \S \ref{Chiral_anomaly_in_non-abelian_gauge_theories}, it is straightforward to see that this process is described by a triangle diagram having the axial isospin current anomaly given by (\ref{axial_vector_current_anomaly_QED}), multiplied by 
	\begin{eqnarray}
	\label{}
	\frac{1}{2}\rm{tr} [\tau^a\{ Q, Q \}] = \rm{tr} [\tau^a Q^2]
	\end{eqnarray}
where $Q$ is the matrix of quark electric charges:
	\begin{eqnarray}
	\label{}
	Q= \frac{1}{3} \left( \begin{array}{cc} 2 & 0 \\ 0 & -1 \\ \end{array} \right)
	\end{eqnarray}
The trace runs over flavours and colors; the involved matrices do not depend on color, therefore we just get from it a factor $3$. 
Since $\rm{tr} [\tau^1 Q^2] = \rm{tr} [\tau^2 Q^2] =0$, the axial isospin current reads:
	\begin{eqnarray}
	\label{}
	\partial_{\mu} A^{\mu 3} = - \frac{e^2}{16\pi^2} \epsilon^{\alpha\beta\mu\nu}
	F_{\alpha\beta} F_{\mu\nu} \,
	3 \, \frac{1}{6}
	\end{eqnarray}
If there had not been anomalies, the amplitude for the $\pi^0$ decay would have been $0$, that is, not compatible with experiments. Moreover, the decay rate $\Gamma$ is related to the coefficient of the ABJ anomaly, which can be this way experimentally tested. 
The predicted rate
	\begin{eqnarray}
	\label{}
	\Gamma(\pi^0\rightarrow2\gamma) = \left( \frac{N_c}{3} \right)^2
	\frac{\alpha^2 m_{\pi}^3}{64\pi^3 f_{\pi}^2} = 7.73 \; \rm{eV}
	\end{eqnarray}
where $f_{\pi}=92.4$ MeV is in very good agreement with the measured value $\Gamma = 7.7 \pm 0.6$ eV \cite{Yao:2006px}.


%
%

\subsection{Anomaly cancellation in the standard model}
\label{Anomaly_cancellation_in_the_standard_model}

We have demonstrated above that the presence of anomalies is not forbidden, on the contrary these diagrams can sometimes be useful. 
But they also spoil the gauge invariance of the theory; therefore the anomalous terms must cancel each other in consistent gauge theories.
Some models anomaly free can be found in \cite{Gross:1972pv,Georgi:1972bb}, while the general features of a gauge theory free of anomalies are analyzed in \cite{Bouchiat:1972iq}. 

We will show that the electro-weak model first proposed by Glashow-Weinberg-Salam \cite{Glashow:1961tr,Weinberg:1967tq,Salam:1968} is not anomaly free, while the whole SM $\mathcal{SU}(3) \times \mathcal{SU}(2) \times \mathcal{U}(1)$, in which quarks are weighted by the color factor 3, does. This is a remarkable proof of internal consistency of the SM.

\subsubsection{Electroweak anomalies}

Let us consider now the only electroweak sector, that is the gauge group $\mathcal{SU}(2) \times \mathcal{U}(1)$. 
To evaluate its anomalies it is easiest to work in the basis of the gauge bosons before the mixing leading to the photon and $Z^0$ definition - see equation (\ref{B_W_mixing}).
We therefore have four generators, three $\tau_i$ from $\mathcal{SU}(2)$, which are proportional to the Pauli matrices $\sigma_i$, and one from the weak hypercharge $Y$:
	\begin{eqnarray}
	\label{}
	\tau_i &=& \frac{\sigma_i}{2} \hspace{1.5cm}  i=1-3 \cr\cr\cr
	Y &=& \frac{1_{2 \times 2}}{2}
	\end{eqnarray}
%
The Pauli matrices obey to 
	\begin{eqnarray}
	\label{Pauli_matrices_properties}
	\{\sigma_i,\sigma_j\} = 2\delta_{ij} \hspace{1.5cm}
	\rm{tr}[\sigma_i] = 0  \hspace{1.5cm}
	\sigma_i \sigma_j = i \epsilon_{ijk} \, \sigma_k + \delta_{ij} 1_{2\times 2}
	\end{eqnarray}
where $ \epsilon_{ijk}$ is the antisymmetric Levi-Civita symbol.
Let us analyze now all the possible combinations. Equations (\ref{Pauli_matrices_properties}) imply that
	\begin{eqnarray}
	\label{3_SU(2)_generators_trace}
	\rm{tr}[\tau_i \{ \tau_j , \tau_k \}] = \frac{1}{4} \delta_{jk} \; \rm{tr}[\sigma_i] = 0 
	\end{eqnarray}
Considering then that every member of a $\mathcal{SU}(2)$ multiplet has the same hypercharge, we get:
	\begin{eqnarray}
	\label{taui_y_y_trace}
	\rm{tr}[\tau_i \{ Y , Y\}] \propto \rm{tr}[\tau_i] = 0 
	\end{eqnarray}
In general, any anomalies containing one only $\mathcal{SU}(2)$ boson are proportional to $\rm{tr} (\tau_i)$ ans therefore vanish. There are therefore only three terms which do not automatically vanish. The first concerns the left-handed particle hypercharges:
	\begin{eqnarray}
	\label{Y_tauj_taui_trace}
	\rm{tr}[Y \{ \tau_j , \tau_k \}] = \frac{1}{2}  \delta_{jk} \; \rm{tr}[Y]_L = 
	\frac{1}{2}   \delta_{jk} \left( \sum_{leptons} Y + \sum_{quarks} Y \right)_L 
	\end{eqnarray}
From which follows the condition to have an anomaly-free model
	\begin{eqnarray}
	\label{u2u2Y_anom_canc}
	3 Y_{q_L} + Y_{l_L} = 0
	\end{eqnarray}
where the factor 3 arises summing on the three quark colors.
%
%
%
%
%
%
%
%
%
%
%
The second lead to the same condition by using the third equation in (\ref{Pauli_matrices_properties}) in the trace
	\begin{eqnarray}
	\label{taui_y_tauj_trace}
	\rm{tr} [\tau_i \{ Y , \tau_j  \}] &=& 2 \, \rm{tr} [\tau_i \tau_j Y ]  
	\propto \rm{tr} [(i \, \epsilon_{ijk} \, \sigma_k + \delta_{ij} 1_{2\times 2} ) Y ]  
	\propto \delta_{jk} \rm{tr} [ Y ] \cr\cr
	&\Longrightarrow& \rm{tr} [\tau_i \{ Y , \tau_j  \}] = \frac{1}{2} \delta_{jk}  \rm{tr} [ Y ]_L \
	\end{eqnarray}
The third term is $\rm{tr} [YYY] $ and lead to a different condition:
	\begin{eqnarray}
	\label{}
	\rm{tr} [Y^3] = 0 \hspace{0.5cm} \Longrightarrow \hspace{0.5cm}  \sum Y^3 = 0
	\end{eqnarray}
which becomes, if written explicitely for the first particle generation:
	\begin{eqnarray}
	\label{YYY_anom_canc}
	6 Y_{q_L}^3 + 2 Y_{l_L}^3 - 
	\left( 3 Y_{u_R}^3 + 3 Y_{d_R}^3 + Y_{l_R}^3
	\right) =0
	\end{eqnarray}
\noindent
In conclusion, the two equations (\ref{u2u2Y_anom_canc}) and (\ref{YYY_anom_canc}) are enough to ensure the anomaly cancellation in the electro-weak sector of the SM. 

\subsubsection{Color anomalies}

Let us consider now also the color interactions; the
$\mathcal{SU}(3)$ generators are the eight Gell-Mann matrices $\lambda^a$, having the properties:
	\begin{eqnarray}
	\label{Gell-Mann_metrices_propetries}
	\rm{tr} (\lambda_a) &=& 0 \cr\cr
	\rm{tr} (\lambda_a \lambda_b) &=& 2 \delta_{ab} \cr\cr
	\{\lambda_a , \lambda_b\} &=& \frac{4}{3} \, \delta_{ab} \, 1_{3 \times 3}
	+ 2 \, d_{abc} \, \lambda_c
	\end{eqnarray}
where $d_{abc}$ is the totally symmetric symbol defined as:
	\begin{eqnarray}
	\label{}
	d_{118} &=& d_{228} = d_{338} = \frac{1}{\sqrt{3}} \cr\cr
	d_{448} &=& d_{558} = d_{668} = d_{778} = \frac{1}{2} d_{888} = - \frac{1}{2\sqrt{3}} \cr\cr
	d_{146} &=& d_{157} = - d_{247} = d_{256} = d_{344} = d_{355} 
	= - d_{366} = - d_{377} =  \frac{1}{2} \hspace{0.5cm}
	\end{eqnarray}
QCD is a left-right symmetric theory, such as QED, and therefore its anomalies cancel each other. 
Also, traces containing only one $\mathcal{SU}(2)$ or $\mathcal{SU}(3)$ bosons are proportional to \rm${tr} (\tau_i)$ and $\rm{tr} (t_a)$ and therefore vanish.
%
%
The only terms which survives is
	\begin{eqnarray}
	\label{Y_lambdaj_lambdai_trace}
	\rm{tr}[Y \{ \lambda_j , \lambda_k \}] = \frac{4}{3}  \delta_{jk} \; \rm{tr}[Y] =
	\sum_q (-1)^l  \, \frac{4}{3}  \delta_{jk} Y_q
	\end{eqnarray}
where the sum runs over left- and right-handed quarks and $l$ is 1 for left-handed and 0 for right-handed quarks. 
To get anomaly cancellation therefore the model must satisfy
	\begin{eqnarray}
	\label{}
	2 Y_{q_L} - Y_{u_R} - Y_{d_R} =0
	\end{eqnarray}
%
%
%
%
%
%
%
%
%


\subsubsection{Gravitational anomaly}

There is also a gravitational anomaly with one $\mathcal{U}(1)$ which is proportional to $\rm{tr}[Y]$ and leads therefore to the condition
	\begin{eqnarray}
	\label{ggY_anom_canc}
	\rm{tr} [Y] &=& \sum Y = 0 \cr\cr
	&\Longrightarrow& \hspace{0.5cm} 
	2 Y_{l_L} - Y_{l_R} + 6 Y_{q_L} - 3 Y_{u_R} - 3 Y_{d_R} = 0 
	\end{eqnarray}
summed over left- and right-handed leptons and quarks.

\it In conclusion\rm, the five hypercharges we have for any particle generations must satisfy four conditions to lead to the anomaly cancellation in the SM:
	\begin{eqnarray}
	\label{}
	3 Y_{q_L} + Y_{l_L} &=& 0 \cr\cr
	6 Y_{q_L}^3 + 2 Y_{l_L}^3 - \left( 3 Y_{u_R}^3 + 3 Y_{d_R}^3 + Y_{l_R}^3 \right) &=& 0 \cr\cr
	2 Y_{q_L} - Y_{u_R} - Y_{d_R} &=& 0 \cr\cr
	2 Y_{l_L} - Y_{l_R} + 6 Y_{q_L} - 3 Y_{u_R} - 3 Y_{d_R} &=& 0 
	\end{eqnarray}
These equations can be recasted in terms of the electric charge using the electric charge definition (\ref{electric_ch_sm}), from which follows
	\begin{eqnarray}
	\label{Hyperch_fz_Q_tau3}
	Y=2(Q - \tau_3)
	\end{eqnarray}
and the $\tau_3$ is tracelessness, from which $\rm{tr} [Y] = 2 \rm{tr} [Q] $.

\def \sec-SU5{Grand unification and $\mathcal{SU}(5)$}
\section{\sec-SU5}
\label{sec-SU5}
\markboth{Chapter \ref{chap-Electric_charge_quantization}. ~ \chap-Electric_charge_quantization}
                    {\S \ref{sec-SU5} ~ \sec-SU5}

%
%
%
%
%
%
Grand Unification (or Unified) Theories (GUTs) are based on the intriguing idea 
that at extremely high energies, of order $10^{16} \, GeV$, the three forces involved in the SM are unified in a single group with one coupling constant. At low energies this group is broken in the familiar 
$\mathcal{SU}(3) \times \mathcal{SU}(2) \times \mathcal{U}(1)$. 

Leptons and quarks are tipically put together in GUT multiplets; as a consequence, quantization rules for the electric charge naturally arise in these theories. Some models trying to evade this feature can be found in literature - see for example \cite{Okun:1983vw} 
- but they need the addition of several new particles. 

It has been shown by Georgi and Glashow in 1974 that the simplest grand unified model including all the SM features is $\mathcal{SU}(5)$ \cite{Georgi:1974sy}.
Indeed the unified group should:
	\begin{itemize}
		\item Be at least of rank 4 because the SM has four mutually commuting generators,
		one from $\mathcal{U}(1)$, one from $\mathcal{SU}(2)$ 
		and two from $\mathcal{SU}(3)$.
		\item Have complex representations, because in the SM fermions are not equivalent to 
		their complex conjugates, that is they tranform differently.
		\item Take in account of the existence of both integer and fractional charges.
	\end{itemize}
Besides having all the features written above, $\mathcal{SU}(5)$ has the anomaly-free representation $\bf \overline{5} +10 $ \rm, in which the observed quarks and leptons fit neatly getting the correct quantum numbers. In the following this representation and its particle content are analyzed and the electric charge quantization rule is shown.
\subsection{Representations and particle content - one family case}
%
%
%
%
%
\label{One_family_case}
Every lepton family entering the SM contains 15 chiral modes; let us consider for semplicity one lepton family $e$ and its chiral modes, which are listed below with their transformation properties under $\mathcal{SU}(3) \times \mathcal{SU}(2) \times \mathcal{U}(1) $:
	\begin{eqnarray}
	\label{lepton_modes}
	(\nu_e \, , \, e^-)_L \; &:& \; ( \bf 1, 2 , -1 \rm ) \cr\cr
	(e^+)_L \; &:& \; ( \bf 1 , 1 , 2 \rm ) \cr\cr
	(u_{\alpha} \, , \, d_{\alpha})_L \; &:& \; \left( \bf 3 \, ,\, 2 \,,\, \frac{1}{3}\rm \right) \cr\cr
	(u^{c \, \alpha})_L \; &:& \; \left( \bf \overline{3} \,,\, 1 \,,\, - \frac{4}{3}\rm \right)\cr\cr
	(d^{c \, \alpha})_L \; &:& \; \left( \bf \overline{3} \,,\, 1 \,,\, \frac{2}{3}\rm \right) 
	\end{eqnarray}
The $\mathcal{U}(1) $ transformation properties will be omitted in the following. The superscript $c$ indicates the charge conjugate field:
	\begin{eqnarray}
	\label{}
	\psi^c = C \gamma ^0 \, \psi^* = C \, \overline{\psi}\,^T
	\end{eqnarray}
$\mathcal{SU}(5)$ does not have a 15-dimensional representation, thus the fermion content is split into the sum of a 5-dimensional and a 10-dimensional representation. 
The \bf 5 \rm is assumed to be right handed and the \bf 10 \rm to be left handed, so that the anomaly-free combination $\bf \overline{5} +10$ is left handed and contains all the modes listed in (\ref{lepton_modes}). 

The fermion content of this representation is therefore:
	\begin{eqnarray}
	\label{5_repr}
	\mathbf{5} = (\mathbf{3,1}) \oplus (\mathbf{1,2}) : \psi_5=
	\left( \begin{array}{c}
	d^1 \\
	d^2 \\
	d^3 \\
	e^c =e^+\\
	\nu_e^c
	\end{array} \right)_R
	\end{eqnarray}
	\begin{eqnarray}
	\label{anti_5_repr}
	\mathbf{\overline{5}} = (\mathbf{\overline{3},1}) \oplus (\mathbf{1,\overline{2}}) : \psi_{\overline{5}} = 
	\left( \begin{array}{c}
	d^{c1} \\
	d^{c2} \\
	d^{c3} \\
	e^- \\
	\nu_e
	\end{array} \right)_L
	\end{eqnarray}
	\begin{eqnarray}
	\label{10_repr}
	\mathbf{10} = (\mathbf{3,2}) \oplus (\mathbf{\overline{3},1}) \oplus (\mathbf{1,1}) : \psi_{10}=
	\left( \begin{array}{ccccc}
	0 & u^{c3} & -u^{c2} & u_1 & d_1 \\
	-u^{c3} & 0 & u^{c1} & u_2 & d_2 \\
	u^{c2} & -u^{c1} & 0 & u_3 & d_3 \\
	-u_{1} & -u_{2} & -u_3 & 0 & e^+ \\
	-d_{1} & -d_{2} & -d_3 & -e^+ & 0
	\end{array} \right)_L
	\end{eqnarray}
The gauge mesons transform according to the adjoint representation of $\mathcal{SU}(5)$, which has dimension $5^2-1=24$ and can be decomposed according to the $\mathcal{SU}(3) \times \mathcal{SU}(2)$ quantum numbers as:
	\begin{eqnarray}
	\label{adj_su(5)_repr}
	\mathbf{24} = (\mathbf{8,1}) \oplus (\mathbf{1,3}) \oplus (\mathbf{1,1}) \oplus (\mathbf{3,2})
	\oplus (\mathbf{\overline{3},2}) 
	\end{eqnarray}
The gauge bosons are identified as follows:
	\begin{itemize}
	\item $(\mathbf{8,1})$ are the eight colored bosons of $\mathcal{SU}(3)_c$;
	\item $(\mathbf{1,3}) \oplus (\mathbf{1,1})$ are the four $\mathcal{SU}(2) \times \mathcal{U}(1)$
		bosons, that are $\gamma, \, W^{\pm} \; \rm {and} \; Z^0$;
	\item $(\mathbf{3,2}) \oplus (\mathbf{\overline{3},2})$ are new superheavy gauge bosons 
		which couple the quarks to the leptons and mediate the proton decay. 
		They are usually denoted as the $X,Y$ bosons.
	\end{itemize}
%

%
%
%
%
%
%
%
%
%
%
%
%
%
\subsection{Three families generalization}
\label{Three_families_generalization}
In the one family approximation gauge and mass eigenstates coincide. To generalize to  three families we have to replace the fermion fields with gauge eigenstates, which are vectors in the three-dimensional space of the family index $A= e \, , \, \mu \, , \, \tau$:
	\begin{eqnarray}
	\label{3_family_gen_su(5)_repr}
	e \rightarrow e'_A &=& \delta_{AB} \; e_B \;\;\;\;\;\;\; e_B=(e,\mu,\tau) \cr\cr
	\nu_e \rightarrow e\nu'_A &=& \delta_{AB} \; \nu_B \;\;\;\;\;\;\; \nu_B=(\nu_1,\nu_2,\nu_3) \cr\cr
	u \rightarrow p'_A &=& U^{\dagger}_{AB} \; p_B  \;\;\;\;\;\;\; p_B=(u,c,t) \cr\cr
	d \rightarrow n'_A &=& V^{\dagger}_{AB} \; n_B  \;\;\;\;\;\;\; n_B=(d,s,b) \cr\cr
	u^c \rightarrow p'^c_A &=& U^{\dagger c}_{AB} \; p^c_B  \;\;\;\;\;\;\; p_B=(u^c,c^c,t^c) \cr\cr
	d^c \rightarrow n'^c_A &=& V^{\dagger c}_{AB} \; n^c_B  \;\;\;\;\;\;\; n_B=(d^c,s^c,b^c)
	\end{eqnarray}
The basis has been chosed such that lepton-gauge and mass eigenstate $e_A$ coincide. For neutrinos we should generally have $T^{\dag}_{AB}$ instead of $\delta_{AB}$, but if we assume that neutrinos are massless, any linear combination of the degenerate fields can be taken to be their mass eigenstates. 

If we wish to work only with left handed fermions, the transformations for left- and right- handed particles should be different. This is why we wrote down explicitly the last two equations in (\ref{3_family_gen_su(5)_repr}) where
	\begin{eqnarray}
	\label{}
	U_{AB} \neq U^c_{AB} \;\;\;\;\;\;\; V_{AB} \neq V^c_{AB}
	\end{eqnarray}
\subsection{Charge quantization}
\label{Charge_quantization}
The charge quantization arises in the $\mathcal{SU}(5)$ scenario because this group is simple and non abelian. 
$Q$ is an additive quantum number and has therefore to be some linear combination of the four diagonal $\mathcal{SU}(5)$ generators.
%
%

Among the 24 $\mathcal{SU}(5)$ generators, eleven are just the generalization in five dimension of the  $\mathcal{SU}(3)$ and $\mathcal{SU}(2)$ generators, namely the Gell-Mann and Pauli matrices:
	\begin{eqnarray}
	\label{SU(3)_SU(2)_gen_5_dim}
	L^a &=& \left( \begin{array}{cc}
	\frac{\lambda^a}{2} & 0 \\ 0 & 0 
	\end{array} \right);
	\;\;\;\;\;\;\;\;\; a=1,2,...,8 \cr\cr\cr
	L^{b} &=& \left( \begin{array}{cc}
	0 & 0 \\ 0 & \frac{\sigma^i}{2}
	\end{array} \right);
	\;\;\;\;\;\;\;\;\; b=9,10,11 \;\;\;\;\; i=1,2,3
	\end{eqnarray}
from which we get three diagonal generators:
	\begin{eqnarray}
	\label{primi_3_gen_diag_SU(5)}
	L^3 &=& \frac{1}{2}\left( \begin{array}{ccccc}
	1 & 0 & 0 & 0 & 0 \\ 
	0 & -1 & 0 & 0 & 0 \\
	0 & 0 & 0 & 0 & 0 \\
	0 & 0 & 0 & 0 & 0 \\
	0 & 0 & 0 & 0 & 0 \\
	\end{array} \right)
	\cr\cr\cr
	L^8 &=& \frac{1}{2 \cdot \sqrt{3}} \left( \begin{array}{ccccc}
	1 & 0 & 0 & 0 & 0 \\ 
	0 & 1 & 0 & 0 & 0 \\
	0 & 0 & -2 & 0 & 0 \\
	0 & 0 & 0 & 0 & 0 \\
	0 & 0 & 0 & 0 & 0 \\
	\end{array} \right)
	\cr\cr\cr
	L^{11} &=& \frac{1}{2} \left( \begin{array}{ccccc}
	0 & 0 & 0 & 0 & 0 \\ 
	0 & 0 & 0 & 0 & 0 \\
	0 & 0 & 0 & 0 & 0 \\
	0 & 0 & 0 & 1 & 0 \\
	0 & 0 & 0 & 0 & -1 \\
	\end{array} \right)
	\end{eqnarray}
The fourth diagonal $\mathcal{SU}(5)$ generator is a traceless matrix which commutes with the others;  since the identity matrix commutes both with $\mathcal{SU}(3)$ and $\mathcal{SU}(2)$ generators, this fourth diagonal generator is split in two diagonal blocks, both proportional to the identity in their dimension, that is
	\begin{eqnarray}
	\label{}
	L^{12}=  \left( \begin{array}{cc}
	\alpha \cdot 1_{3\times3} & \mathbf{0} \\
	\mathbf{0} & \beta \cdot 1_{2\times 2}
	\end{array} \right) \cr\cr
	\end{eqnarray}
From the traceless and the normalization condition $Tr L^a L^b = 2 \delta ^{ab}$ we get:
	\begin{eqnarray}
	\label{}
	\begin{array}{c} 3 \alpha + 2\beta = 0
	\\ 3 \alpha^2+2\beta^2=2
	\end{array}\ \Biggl\} 
	\Rightarrow \alpha= -2 \cdot \sqrt{\frac{1}{15}}
	\hspace{0.5cm} , \hspace{0.5cm} \beta= 3 \cdot \sqrt{\frac{1}{15}}
	\end{eqnarray}
and therefore
	\begin{eqnarray}
	\label{identity_SU(5)}
	L^{12}=  \frac{1}{\sqrt{15}} \left( \begin{array}{ccccc}
	-2 & 0 & 0 & 0 & 0 \\
	0 & -2 & 0 & 0 & 0 \\
	 0 & 0 & -2 & 0 & 0 \\
	 0 & 0 & 0 & 3 & 0 \\
	 0&0&0&0&3
	\end{array} \right)
	\end{eqnarray}
As stated before, $Q$ must be a linear combination of the diagonal generators in (\ref{primi_3_gen_diag_SU(5)}) and (\ref{identity_SU(5)}), that is
	\begin{eqnarray}
	\label{}
	Q = \alpha L^3 + \beta L^8 + \gamma L^{11} + \delta L^{12}
	\end{eqnarray}
Let us now apply $Q$ to the fermion representations in (\ref{5_repr}):
	\begin{eqnarray}
	\label{}
	Q \psi_5 &=& 
	\left( \begin{array}{ccccc}
	-\frac{1}{3} & 0 & 0 & 0 & 0 \\
	0 & -\frac{1}{3} & 0 & 0 & 0 \\
	 0 & 0 & -\frac{1}{3} & 0 & 0 \\
	 0 & 0 & 0 & 1 & 0 \\
	 0&0&0&0&0
	\end{array} \right)
	\left( \begin{array}{c}
	d^1 \\
	d^2 \\
	d^3 \\
	e^c =e^+\\
	\nu_e^c
	\end{array} \right)_R  = \cr\cr
	&=& \left( \begin{array}{ccccc}
	\frac{\alpha}{2} + \frac{\beta}{2\sqrt{3}} -2\frac{\delta}{\sqrt{15}} & 0 & 0 & 0 & 0 \\
	0 & - \frac{\alpha}{2} + \frac{\beta}{2\sqrt{3}} -2\frac{\delta}{\sqrt{15}} & 0 & 0 & 0 \\
	 0 & 0 & - \frac{\beta}{2\sqrt{3}} -2\frac{\delta}{\sqrt{15}} & 0 & 0 \\
	 0 & 0 & 0 & \frac{\gamma}{2} + 3 \frac{\delta}{\sqrt{15}} & 0 \\
	 0&0&0&0&- \frac{\gamma}{2} + 3 \frac{\delta}{\sqrt{15}} 
	\end{array} \right) \cdot \cr\cr
	&\cdot& \left( \begin{array}{c}
	d^1 \\
	d^2 \\
	d^3 \\
	e^c =e^+\\
	\nu_e^c
	\end{array} \right)_R
	\end{eqnarray}
Therefore we can deduce that the electric charge is given by:
	\begin{eqnarray}
	\label{}
	Q = \frac{1}{2} \left( L^{11} + \sqrt{\frac{5}{3}} L^{12} \right) =
	\frac{1}{3}\left( \begin{array}{ccccc}
	-1 & 0 & 0 & 0 & 0 \\
	0 & -1 & 0 & 0 & 0 \\
	 0 & 0 & -1 & 0 & 0 \\
	 0 & 0 & 0 & 3 & 0 \\
	 0&0&0&0&0
	\end{array} \right)
	\end{eqnarray}
From the traceless of $Q$ we get, for $n_c=3$ colors
	\begin{eqnarray}
	\label{}
	n_c \cdot Q_d + Q_e + Q_{\nu} = 0 \Rightarrow Q_d = - \frac{1}{3} Q_{e^+}
	\end{eqnarray}
Thus the electric charge quantization arises in $\mathcal{SU}(5)$ as a consequence of having three quark colors. \\

\def \sec-Dirac_monopoles{Magnetic monopoles}
\section{\sec-Dirac_monopoles}
\label{sec-Dirac_monopoles}
\markboth{Chapter \ref{chap-Electric_charge_quantization}. ~ \chap-Electric_charge_quantization}
                    {\S \ref{sec-Dirac_monopoles} ~ \sec-Dirac_monopoles}

%
The theory of magnetic monopoles was first formulated by Dirac in 1931 \cite{Dirac:1931kp}. As it is shown in the following, it is consistent with quantum theory only if the electric charge is quantized.

\def \sec-Dir_clas_mon{Dirac classic theory of magnetic monopoles}
\subsection{\sec-Dir_clas_mon}
\label{sec-Dir_clas_mon}
\markboth{Chapter \ref{chap-Electric_charge_quantization}. ~ \chap-Electric_charge_quantization}
                    {\S \ref{sec-Dir_clas_mon} ~ \sec-Dir_clas_mon}

%
The original Dirac's theory is a straighforward generalization of the classical electromagnetism, which is modified to include magnetic sources. Classical electromagnetism is based on the \it Maxwell's equations\rm :
	\begin{eqnarray}
	\label{Maxwell_eqs}
	\nabla \cdot \mathbf{E} &=& \rho \hspace{2.5cm} \nabla \cdot \mathbf{B} = 0 \cr\cr
	\nabla \times \mathbf{E} &=& - \frac{\partial \mathbf{B}}{\partial t} \hspace{1.6cm}
	\nabla \times \mathbf{B} = \mathbf{j} + \frac{\partial \mathbf{E}}{\partial t} 
	\end{eqnarray}
the solutions of which, $\mathbf{E}$ and $\mathbf{B}$, can be parametrized in terms of the vector and scalar potential $\mathbf{A}$ and $\phi$ as:
	\begin{eqnarray}
	\label{clas_eqs_potenziali}
	\mathbf{E} = -\nabla \phi - \frac{\partial \mathbf{A}}{\partial t} \hspace{2cm}
	\mathbf{B} = \nabla \times \mathbf{A}
	\end{eqnarray}
It is worth to stress that we can write the second equation in (\ref{clas_eqs_potenziali}) only in consequence of the Maxwell's equation stating $\nabla \cdot \mathbf{B} = 0$, because for any arbitrary vector \bf V \rm we have $\nabla \cdot (\nabla \times \mathbf{V} = 0)$. \\
Maxwell's equations can be written in the covariant form:
	\begin{eqnarray}
	\label{maxwell_covarianti}
	\partial_{\nu} F^{\mu \nu} &=& -j^{\mu} \cr\cr
	\partial_{\nu} \tilde F^{\mu \nu} &=& 0
	\end{eqnarray}
in terms of the electromagnetic field tensor $F_{\mu \nu}$:
	\begin{eqnarray}
	\label{}
	 F_{\mu\nu} = \partial_{\mu}A_{\nu} - \partial_{\nu}A_{\mu}
	 \;\;\;\;\,\;\;\;\;\;\;\;\;\;\;\;
	A_{\mu}=(\phi, \mathbf{A})
	\end{eqnarray}
and its dual $\tilde F^{\mu\nu}$:
	\begin{eqnarray}
	\label{}
	\tilde F^{\mu\nu} = \frac{1}{2} \epsilon^{\mu\nu\rho\sigma} F_{\rho\sigma}
	\end{eqnarray}
%
%
%
%
%
The only source in (\ref{maxwell_covarianti}) is the electric current $j^{\mu} = (\rho, \mathbf{j})$; in its absence, that is in vacuum, Maxwell's equations are symmetric under the duality transformation:
	\begin{eqnarray}
	\label{duality_transf}
	F^{\mu \nu} \rightarrow \tilde F^{\mu \nu}  \;\;\;\;\,\;\;\;\;\;\;\;\;\;\;\;\;\;\;\;
	\tilde F^{\mu \nu} \rightarrow - F^{\mu \nu} 
	\end{eqnarray}
This symmetry is broken by the current $j^{\mu}$, but it can be restored introducing the magnetic current $k^{\mu} = (\sigma, \mathbf{k})$ in eqs.(\ref{maxwell_covarianti}), so that the covariant Maxwell's equations become:
	\begin{eqnarray}
	\label{maxwell_covarianti_sorgente_magnetica}
	\partial_{\nu} F^{\mu \nu} &=& -j^{\mu} \cr\cr
	\partial_{\nu} \tilde F^{\mu \nu} &=& k^{\mu}
	\end{eqnarray}
and the duality symmetry holds if the substitution in (\ref{duality_transf}) are made together with:
	\begin{eqnarray}
	\label{}
	j^{\mu} \rightarrow k^{\mu} \;\;\;\;\,\;\;\;\;\;\;\;\;\;\;\;\;\;\;\;
	k^{\mu} \rightarrow - j^{\mu}
	\end{eqnarray}
In classic electromagnetism the current $j^{\mu}$ is produced by electrically charged particles; similarly the magnetic current $k^{\mu}$ can exist only if there are magnetically charged particles called magnetic monopoles.

\def \sec-quantum_mon{Quantum theory of magnetic monopoles and electric charge quantization}
\subsection{\sec-quantum_mon}
\label{sec-quantum_mon}
\markboth{Chapter \ref{chap-Electric_charge_quantization}. ~ \chap-Electric_charge_quantization}
                    {\S \ref{sec-quantum_mon} ~ \sec-quantum_mon}

%
%
To work out the quantum behaviour of Dirac's magnetic monopoles, we recall equations (\ref{maxwell_covarianti_sorgente_magnetica}) and their solutions for point electric charge/magnetic monopole fixed at the origin:
	\begin{eqnarray}
	\label{point_charge_monopole_origin}
	\nabla \cdot \mathbf{E} = 4\pi q \; \delta^3(r) \;\;\;\;\,\;\;\;\;\; \rightarrow \;\;\;\;\,\;\;\;\;\;
	\mathbf{E} =q \frac{\mathbf{r}}{r^3}  \cr\cr
	\nabla \cdot \mathbf{B} = 4\pi g \; \delta^3(r) \;\;\;\;\,\;\;\;\;\; \rightarrow \;\;\;\;\,\;\;\;\;\;
	\mathbf{B} =g \frac{\mathbf{r}}{r^3} 
	\end{eqnarray}
Let us now consider the surface integral for $\mathbf{B}$ over any closed surface $\mathbf{S}$ containing the monopole:
	\begin{eqnarray}
	\label{int_sup_B_monopole}
	\oint _S  \mathbf{B} \cdot d\mathbf{S} = 4 \pi \cdot g 
	\end{eqnarray}
This integral is not zero because $\nabla \cdot \mathbf{B} \neq0$; as a consequence, we have for arbitrary $\mathbf{x}$, $\mathbf{B} (\mathbf{x}) \neq \nabla \times \mathbf{A} (\mathbf{x})$ as we stressed below eqs. (\ref{clas_eqs_potenziali}). 

A trick to evade this condition can be found considering the field generated by a long and thin solenoid placed along the $z$-axis with its positive pole placed at the origin. The solenoid's field is:
	\begin{eqnarray}
	\label{solenoid_field_Dirac}
	\mathbf{B}_{sol} = \frac{g}{4\pi r^3}\mathbf{r} + g \theta(-z) \delta(x) \delta(y) \mathbf{\hat z}
	\end{eqnarray}
The solenoid's field has no magnetic sources, therefore we can write:
	\begin{eqnarray}
	\label{}
	\mathbf{B}_{sol} = \nabla \times \mathbf{A}
	\end{eqnarray}
Comparing (\ref{point_charge_monopole_origin}) and (\ref{solenoid_field_Dirac}) we see that 
	\begin{eqnarray}
	\label{B_mon_rot_A}
	\mathbf{B}_{monopole} = \mathbf{B} 
	&=& \mathbf{B}_{sol} - g \theta(-z) \delta(x) \delta(y) \mathbf{\hat z} \cr\cr
	&=&  \nabla \times \mathbf{A}  - g \theta(-z) \delta(x) \delta(y) \mathbf{\hat z} 
	\end{eqnarray}
that is, it is possible to define $\mathbf{A}$ such that $\mathbf{B} = \nabla \times \mathbf{A}$ everywhere except a line going from the origin to infinity, which is called \it Dirac string\rm. 

The quantization of electric charge follows requiring that the solution of the system with the monopole is mono-valued when integrating around a loop. 
%
%

The solution of (\ref{point_charge_monopole_origin}) is a plane wave:
	\begin{eqnarray}
	\label{}
	\psi = C \cdot \exp^{i \, (\mathbf{p \cdot r} - E \, t )}
	\end{eqnarray}
with the substitution:
	\begin{eqnarray}
	\label{}
	\mathbf{p} \rightarrow \mathbf{p} - q \mathbf{A} \hspace{1.2cm}  \Rightarrow \hspace{1.2cm} 
	\psi \rightarrow \psi \cdot \exp^{-i q \, (\mathbf{A} \cdot \mathbf{r})}
	\end{eqnarray}
which leads to an additional phase factor. 

The wave function is mono-valued when performing an integration around a loop only if its phase factor has the value $\exp^{i 2\pi n} = 1$, which is achieved if:
	\begin{eqnarray}
	\label{}
	2 \pi n = q \oint \mathbf{A} \cdot d \mathbf{l} 
	\end{eqnarray}
Using (\ref{int_sup_B_monopole}) and (\ref{B_mon_rot_A}) we finally have:
	\begin{eqnarray}
	\label{}
	q \oint \mathbf{A} \cdot d \mathbf{l} = q \int \mathbf{B} \cdot d \mathbf{S} 
	\hspace{1.cm} \Rightarrow \hspace{1.cm} 
	2 \pi n = q \, 4 \pi g
	\end{eqnarray}
The quantization condition induced by the presence of the Dirac monopole is therefore:
	\begin{eqnarray}
	\label{Dirac_quant_mon}
	q=\frac{n}{2 g}
	\end{eqnarray}
Nevertheless, electrodynamics is perfectly consistent also without monopoles and therefore it does not require their existence.

\def \sec-H_P_mon{'t Hooft-Polyakov monopoles}
\subsection{\sec-H_P_mon}
\label{sec-H_P_mon}
\markboth{Chapter \ref{chap-Electric_charge_quantization}. ~ \chap-Electric_charge_quantization}
                    {\S \ref{sec-H_P_mon} ~ \sec-H_P_mon}

It can be shown that a gauge theory coupled to scalar fields possesses monopole solutions, which are called \it 't Hooft-Polyakov monopoles \rm \cite{Hooft:1974qc}.

The simplest example of non-abelian gauge theory having monopole solutions is the $\mathcal{SO}(3)$ Georgi and Glashow model \cite{Georgi:1972cj}, which is out by experiments. Nevertheless, the mathematical procedure developed for $\mathcal{SO}(3)$ can be applied to any compact non-abelian groups which break to a lower one containing some abelian factor, such as $\mathcal{SU}(5)$.

Th $\mathcal{SO}(3)$ lagrangian density is 
	\begin{eqnarray}
	\label{}
	\mathcal{L} = -\frac{1}{4} F_a^{\mu\nu} F_{a\mu\nu} 
	+ \frac{1}{2} D^{\mu} \phi^a D_{\mu} \phi^a - V(\phi)
	\end{eqnarray}
where
	\begin{eqnarray}
	\label{}
	F^a_{\mu\nu} &=& \partial_{\mu} A_{\nu}^a -   \partial_{\nu} A_{\mu}^a - e \epsilon^{abc}
	A_{\mu}^b A_{\nu}^c \cr\cr
	(D_{\mu} \phi)^a &=& \partial_{\mu} \phi^a - e \epsilon^{abc} A_{\mu}^b \phi^c \cr\cr
	V(\phi) &=& \frac{\lambda}{4} \left(\phi \cdot \phi - \frac{\mu^2}{\lambda} \right)^2
	\end{eqnarray}
This problem has a smooth solution with finite energy located in $r=0$ which behaves like a Dirac monopole at large distances since imposing total finite energy implies
	\begin{eqnarray}
	\label{}
	\mathbf{B} \sim \frac{-1}{e} \, \frac{\mathbf{r}}{r^3}
	\end{eqnarray}
which has the same form of the last of (\ref{point_charge_monopole_origin}). Therefore the field $\mathbf{B}$ is generated by a monopole having a magnetic charge
	\begin{eqnarray}
	\label{}
	g = - \frac{1}{e}
	\end{eqnarray}
In this model the constant $e$ is related to the electric charge operator by
	\begin{eqnarray}
	\label{}
	Q = eT_3
	\end{eqnarray}
where $T_3$ is the third component of the weak isospin operatos; its smallest value is $\frac{1}{2}$. When $e = \frac{1}{2}$ we get in modulus the Dirac condition (\ref{Dirac_quant_mon}) with $n=1$.

It can be shown that stable monopole solutions occour for any gauge theories in which a simple group is broken down to a smaller group containing an explicit $\mathcal{U} (1)$ factor. 
This fact is compatible with the reasonable idea that electric charge quantization and the existence of monopoles are related and that the eletric charge quantization folows from the SSB of a simple gauge group.

\subsubsection{Experimental bounds}

Monopoles have many astrophysical and cosmological effects which depend on their average flux in the universe, defined as
	\begin{eqnarray}
	\label{}
	\langle F_M \rangle =  \frac{n_M \beta_M}{4\pi}
	\end{eqnarray}
where $n_M$ is the monopoles number density and $\beta_M$ stands for their velocity. No monopoles have been observed so far, therefore astrophysics and cosmology observations lead to only upper limits on $\langle F_M \rangle$.

The most stringent mass-independent limit on $\langle F_M \rangle$ is the \it Parker limit \rm which is based upon the survival of the galactic magnetic fields:
	\begin{eqnarray}
	\label{}
	\langle F_M \rangle \leq 10^{-16} cm^{-2} sr^{-1} sec^{-1}
	\end{eqnarray}
This bound is analyzed considering also its dependence on the monopole mass in \cite{Turner:1982ag}. The larger possible flux of magnetic monopoles compatible with  the survival of the galactic magnetic fields is
	\begin{eqnarray}
	\label{}
	\langle F_M \rangle &\simeq& 10^{-12} cm^{-2} sr^{-1} sec^{-1} \cr\cr
	\mathrm{for} \hspace{0.8cm}
	m_M &\simeq& 10^{19} GeV , \hspace{0.5cm} \beta_M \simeq 3 \cdot 10^{-3}
	\end{eqnarray}
A direct experimental bound has been set by the MACRO collaboration \cite{Ambrosio:2002qq}. The interaction of a magnetic monopole with a nucleon indeed can lead to a baryon-number violating process in which the nucleon decays into a lepton and one or more mesons (\it catalysis of nucleon decay\rm ).
Searching for these events in the MACRO detector lead to the bound:
	\begin{eqnarray}
	\label{}
	\langle F_M \rangle &\leq& 3\cdot 10^{-16} cm^{-2} sr^{-1} sec^{-1} \cr\cr
	\mathrm{for} \hspace{0.8cm}
	1.1 \cdot 10^{-4} &\leq& \beta_M \leq 5 \cdot 10^{-3}
	\end{eqnarray}
%


%% file: files/MCP.tex

\chapter{Milli-Charged Particles and the mirror universe}
\label{The_mirror_universe}
\markboth{Chapter \ref{The_mirror_universe}. ~ The mirror universe}
	{Chapter \ref{The_mirror_universe}. ~ The mirror universe}

The possible existence of particles having electric charge not multiple of $\frac{1}{3} \,e$ has been longly investigated. 
These hypothetical particles as usually referred to as MCP, which stands for Milli- (or Mini-) Charged Particles. 

We have seen in chapter \ref{chap-Electric_charge_quantization} that several advanced theories vinculate the electric charges of the known elementary particles to be integer multiples of a fixed unity; nevetheless, it is possible to add new particles with unquantized charge to the model under some special conditions.

For instance, Holdom \cite{Holdom:1985ag} realized in 1985 that in a theory gauged by the group $\mathcal{U} (1) \times \mathcal{U} (1)$ there can be millicharged interactions; this scenario is compatible with some gauge theories as well as with the constraints arising from the anomalies cancellation since the interaction takes place via a vector current which is automatically anomaly-free.

We will explain in this chapter how the presence of a second group $\mathcal{U} (1)$ induces milli-charged interactions; then we will introduce a SM extension - called \it mirror universe \rm - where this mechanism can be naturally introduced.

\def \sec-Holdom_s_mechanism{Paraphotons and millicharges}
\section{\sec-Holdom_s_mechanism}
\label{sec-Holdom_s_mechanism}
\markboth{Chapter \ref{The_mirror_universe}. ~ The mirror universe}
                    {\S \ref{sec-Holdom_s_mechanism} ~ \sec-Holdom_s_mechanism}

The mechanism by which millicharged particles are introduced is based on a lagrangian density containing two photons kinetically mixed. Two transformations are performed, a non-unitary one to diagonalize the photon kinetic term and a unitary one to make ordinary matter coupled to one only photon. By mean of these transformations, particles which were coupled to the only non-ordinary photon get a small charge $\epsilon$ with respect to the ordinary photon \cite{Holdom:1985ag,Foot:2000vy}

Let us start with a generic lagrangian density for a model gauged by the group $\mathcal{U} (1) \times \mathcal{U} (1) \times \ldots$. It contains a mass term plus the interactions between the photons and the fermions and a kinetic term which we assume to be non-diagonalized:
        \begin{equation}
        \label{L_Holdom}
        \mathcal{L} =
        \mathcal{L}_{kin} + \mathcal{L}_{mass} + \mathcal{L}_{int}
        = -\frac{1}{4}F^T M_F F + \frac{1}{2} A^T M_A A + \ldots
        \end{equation}
In the following we will only consider the photons' kinetic and mass term for sake of semplicity.
As stated before, we will assume a diagonal mass matrix and a kinetic term having small quantities $\epsilon$ out of the diagonal:
        \begin{equation}
        \label{Holdom_lagrangian_matrices}
        M_F = \pmatrix{
        1 & \epsilon \cr
        \epsilon & 1 \cr
        } \hspace{1.5cm}
        M_A = \pmatrix{
        m_1^2 & 0 \cr
        0 & m_2^2 \cr
        }
        \end{equation}
With this choice of the matrices, the lagrangian in eq.(\ref{L_Holdom}) contains the interaction term for the photons:
	\begin{eqnarray}
	\label{kin_mix_term}
	\mathcal{L}_{int} = - \frac{\epsilon}{2} F'_{\mu\nu} F^{\mu\nu}  
	\end{eqnarray}
which is gauge invariant and renormalizable; it can exist at tree level and may be induced in GUTs \cite{Berezhiani:2005ek,Glashow:1985ud}.

To get a canonical kinetic term, we transform the kinetic matrix $M_F$ to a diagonal matrix $D$ via:
        \begin{equation}
        M_F = O D O^T \hspace{1.5cm}
        O^T O = 1
        \end{equation}
which leads to a canonical kinetic term under the non-unitary transformation of the fields:
        \begin{equation}
        F'=D ^{\frac{1}{2}} O^T F \hspace{1.5cm}
        A'=D ^{\frac{1}{2}} O^T A 
        \end{equation}
Applying these transformations the lagrangian density becomes:
        \begin{equation}
        \label{Holdom_can_lagrangian}
        \mathcal{L} = -\frac{1}{4} F'\,^T \: 1_{2 \times 2} F' + \frac{1}{2} A'\,^T M'_A A' 
        \end{equation}
where
        \begin{equation}
        M'_A=D^{-\frac{1}{2}} O^T M_A O D^{-\frac{1}{2}}
        \end{equation}
An explicit working out of these matrices gives:
	\begin{eqnarray}
	\label{}
	O =  \left( \begin{array}{cc}
	-\frac{1}{\sqrt{2}} & \frac{1}{\sqrt{2}}\\
	\frac{1}{\sqrt{2}} & \frac{1}{\sqrt{2}}
	\end{array} \right) 
	\hspace{1.5cm}
	D^{-\frac{1}{2}} \simeq \left( \begin{array}{cc}
	1+\frac{\epsilon}{2} & 0 \\
	0 & 1-\frac{\epsilon}{2}
	\end{array} \right)
	\end{eqnarray}
Let us now insert the physics in the model. Since the ordinary photon is massless, we assume that both $A_1$ and $A_2$ are massless and therefore state $m_1 = m_2 = 0$.

Any pair of orthogonal combinations of the $A_{\mu}'$ fields will keep canonical kinetic term. 
The theory can contain particles coupled to one only photon or both, but since QED has been tested to very high precision and its previsions are in very good agreement with the experimental results, 
we want ordinary particles - electron, muon and so on - to couple to one only photon. 
We therefore re-define the fields such that one photon - which we will call $\tilde A^{\mu}_1$ - will satisfy this condition. The field $\tilde A^{\mu}_1$ therefore represents ordinary QED photon, while the  orthonormal combination $\tilde A^{\mu}_2$ will be called \it paraphoton \rm and will be sterile with respect to the ordinary matter.

If the model contains ordinary particles coupled to $A^{\mu}_1$ with charge $e'_1$ - which we will call $f_1$ - and para-particles coupled to $A^{\mu}_2$ with charge $e'_2$ - which we will call $f_2$ - and the interaction lagrangian density is of the usual form:
	\begin{eqnarray}
	\label{}
	\mathcal{L}_{int} = - e'_1 \bar f_1 \gamma_{\mu} f_1 A^{\mu}_1 - 
	e'_2 \bar f_2 \gamma_{\mu} f_2 A^{\mu}_2
	\end{eqnarray}
performing the transformations we introduced above we get:
	\begin{itemize}
	\item ordinary particles $f_1$ couple to the only $\tilde A^{\mu}_1$ with the usual electric charge 
	$e_1 \propto e'_1$; as a consequence the electric charges of ordinary particles 
	keep the same ratio they have in the SM scenario and - for istance - the
	electron to proton charge ratio is not modified;
	\item para-particles $f_2$ couple to $\tilde A^{\mu}_2$ with charges $e_2 \propto e'_2$
	but also to the ordinary photon $\tilde A^{\mu}_1$ with charge $\epsilon$.
	\end{itemize}
In conclusion: paraparticles, which were coupled to the only paraphoton in the lagrangian (\ref{L_Holdom}), via this mechanism gain a small electric charge which makes them coupled also to ordinary photon. This small charge is equal to $\epsilon$, which is the kinetic mixing parameter.


\def \sec-mir-univ{The mirror universe}
\section{\sec-mir-univ}
\label{sec-mir-univ}
\markboth{Chapter \ref{The_mirror_universe}. ~ The mirror universe}
                    {\S \ref{sec-mir-univ} ~ \sec-mir-univ}


We introduced in \S \ref{sec-Parity_violation_weak_interactions} that elementary particles experience left-handed weak interactions. But it has been suggested since 1956 by Lee and Yang \cite{Lee:1956qn} that there may exist right-handed particles restoring the left-right symmetry. 
Ten years later Okun and Pomeranchuk \cite{Okun:1966} suggested that these particle may come from a hidden mirror sector, which is an exact copy of our one but where weak interactios are right-handed.

The mirror theory is based on the gauge group $G \times G'$, where $'$ marks quantities of the mirror sector. 
If it is not differently specified, we will assume below that the discrete symmetry $G \leftrightarrow G'$,  which interchanges corresponding fields of $G$ and $G'$ and is called \it mirror parity\rm, is valid; this symmetry guarantees that two particle sectors have identical lagrangians with the same interactions and coupling constants. 
If the electro-weak symmetry scale is different in the two worlds, this parity will be spontaneously broken and this can lead to different physics in the mirror sector \cite{Akhmedov:1992hh,Berezhiani:1995am,Berezhiani:2000gh,Gianfagna:2004je}.

The two worlds communicate through the gravity, but there may also be mini-charged interactions like the ones we introduced in \S \ref{sec-Holdom_s_mechanism}. Other possible interactions between the two worlds have been investigated, for instance the mixing of ordinary and mirror (sterile) neutrinos \cite{Foot:1991py,Akhmedov:1992hh} or ordinary and mirror neutron \cite{Zurab_Bento_last_minute}, a quartic interaction between the two worlds' Higgses \cite{Berezhiani:2003xm,Berezhiani:1995am} and so on - see \cite{Berezhiani:2003xm} for more references.

An interesting feature of the millicharged mirror world is that \it every \rm particle contained in the SM has its mirror counterpart, which has, with respect to the ordinary photon, the same electric charge of its standard corresponding but scaled by a factor $\epsilon$.
Hence, the milli-charged mirror sector can not give rise to exotic unwelcome phenomena, such as violation of the electric charge conservation or atoms having the same number of mirror protons and electrons but nevertheless globally charged. Moreover, the conditions which ensure the anomaly cancellation in the SM - see \S \ref{Anomaly_cancellation_in_the_standard_model} - if satisfied in the ordinary sector, will be automatically satisfied in the whole theory.

\def \sec-mir-SM{Mirror standard model and mirror symmetry}
\subsubsection{\sec-mir-SM}

Let us consider the mirror principle applied to the SM; the whole gauge group for this case will be $G \times G' = \mathcal{SU}(3) \times \mathcal{SU}(2) \times \mathcal{U}(1) \times \mathcal{SU}(3)' \times \mathcal{SU}(2)' \times \mathcal{U}(1)'$.
Also the lagrangian is the same than in the SM; the mirror symmetry introduced above, which interchanges all ordinary particles with their mirror partners, induces also the lagrangian interchange
	\begin{eqnarray}
	\label{}
	\mathcal{L} \leftrightarrow \mathcal{L}'
	\end{eqnarray}
such that the total lagrangian $\mathcal{L} + \mathcal{L}'$ remains invariant.

The SM particles content has been already introduced in \S \ref{sec-Construction_of_the_SM} - see equations (\ref{SM-leptons}), (\ref{SM-left-q}) and (\ref{SM-right-q}).
The mirror sector contains the same particles, which will be marked by the usual $'$; also the barion and lepton numbers $B',L',L_i' \;\; i=e,\mu,\tau$ are assigned in analogy with the ordinary ones.

When we introduced the SM in chapter \ref{chap-The_standard_model} we stressed that weak interactions are not symmetric between the left and the right components: parity $P: f_L \rightarrow f_R$ and $CP: f_L \rightarrow f_R^c$ simmetries are both broken%
%
%

Nevertheless, it is possible to preserve the symmetry between left- and right-handed components in the mirror scenario. This symmetry - sometimes called \it matter parity \rm- is made through the transformations \cite{Foot:1991bp}:
	\begin{eqnarray}
	\label{}
	x &\rightarrow& - x  \cr\cr 
	f_L, f_L^c &\rightarrow& \gamma_0 f'^c_R, \gamma_0 f_R' \cr\cr
	f_R^c, f_R &\rightarrow& \gamma_0 f'_L, \gamma_0 f'^c_L \cr\cr
	\phi &\rightarrow& \phi'^c \cr\cr
	B^{\mu},W_i^{\mu},G_j^{\mu} &\rightarrow& B'_{\mu},W'_{i \, \mu},G'_{j \, \mu} \cr\cr
	i=1,2,3 \hspace{0.2cm} &,& \hspace{0.2cm} j = 1, \ldots , 8
	\end{eqnarray}
It is worth to stress that if we require the validity of this symmetry, the introduction of the mirror world does not add any new parameter to the SM; the matter symmetry validity indeed 
%
%
also implies that the gauge couplings and the Higgs potential are the same for the two sectors, while the Yukawa couplings must satisfy the property:
	\begin{eqnarray}
	\label{}
	Y_{e,d,u} = Y'^*_{e,d,u}
	\end{eqnarray}
Through these transformations, the left-handed particle physics we have in the standard model has an identical but right-handed counterpart in the mirror sector. The left-right symmetry is therefore restored in the two sectors theory.

Generalizations to supersymmetric and grand unified models can be found in literature - see for instance \cite{Berezhiani:2003xm,Berezhiani:2005ek,Berezhiani:1998kkk}. 
From grand unified mirror models we can get a natural understanding of the smallness of the mixing parameter $\epsilon$. Indeed, cosmology, astrophysics and experiments bound $\epsilon$ to have small values (of order $\lsim 10^{-6}$, see \S \ref{sec-mirror_mat_DM}); but if the theory does not provide any reasons for this smallness, it would be natural to expect $\epsilon \sim 1$.

A kinetic mixing term like (\ref{kin_mix_term}) is forbidden in GUTs, like for instance $\mathcal{SU} (5) \times \mathcal{SU} (5)'$ which do not contain abelian factors \cite{Berezhiani:1998kkk}. 
However, given that both $SU(5)$ and $SU(5)'$ symmetries are broken down to their $SU(3)\times SU(2)\times U(1)$ subgroups by the Higgs 24-plets $\Phi$ and $\Phi'$, such a term could emerge from the higher order effective operator 
	\begin{eqnarray}
	\label{GGpr}
	\mathcal{L}= -\frac{\zeta}{M^2} (G^{\mu\nu} \Phi) (G'_{\mu\nu} \Phi') 
	\end{eqnarray}
where $G_{\mu\nu}$ and $G'_{\mu\nu}$ are field-strength tensors respectively of $SU(5)$ and $SU(5)'$, and $M$ is some cutoff scale which can be of the order of $M_{P}$ or so. 
The operator (\ref{GGpr}) can be effectively induced with $\epsilon \sim \zeta (\langle\Phi\rangle/M)^2$ ($\zeta \sim \alpha / 3\pi \sim 10^{-3} $) by loop-effects involving some heavy fermion or scalar fields in the mixed representations of $SU(5)\times SU(5)'$. 
It can be shown that, taking the GUT scale as $\langle\Phi\rangle\sim 10^{16}$ GeV and $M\sim M_{P}$, 
the strength of kinetic mixing term (\ref{kin_mix_term}), could vary vary from $\epsilon \sim 10^{-10}$ to $10^{-8}$. 

Finally there is a last remark: also if the micro-physics is the same but with opposite chiralities in the two sectors, these sectors may have different macroscopic evolutions. In particular, as we will see more in detail in \S \ref{mirror_cosmology}, BBN bounds the mirror sector to a temperature lower than ordinary world.

\def \sec-mirror_cosmology{Mirror cosmology: thermodynamics}
\section{\sec-mirror_cosmology}
\label{mirror_cosmology}
\markboth{Chapter \ref{The_mirror_universe}. ~ The mirror universe}
                    {\S \ref{mirror_cosmology} ~ \sec-mirror_cosmology}

We introduced in \S \ref{sec-std-BBN} the criterion through which, by comparing experimental measures and theoretical calculations of the primordial abundance of light elements, the number of extra neutrinos is bounded to 
	\begin{eqnarray}
	\label{N_nu_bound}
	\Delta N_{\nu} < 1  \hspace{1cm} \mathrm{at} \;  95\% \; \mathrm{C. L.}
	\end{eqnarray}
Since the mirror universe particle content would lead to 
	\begin{eqnarray}
	\label{N_nu_mir}
	\Delta N_{\nu} = \frac{\Delta g}{2 \cdot \frac{7}{8}} = \frac{10.75}{1.75} \simeq 6.14
	\end{eqnarray}
it is straightforward that  the mirror sector can exist only if there is some mechanism which reduces this number. This is naturally achieved if the mirror world temperature is lower than the ordinary one - see equation (\ref{g_tot__def}); in particular equation (\ref{N_nu_bound}) is satisfied if
	\begin{eqnarray}
	\label{T_bound}
	\frac{T'}{T} < 0.64
	\end{eqnarray}
As stated above, it is possible that the two sectors, in spite of having the same microphysics, may have different macroscopic evolutions. In effect, the mirror sector temperature can be lower than the ordinary one if:
	\begin{itemize}
	\item The mirror sector has the inflationary reheating temperature lower than the ordinary one,
	as it happens in certain models \cite{BDM,Venya,KST};
	\item Interactions between ordinary and mirror particles are so "weak" that the worlds do not come 
	in thermal equilibrium.
	\end{itemize}
The expressions for entropy and energy density we introduced in \S \ref{sec-cosm-E-S} can be easily generalized to the mirror world \cite{Berezhiani:2000gw}:
	\begin{eqnarray}
	\label{mirror_entropy}
	s' (T') &=& \frac{2\pi^2}{45}\: q'_{tot}(T)\: T'^3 
	\end{eqnarray}
	\begin{eqnarray}
	\label{mirror_q_tot__def}
	q'_{tot}(T')&\equiv& \sum_{\tiny{bosons}} q'_b (T') \left(\frac{T'_b}{T'}\right)^3 +
        \frac{7}{8}\sum_{\tiny{fermions}} q'_f (T')
        \left(\frac{T'_f}{T'}\right)^3
	\end{eqnarray}
	\begin{eqnarray}
	\label{mirror_energy}
	\rho' (T')&=& \frac{\pi^2}{30}\: g'_{tot}(T')\: T'^4 
	\end{eqnarray}
	\begin{eqnarray}
	\label{mirror_g_tot__def}
	g'_{tot} (T')&\equiv& \sum_{\tiny{bosons}} g'_b (T') \left(\frac{T'_b}{T'}\right)^4 +
        	\frac{7}{8}\sum_{\tiny{fermions}} g'_f (T')
        	\left(\frac{T'_f}{T'}\right)^4
        	\end{eqnarray}
If the two sectors communicate only through gravity and they both expand adiabatically, they evolve with separately conserved entropies and therefore the ratio $x$, defined as:
	\begin{eqnarray}
	\label{x_definition}
	x \equiv \left(\frac{s'}{s}\right)^{\frac{1}{3}}
	\hspace{0.5cm} \Longrightarrow \hspace{0.5cm}
	\frac{T'(t)}{T(t)}=
	x \cdot \left[ \frac{q(T)}{q'(T')}
	\right]^{\frac{1}{3}}
	\end{eqnarray}
is time independent%
\footnote{Other interactions may lead to entropy exchange between the two sectors. This is analyzed more in detail in chapter \ref{chap-mir-MCP} for minicharged interactions. Of course entropy exchange imply that the parameter $x$ is not a constant anymore but a function of time (or temperatures).}.

The universe evolution in the Friedmann equations - see (\ref{friedmann1}) - is determined by the total energy density $\bar \rho = \rho + \rho'$, which enters the Hubble constant as
	\begin{eqnarray}
	\label{}
	H = \sqrt{\frac{8 \pi }{3} G \bar \rho}
	\end{eqnarray}
which can be recasted using the definitions (\ref{energy}) and (\ref{mirror_energy}) as
	\begin{eqnarray}
	\label{g_bar_def}
	H = 1.66 \sqrt{\bar g (T)} \frac{T^2}{M_P} = 
	1.66 \sqrt{\bar g' (T')} \frac{T'^2}{M_P}
	\end{eqnarray}
where
	\begin{eqnarray}
	\label{g_tot_with_mirror}
	\overline{g}(T) = g (T) (1 + \alpha x^4) \hspace{1.5cm}
	\overline{g}'(T') = g' (T')\left(1 + \frac{1}{\alpha x^4}\right) 
	\end{eqnarray}
and the factor $\alpha (T,T')$, defined in the following way
	\begin{eqnarray}
	\label{beta_factor}
	\alpha=\frac{g'(T')}{g(T)} \cdot 
	\left[\frac{q(T)}{q'(T')}\right]
	^{\frac{4}{3}}
	\end{eqnarray}
takes into account that for $T \neq T'$ the relativistic particle content of the two worlds can be different.

It can be easily seen that the bound (\ref{N_nu_bound}) on the extra-neutrino number can be expressed as a bound on $x$. If we assume indeed $ g \simeq g'$ in (\ref{x_definition}) and recall the condition on the mirror temperature in (\ref{T_bound}) we have 
	\begin{eqnarray}
	\label{x_bound}
	x \simeq \frac{T'}{T} < 0.64
	\end{eqnarray}
The link between $x$ and $\Delta N_{\nu}$ can be summerized in the formula
	\begin{eqnarray}
	\label{x_N_nu}
	\Delta N_{\nu} = 6.14 \, x^4
	\end{eqnarray}
which is valid in the limit $\alpha \simeq 1$.
This means that the bound on $x$ we reported in equation (\ref{x_bound}) has a very weak dependence on $\Delta N_{\nu}$, but also that any changes on $x$ has a great impact on $\Delta N_{\nu}$. Some consequences of this phenomenon are analyzed in \S \ref{sec-E-exch}.

In conclusion, the presence of the mirror sector having $x < 0.64$ has pratically no impact on the standard BBN. But the impact of the ordinary world for the mirror BBN is much stronger, as we will discuss further in the following chapter.

\def \sec-mirror_mat_DM{Mirror matter as dark matter}
\section{\sec-mirror_mat_DM}
\label{sec-mirror_mat_DM}
\markboth{Chapter \ref{The_mirror_universe}. ~ The mirror universe}
                    {\S \ref{sec-mirror_mat_DM} ~ \sec-mirror_mat_DM}


Cosmological observations indicate that the Universe is nearly flat, with the energy density very close to the critical one: $\Omega_{\rm tot} = 1$. The non-relativistic matter in the Universe consists of two components, baryonic (B) and dark (D). 
\it Dark Matter \rm (DM) is non-luminous and non-absorbing matter which interacts with the ordinary one mainly via gravity; its existence is now well established, since it has measurable effects on the velocity of various luminous objects and on the cosmic microwave background anisotropies \cite{Yao:2006px}.

Recent data fits imply \cite{Tegmark:2006az}:
	\begin{eqnarray}
	\label{Om}
	\Omega_{\rm B} h^2 = 0.0222 \pm 0.0007 \hspace{1.5cm}
	\Omega_{\rm D} h^2 = 0.105 \pm 0.004
	\end{eqnarray}
%
where $h=0.73\pm 0.02$ is the Hubble parameter. Hence, matter gives only a fraction of the total energy density: 
	\begin{eqnarray}
	\label{}
	\Omega_{\rm M} = \Omega_{\rm B} +\Omega_{\rm D} = 0.24 \pm 0.02
	\end{eqnarray}
while the rest is attributed to dark energy (cosmological term): 
	\begin{eqnarray}
	\label{}
	\Omega_\Lambda = 0.76 \mp 0.02
	\end{eqnarray}
The closeness of $\Omega_{\rm D}$ and $\Omega_{\rm B} $
($\Omega_{\rm D}/\Omega_{\rm B} = 4.7 \pm 0.3$) 
gives rise to a painful problem, called {\it Fine Tuning} problem. 
Both $\rho_{\rm D}$ and $\rho_{\rm B}$ scale as $\sim a^{-3}$ with the Universe expansion, and their ratio should not  be dependent on time. Why then these two fractions are comparable, if they have a drastically different nature and different origin?

An answer to this question is found assuming that DM is made of mirror baryons.
We can indeed define the \it baryon mass density \rm $\rho_{\rm B} =m_{\rm B} n_{\rm B}$,     
where $m_{\rm B} \simeq 1$ GeV is the nucleon mass, and 
$n_{\rm B}$ is the baryon number density. 
Thus we have  
	\begin{eqnarray}
	\label{}
	\Omega_{\rm B} h^2 = 2.6 \times 10^8 \, \frac{n_{\rm B}}{s}
	\end{eqnarray}
where $s$ is as usual the entropy density; equation (\ref{Om}) translates in 
$B= n_{\rm B}/s \approx 0.85\times 10^{-10}$, 
in a nice consistence with the BBN bound
$B = (0.5-1) \times 10^{-10}$ \cite{Yao:2006px}.  

The origin of non-zero baryon asymmetry $B$, which 
presumably was produced in a very early universe 
as a tiny difference $n_{\rm B} = n_b - n_{\bar b}$
between the baryon and anti-baryon abundances, is yet unclear.  
The popular mechanisms known as GUT Baryogenesis, Leptogenesis, 
Electroweak Bariogenesis, etc., all are conceptually based on 
out-of-equilibrium processes violating $B(B-L)$ and C/CP 
and they  generically predict $B$   as a function of  
the relevant interaction strengths and CP-viollating phases.

Concerning dark matter, almost nothing is known besides the fact it must be 
constituted by some cold relics with mass $m_{\rm D}$ 
which exhibits enormous spread between different popular candidates    
as e.g. axion ($\sim 10^{-5}$ eV), 
sterile neutrino ($\sim 10$ keV), 
WIMP/LSP ($\sim 1$ TeV), or Wimpzilla  ($\sim 10^{14}$ GeV).  
None of these candidates has any organic  
link with any of the popular baryogenesis schemes.
The respective abundances $n_{\rm D}$ could be produced 
thermally (e.g. freezing-out of WIMPs) or non-thermally  (e.g. axion condensation 
or gravitational preheating for Wimpzillas),  but in no case they are related 
to the CP-violating physics.    
In this view,  the conspiracy between $\rho_{\rm D} = m_{\rm D} n_{\rm D}$
and $\rho_{\rm B} =m_{\rm B} n_{\rm B}$ indeed looks as a big paradox.  

Mirror baryons may shed a new light to the baryon and dark matter coincidence problem: they are invisible in terms of the ordinary photons but interact gravitationally with ordinary matter and thus they constitute a viable dark matter candidate \cite{Berezhiani:2000gw,BDM,Berezhiani:2000gh}. 

In general we can write
	\begin{eqnarray}
	\label{}
	\Omega_D = \Omega'_B + \Omega_x \hspace{1.5cm}
	 \Omega'_B \leq  \Omega_D
	\end{eqnarray}
where $\Omega'_B$ is the mirror baryon density and $\Omega_x$ is some kind of exotic dark matter. We are interested in a situation when the ratio
	\begin{eqnarray}
	\label{beta_DM_def}
	\beta = \frac{\Omega'_B}{\Omega_B} 
	\end{eqnarray}
falls in the range from 1 to few tens. From $\beta$ we can also work out a lower bound on the parameter $x$: since $n'_{\gamma} = x^3 n_{\gamma}$, we can write
	\begin{eqnarray}
	\label{beta_x}
	\beta = x^3 \frac{\eta'}{\eta} 
	\hspace{1cm} \Longrightarrow \hspace{1cm} 
	x^3 = \beta \frac{\eta}{\eta'}
	\end{eqnarray}
Recalling now that $\eta \sim 10^{-9}$ and $\eta' \lsim 10^{-3}$, we get the lower limit
	\begin{eqnarray}
	\label{}
	x > 10^{-2}
	\end{eqnarray}
Let us assume now $\Omega_B = \Omega_D$;
the closeness of $\Omega_D = \Omega_B$ and $\Omega'_B$ arises because of the mirror symmetry. 
Their difference instead can be explained, in a \it symmetric mirror scenario \rm where $m_B = m'_B$, in terms of a $n'_B > n_B$, as one can achieve in certain leptogenesis models - see \cite{Berezhiani:2003xm,Berezhiani:2005ek,Bento:2002sj}.

An other intriguing mechanism leading to $\Omega'_B > \Omega_B$ is based on the possibility that the Higgs Vacuum Expectation Values (VEVs), which is the electro-weak scale of the theory, may be different in the two sectors \cite{BDM} :
	\begin{eqnarray}
	\label{}
	\langle \phi \rangle = v \hspace{1.2cm}
	\langle \phi' \rangle = v' \hspace{1.2cm}
	\frac{v'}{v} = \zeta \neq 1
	\end{eqnarray}
Changing the Higgs VEV implies different weak interaction strenghts and different masses of the weak gauge bosons and of the fermions in the two sectors:
	\begin{eqnarray}
	\label{}
	G'_F = \frac{G_F}{\zeta^2 } \hspace{1.2cm}
	M'_{W,Z} = \zeta M_{W,Z} \hspace{1.2cm}
	m_f = \zeta m'_f 
	\end{eqnarray}
and also a slower change in the confinement scale $\Lambda$, which scales approximately as $\zeta^{0.3}$ \cite{Berezhiani:2000gh,Gianfagna:2004je}. This way, assuming $\zeta \sim 100$ we find $\Lambda'\sim 800 MeV$, which implies
	\begin{eqnarray}
	\label{asym_mir_masses}
	m'_f \sim 100 \, m_f \hspace{1.5cm}
	m'_B \sim 5 \, m_B 
	\end{eqnarray}
This discrepancy on the baryon mass leads to $\Omega'_B \sim 5 \Omega_B$ \cite{AIP_Zurab}.
Consequences on BBN of having heavier mirror fermions will be analyzed in section \ref{chap-mir-MCP}. 
Other effects arise since the Bohr radius scales as $m_e^{-1} \sim 0.01$ and this implies that mirror atoms are much more compact and therefore less collisional than the ordinary ones. Moreover, the hydrogen recombination and photon decoupling in M-sector would occur much earlier the matter-radiation equality epoch, and as a consequence, mirror matter will manifest rather like a cold dark matter. 

Also, mirror nuclei will be all, but hydrogen, unstable since 
the light quark mass difference scales as
	\begin{eqnarray}
	\label{}
	(m'_d-m'_u)\approx \zeta (m_d-m_u)
	\end{eqnarray}
We therefore expect the mirror neutron $n'$ to be heavier than the mirror proton $p'$ by few hundred MeV. 
Clearly, such a large mass difference cannot be compensated by the nuclear binding energy and hence
even bound neutrons will be unstable against $\beta$ decay $n'\to p' e' \bar{\nu}'_e$.
Thus in such an asymmetric mirror world hydrogen will be the only stable nucleus \cite{BDM}. 

In the following, if not differently specified, we will always refer to the \it symmetric mirror model\rm, that is we will assume equal masses for the elementary particles. 
In this scenario mirror matter can be unambigously described by mean of two only free parameters, $x$ and $\beta$.



\def \sec-MCP-bounds{Bounds on MCPs and prospects for their search}
\section{\sec-MCP-bounds}
\label{sec-MCP-bounds}
\markboth{Chapter \ref{The_mirror_universe}. ~ The mirror universe}
                    {\S \ref{sec-MCP-bounds} ~ \sec-MCP-bounds}

Bounds on MCP parameters - mainly charge and mass - have been worked out by several authors - see for instance \cite{Davidson:1991si}. 
Complete and more recent reviews can be found in \cite{Davidson:2000hf} and \cite{Raffelt:1996wa}, where bounds arising from accelerator experiments (AC), BBN, globular clusters (GC), supernova 1987A (SN) and white dwarfs (WD) are reported.  
Also, Dubovsky et al. derived bounds from the Cosmic Microwave Background Radiation (CMBR) \cite{Dubovsky:2003yn}.

We are interested in a scenario where the millicharged particles have mass $m_{\epsilon} \sim m_e = 5 \cdot 10^{5}eV$ or  $m_{\epsilon} \sim 100 \, m_e = 5 \cdot 10^{7}eV$. The $\epsilon$ values ruled out for these masses in the previous references are listed in Table \ref{tab-MCP_bounds}.
	\begin{table}[htdp]
	\caption{BBN bounds fom literature.}
	\begin{center}
	\begin{tabular}{|c|c|c|c|} 
	\hline 
	& BBN & SN & AC 
	\\ \hline
	$m_{\epsilon} \sim m_e$ & $\epsilon \lsim 2 \cdot 10^{-9}$ & 
	$10^{-7} \lsim \epsilon \lsim 10^{-9}$ & $\epsilon \lsim 10^{-5}$
	\\ \hline
	$m_{\epsilon} \sim 100 \,m_e$ & - & - & $\epsilon \lsim 10^{-3}$
	\\ \hline
	\end{tabular}\\ 
	\end{center} 
	\label{tab-MCP_bounds}
	\end{table}
The BBN bound is for Davidson et al. \cite{Davidson:2000hf}:
	\begin{eqnarray}
	\label{}
	\epsilon \leq 2 \cdot 10^{-9}
	\hspace{0.8cm} \mathrm{valid \; when} \hspace{0.8cm} 
	m_{\epsilon} \leq m_e
	\end{eqnarray}
This bound has been calculated for a model containing only MCPs and one paraphoton in the ultrarelativistic (UR) limit, that is assuming massless MCPs and electrons and approximating the interaction rate $\Gamma$ to - see \S \ref{sec-ee->ff}:
	\begin{eqnarray}
	\label{}
	\Gamma_\mathrm{{UR}} = \frac{\zeta_3}{2\pi} \left( \epsilon \alpha \right)^2 T
	\end{eqnarray}
	\begin{figure}[htbp]
	\begin{center}
	 \includegraphics[scale=1.2]{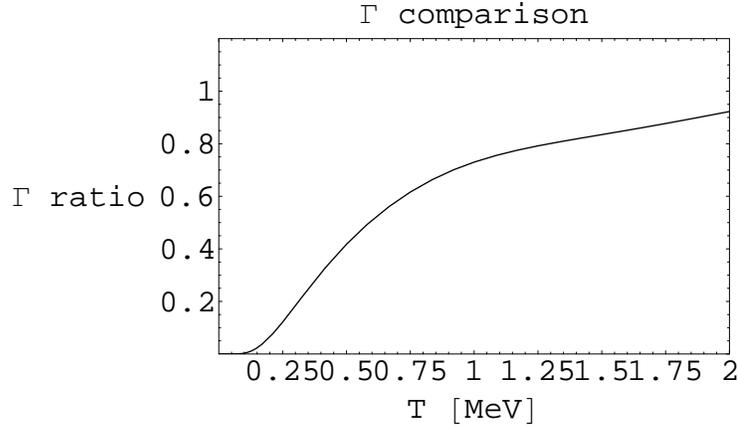}	
	\caption{{\it Ratio of $\Gamma$ numerically calculated and 
	$\Gamma$ in ultrarelativistic particles approximation.}}
	\label{Gamma_cfr_1}
	\end{center}
	\end{figure}
But BBN takes place at temperatures comparable with the electron mass - see \S \ref{sec-std-BBN} - when this approximation is not very good%
\footnote{The procedure we used to numerically work out $\Gamma$ can be found in \S \ref{sec-ee->ff}.}
 as we show in Figure \ref{Gamma_cfr_1}, where the ratio
	\begin{eqnarray}
	\label{}
	\Gamma \; \mathrm{ratio} = \frac{\Gamma_\mathrm{numeric}}{\Gamma_\mathrm{{UR}}}
	\end{eqnarray}
 is plotted. 
 
The bounds we listed above have been calculated for models containing MCPs with or without one paraphoton but are not specific for the millicharged mirror model.
Glashow pointed out \cite{Glashow:1985ud} that a sensitive laboratory test of the mirror model comes from the orthopositronium system: the kinetic mixing leads orthopositronium (o-Ps) to oscillate into mirror orthopositronium with the probability:
	\begin{equation}
	P(o-Ps \to o-Ps') = \sin^2 \omega t,
	\label{eq:osc}
	\end{equation}
where $\omega = 2\pi \epsilon f$ and where $f = 8.7 \times 10^4$ MHz is the contribution to the ortho-para splitting from the one photon annihilation diagram involving orthopositronium.
Since mirror o-Ps are not detected, in experiments they look like \it invisible \rm decays. 
Recent searches by Badertscher et al. gave no evidence for positronium decays into invisible states in a collected sample of about $6 \cdot 10^6$ decays; this way the photon mixing strenght is bounded to $\epsilon \leq 1.55 \cdot 10^{-7}$ at $90 \%$ C.L. \cite{Badertscher:2006fm}.

BBN bounds specific for the mirror model have been worked out in 1987 by Carlson and Glashow \cite{Carlson:1987si}, who derived $\epsilon <  3\cdot 10^{-8}$. Nevetheless, we have today new data and new theoretical hints which render this limit worth to be reanalyzed. 
This will be done in chapter \ref{chap-mir-MCP}.

An other interesting experiment which may have detected millicharged interactions of ordinary and mirror particles is DAMA/NaI \cite{Bernabei:1996vj,Bernabei:2003za}. DAMA is an observatory for rare processes based on the development and use of various kinds of radiopure scintillators located in the underground laboratory of Gran Sasso.
The DAMA/NaI experiment was searching for scattering of WIMPs (weakly interacting massive particles, which are a popular DM candidate) on a target of $100$ kg of radiopure NaI, measuring the recoil energy of the scattered NaI atoms.
The detected signal is expected to have a small annual modulation due to the Earth's motion around the sun: the Earth should be crossed by a larger flux of Dark Matter particles roughly around June 2nd (when its rotational velocity is summed to the one of the solar system with respect to the Galaxy) and by a smaller one roughly around December 2nd (when the two velocities are subtracted), according to
	\begin{eqnarray}
	\label{}
	A \cos \left(2 \pi \, \frac{t-t_0}{T}
	\right)
	\end{eqnarray}
The experiment took data over $7$ annual cycles and was put out of operation in July 2002. The data analysis gave evidence for an annual modulation at $6.3 \sigma $ C.L. with the fitted parameter values $T = (1.00 \pm 0.01)$ year (the expected value for $T$ was $1$ year), $t_0 = 144 \pm 22$ days (the expected value for $t_0$ was 152 days, that is 2 June), while the signal strength is $A = (0.019 \pm 0.003)$ cpd/kg/keV.

Foot proposed to interpret these data in terms of elastic scattering of millicharged mirror atoms on ordinary matter nuclei in the symmetric mirror scenario \cite{Foot:2003iv,Foot:2004ej}. 
Indeed, he showed that DAMA/NaI as well as other DM searching experiments (such as CRESST/Sapphire) are sensitive to this kind of interactions provided that $\epsilon \sim 5 \cdot 10^{-9}$.
We will show in chapter \ref{chap-mir-MCP} that such a value for $\epsilon$ in a symmetric mirror model, where the particle masses are the same than the ordinary ones, is compatible with BBN bounds, also if it may give problems when considering CMB and LSS. 
We will then state reasons for further investigations of the same phenomenon but within an asymmetric mirror model, having $m_f' \sim 100 m_f$, $m_B' \sim m_B$%
\footnote{Where $f$ stands for any fermions and $B$ for any baryons.}%
.







%% file: files/mirror_cosmology.tex


\def \chap-mir-BBN{BBN in presence of the mirror world}
\chapter{\chap-mir-BBN}
\label{chap-mir-BBN}
\markboth{Chapter \ref{chap-mir-BBN}. ~ \chap-mir-BBN}
                    {Chapter \ref{chap-mir-BBN}. ~ \chap-mir-BBN}


BBN in presence of the \it symmetric mirror sector\rm, where the particle masses are the same than the ordinary ones, is analyzed in this section. 
In  \S \ref{sec-Nu-dec-mir} we calculate the equations which link the ordinary and the mirror sector temperatures; then we numerically solve these equations using Mathematica.
Once the temperatures are known, it is possible to work out the total exact number of degrees of freedom (DOFs) in both sectors, which can be, as it is common in the literature, expressed in terms of extra-neutrino number.
We report in \S \ref{sec-Num-calc} the results of these calculations, which have been made again with Mathematica.
Finally, we use the Kawano code for BBN \cite{Kawano:1992ua} to work out the mass fraction of light elements produced in both sectors during the primordial nucleosynthesis; the presence of the other sector indeed, leads in both sectors to the same effects of having more particles. 
These topics can be found in \S \ref{sec-elem-prod-ord} and \S \ref{sec-elem-prod-mir}.
For the moment we do not take in account of the possible existence of MCPs or other interactions - but gravity - between the two secors; millicharged interactions will be instead considered in the following chapter.

\def \sec-Nu-dec-mir{Neutrino decoupling and $e^{\pm}$ annihilation
: working out the mirror and ordinary temperatures}
\section{\sec-Nu-dec-mir}
\label{sec-Nu-dec-mir}
\markboth{Chapter \ref{chap-mir-BBN}. ~ \chap-mir-BBN}
                    {\S \ref{sec-Nu-dec-mir} ~ \sec-Nu-dec-mir}

As already stressed in chapter \ref{The_mirror_universe}, ordinary and mirror sectors have the same microphysics; it is therefore obvious that the neutrino decoupling temperature $T_{D\nu}$ is
the same in both them, that is $T_{D\nu}=T_{D\nu}'$. 

This fact will be used together with the entropy density conservation to find equations the solution of which will give the mirror photon temperature $T'$ corresponding to any values of the ordinary photon one $T$.

\def \sec-Nu-dec-eqs{The equations}
\subsection{\sec-Nu-dec-eqs}
\label{sec-Nu-dec-eqs}
\markboth{Chapter \ref{chap-mir-BBN}. ~ \chap-mir-BBN}
                    {\S \ref{sec-Nu-dec-eqs} ~ \sec-Nu-dec-eqs}

We argumented in \S \ref{mirror_cosmology} that the mirror world must be colder than the ordinary one and therefore the neutrino decoupling takes place before in the mirror sector. 
 If we call $T_{D\nu'}$ the ordinary world temperature when the mirror annihilation takes place, we can split the universe evolution in three phases:

\subsubsection{Phase 1: $T > T_{D\nu'}$}

Photons and neutrinos are in thermal equilibrium in both worlds, that is $T_{\nu} = T \; , \; T_{\nu}' = T'$. 
Using equations (\ref{entropy}), (\ref{q_tot__def}), (\ref{energy}) and (\ref{g_tot__def}) and the corresponding ones for the mirror sector that is (\ref{mirror_entropy}), (\ref{mirror_q_tot__def}), (\ref{mirror_energy}) and (\ref{mirror_g_tot__def}) we are able to calculate the  DOFs number in ordinary or mirror world; but to work out the \it total \rm DOFs number, that is summed on both worlds, we need the mirror temperature $T'$ as a function of the ordinary temperature $T$ or viceversa.

If we neglet 
the entropy exchanges between the sectors we can get both these functions imposing	$x=$const in:
	\begin{eqnarray}
	\label{T'_T_1_1}
	x^3=
	\frac{s'\cdot a^3}{s \cdot a^3}
	=
	\frac{\left[ \frac{7}{8}q_e(T')+ q_{\gamma} 
	+ \frac{7}{8} q_{\nu} 
	\right]T'^3} 
	{\left[  \frac{7}{8}q_e(T)+ q_{\gamma} 
	+ \frac{7}{8} q_{\nu} 
	\right] T^3}
	\end{eqnarray}
where $q_{\nu} = 6$ and $q_{\gamma} = 2$ as introduced in (\ref{ultrarel_DOFs}), while $q_e (T)$  stands for:
	\begin{eqnarray}
	\label{}
	q_e (T) \equiv \frac{7}{8} \; \frac{s_e(T)}{\frac{2\pi^2}{45} T^3} \hspace{1.5cm}
	s_e(T) \equiv \frac{p_e (T) + \rho_e (T)}{T} \cr\cr
	\end{eqnarray}
and $\rho_e (T) $, $ p_e (T) $  are the same than in (\ref{energy_density}) and (\ref{pressure}).

Equation (\ref{T'_T_1_1}) can be numerically resolved and will finally give the function $T'(T)$ for every $T > T_{D\nu'}$.

\subsubsection{Phase 2: $T_{D\nu} \leq T < T_{D\nu'}$}

At $T \simeq T_{D\nu'}$ mirror neutrinos decouple - see \S \ref{sec-nu-dec} - and soon after mirror electrons and positrons annihilate, raising the mirror photon temperature. Nevertheless, ordinary photons and neutrinos still have the same temperature $T$.
We have therefore two equations: the first one is
%
	\begin{eqnarray}
	\label{}
	\frac{22}{21}=\frac{\frac{7}{8}q_{e}(T')+q_{\gamma}}{\frac{7}{8}q_{\nu}}
	\left(\frac{T'}{T_{\nu}'}\right)^3 
	\end{eqnarray}
and its solution gives $T_{\nu}'$ as a function of $T'$; the second is 
%
	\begin{eqnarray}
	\label{}
	x^3=
	\frac{s'\cdot a^3}{s \cdot a^3}
	=
	\frac{\left[ \frac{7}{8}q_e(T')+ q_{\gamma} \right] T'^3+ 
	\frac{7}{8} q_{\nu} 
	T_{\nu}'\,^3} 
	{\left[  \frac{7}{8}q_e(T)+ q_{\gamma} + 
	\frac{7}{8} q_{\nu} \right] T^3}
	\end{eqnarray}
and its solution gives $T'$ as a function of $T$.

\subsubsection{Phase 3: $T \leq T_{D\nu}$}

At $T\simeq T_{D\nu}$ ordinary neutrinos decouple and soon after ordinary electrons and positrons anihilate; therefore we need one more equation to work out the ordinary neutrino temperature $T_{\nu}$ as a function of the ordinary photon one $T$:
	\begin{eqnarray}
	\label{}
	\frac{22}{21}=\frac{\frac{7}{8}q_{e}(T')+q_{\gamma}}{\frac{7}{8}q_{\nu}}
	\left(\frac{T'}{T_{\nu}'}\right)^3 
	\end{eqnarray}
	\begin{eqnarray}
	\label{}
	\frac{22}{21}=\frac{\frac{7}{8}q_{e}(T)+q_{\gamma}}{\frac{7}{8}q_{\nu}}
	\left(\frac{T}{T_{\nu}}\right)^3 
	\end{eqnarray}
	\begin{eqnarray}
	\label{}
	x^3=
	\frac{s'\cdot a^3}{s \cdot a^3}
	=
	\frac{\left[ \frac{7}{8}q_e(T')+ q_{\gamma} \right] T'^3+ 
	\frac{7}{8} q_{\nu} 
	T_{\nu}'\,^3} 
	{\left[  \frac{7}{8}q_e(T)+ q_{\gamma} \right] T^3 + 
	\frac{7}{8} q_{\nu} T_{\nu}^3}
	\end{eqnarray}
Once both ordinary and mirror photon temperatures are known, it is straightforward to calculate the total energy and entropy densities; then, reversing equations (\ref{entropy}) and (\ref{energy}), we can work out the entropic $(q)$ and energetic $(g)$ total number of degrees of freedom. Calculations have been made for several different values of $x$, as reported in the following section.

	\begin{figure}[htbp]
	\begin{center}
	 \includegraphics[scale=0.8]{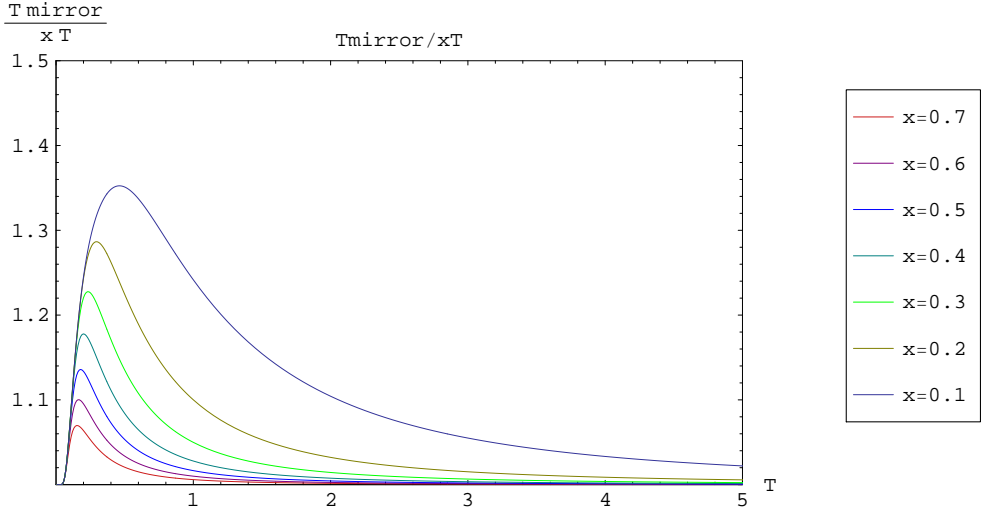}	
	\caption{{\it 	
	$\frac{T'}{xT}$ for several values of $x$. The asymptotic value for this ratio is $1$
	for both high and low temperatures, as expected.}}
	\label{Tm_xT_fig}
	\end{center}
	\end{figure}

\def \sec-Num-calc{Numerical solution}
\subsection{\sec-Num-calc}
\label{sec-Num-calc}
\markboth{Chapter \ref{chap-mir-BBN}. ~ \chap-mir-BBN}
                    {\S \ref{sec-Num-calc} ~ \sec-Num-calc}


The equations we introduced above have been numerically solved with Mathematica.
%
Since $x$ is a free parameter in our theory, several values have been used for it - from $0.1$ to $0.7$ with step $0.1$ or less, see \S \ref{mirror_cosmology} - to study its effect on the light elements production. 
In the extreme cases
	\begin{eqnarray}
	\label{}
	T \gg T_{D\nu'}=\frac{T_{D\nu}}{x} 
	\hspace{1cm} \mathrm{or} \hspace{1cm}
	T \ll T_{ann \: e^{\pm}}\simeq 1MeV
	\end{eqnarray}
%
we expect $q(T)\simeq q'(T')$; therefore in this limit - see (\ref{x_definition}) - 
the ratio of mirror and ordinary photon temperature should be $x$, that is $\frac{T'}{xT} = 1$.
Instead, when 
	\begin{eqnarray}
	\label{}
	T_{D\nu'} \gsim T \gsim T_{ann \: e^{\pm}}
	\end{eqnarray}
we expect $q'(T') \leq q(T)$ because the $e^{\pm}$ annihilation takes place before in the mirror world, leading to a decrease of $q'(T')$ and a corresponding increase of $T'$ in order to keep constant the entropy densities ratio; thus we expect 
$\frac{T'}{xT} > 1$. 
Later on, when the ordinary electrons-positrons annihilate, they make even $T$ increase and thus the ratio $\frac{T'}{xT}$ decrease to the asympotic value $1$.
These remarks have been numerically verified; the ratio $\frac{T'}{xT}$ is plotted in Figure \ref{Tm_xT_fig}.

\def \sec-DOFs-num{The number of degrees of freedom}
\section{\sec-DOFs-num}
\label{sec-DOFs-num}
\markboth{Chapter \ref{chap-mir-BBN}. ~ \chap-mir-BBN}
                    {\S \ref{sec-DOFs-num} ~ \sec-DOFs-num}

%
%
Using equations from (\ref{energy_density}) to (\ref{entropy_generic}) to reverse (\ref{entropy}) and (\ref{energy}), we can work out the total number of entropic ($\bar q$) and energetic ($\bar g$) DOFs at any temperatures $T$. 
Of course, we can apply the same procedure to work out the \it standard\rm , that is in absence of the mirror world, total $q$ and $g$, as well as $\bar q'$ and $\bar g'$ and the influence of a world with the other one. 

In general we expect $\bar q$ ($\bar g$) to have a cubic (quartic) dependence on $x$ - see equation (\ref{g_tot_with_mirror}) - when the temperature is not close to the $\nu$ decoupling and the $e^{\pm}$ annihilation phases, that is, when $\alpha=1$ - see equation (\ref{beta_factor}).

Numerical calculations have been made using Mathematica; in Table \ref{tab-DOFs} some values are reported for special temperatures and several values of $x$. As expected, the DOFs number is always higher than the standard and increases with $x$. Moreover, the mirror sector values are higher than the ordinary ones of a factor of order $x^{-3}$ (for $\bar q$) or $x^{-4}$ (for $\bar g$).

\begin{figure}[htbp]
	\begin{center}
	 \includegraphics[scale=0.6]{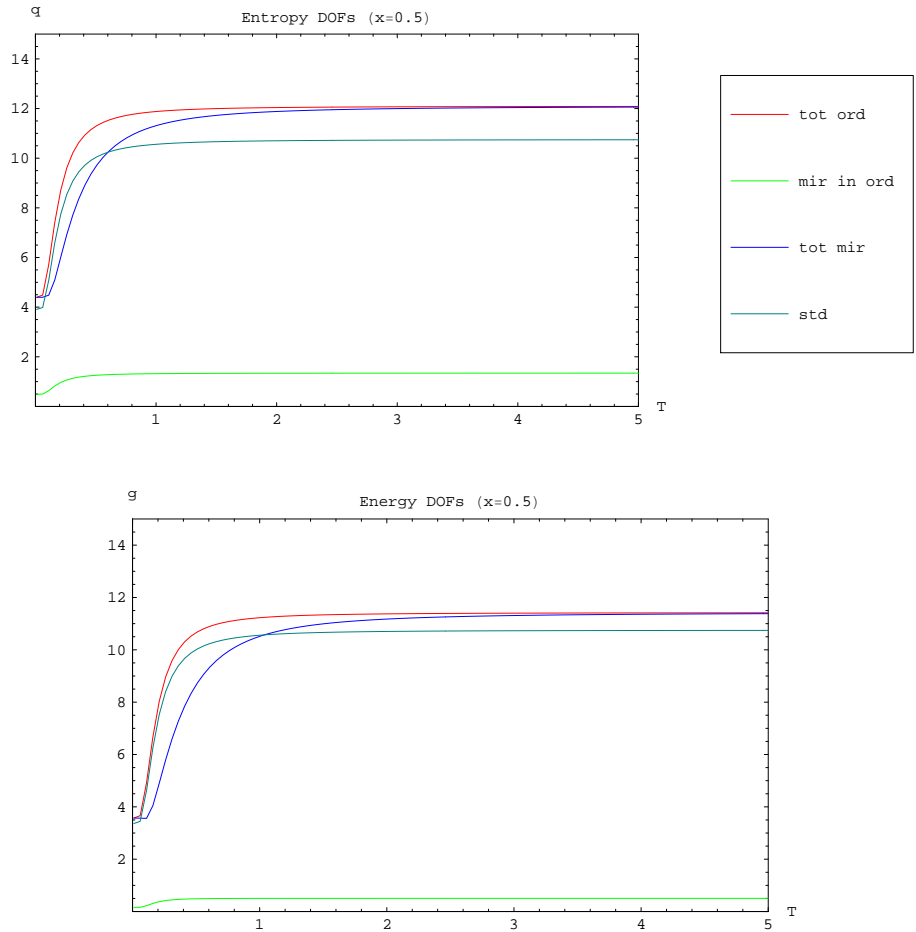}	
	\caption{{\it In this figure we plot: standard $q$ and $g$ (in dark green), 
	mirror contribution to total ordinary $q$ and $g$ (in light green), 
	total $\bar q$ and $\bar g$ in mirror (in blue) and ordinary (in red) sectors, 
	all for $x=0.5$. Mirror $\bar q$ and $\bar g$ (blue lines) have been multiplied respectively
	by $x^3$ and $x^4$ to make them comparable with the ordinary values 
	since $\frac{T'}{xT} \sim 1$. 
	}}
	\label{DOF_tot_x_05}
	\end{center}
	\end{figure}

In Figure \ref{DOF_tot_x_05} $q$, $\bar q$, $\bar q'$ and the ordinary DOFs number increase due to the mirror sector are plotted for the intermediate value $x=0.5$; the second plot in the same figure shows the corresponding $g$s.
In the figure $\bar q'$ ($\bar g'$) has been scaled by a factor $x^3 $ ($x^4$); this way the asymptotic values are the same than the ordinary sector ones because at the extremes $\frac{T'}{T}=x$ (see Figure \ref{Tm_xT_fig}). 
We can see that $\bar q$, $\bar g$, $\bar q'$ and $\bar g'$ have similar trends, but, due to $T'<T$, mirror DOFs begin to decrease before.

\begin{figure}[htbp]
	\begin{center}
	 \includegraphics[scale=0.5]{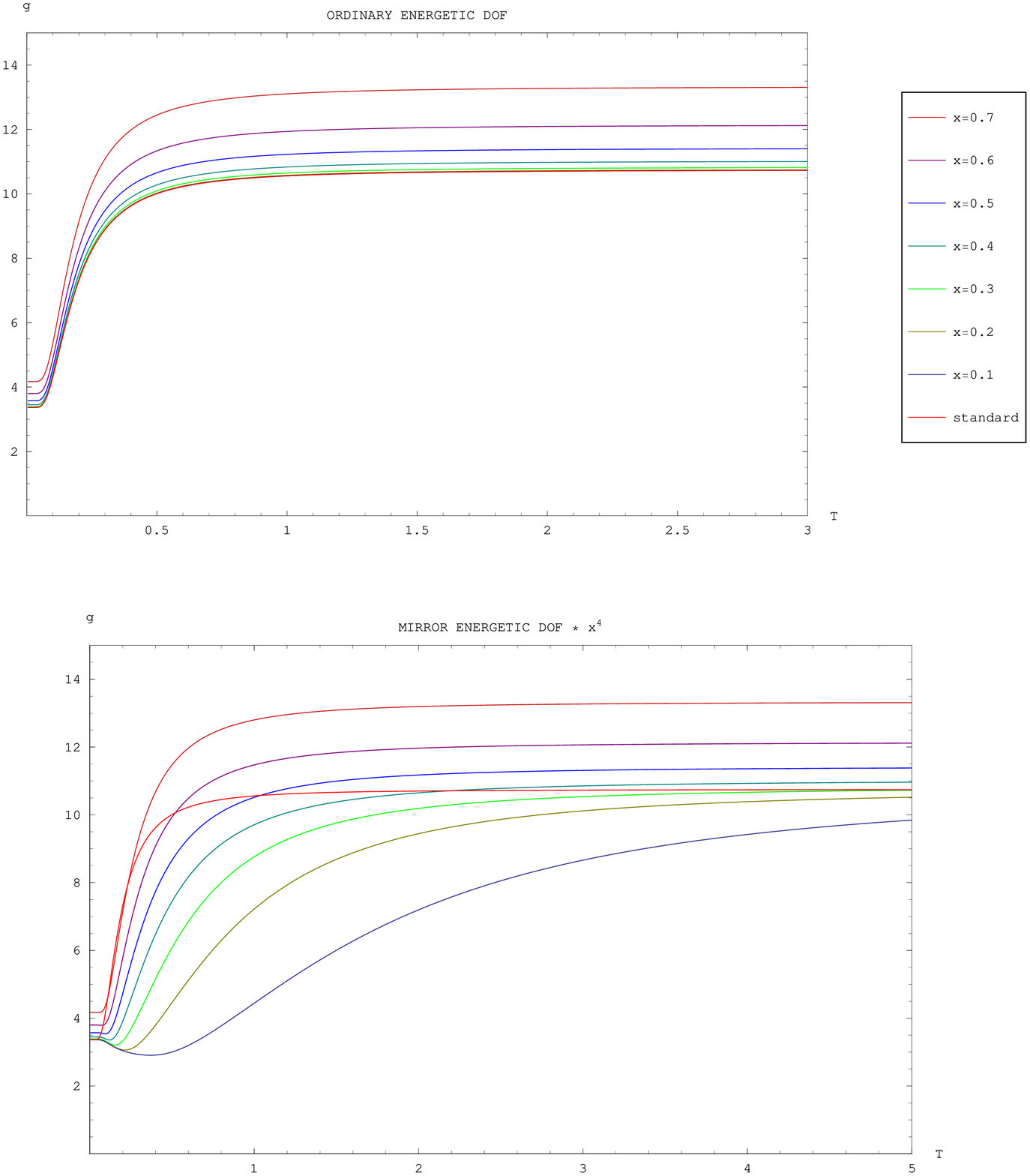}	
	\caption{{\it $\overline{g}$ in ordinary and mirror sectors 
	for several values of $x$ compared with the
	standard.}}
	\label{ord_mir_gDOF}
	\end{center}
	\end{figure}

In Figure \ref{ord_mir_gDOF} $\bar g$ in ordinary and mirror sectors are plotted in comparison with the standard for several $x$ values (from $0.1$ to $0.7$ with step $0.1$). The foreseen quartic dependence of $g$ on $x$ at the extremes is proved correct. 
We note that the plot shape does not change with $x$ in the ordinary sector, while it does in the mirror one. This sector evolves with the temperature $T' \sim xT < T$; therefore the number of DOFs converges later for lower $x$s.

	\begin{table}[htdp]
	\caption{Standard and not standard total DOFs number for several $x$ values 
	in both ordinary and mirror sector.}
	\begin{center}
	\begin{tabular}{|c|c|c|c|c|c|} 
	\hline 
	$T (MeV) $ & $T_1 = 0.005$ & $T_2 = 0.1$ & $T_3 = 0.5$ & $T_4 = 1$ & $T_5 = 5$
	\\ \hline \hline
	$q$ {\footnotesize (standard)} & 3.91 & 4.78 & 10.0 & 10.6 & 10.74
	\\ \hline
	$g$ {\footnotesize (standard)} & 3.36 & 4.30 & 10.0 & 10.6 & 10.74 
	\\ \hline \hline
	\multicolumn{6}{|c|}{$\mathbf{x = 0.1}$} 
	\\ \hline \hline
	$T' (MeV)$ & 0.0005 & 0.0107 & 0.0675 & 0.124 & 0.511 
	\\ \hline
	$\bar q$ & 3.91 & 4.78 & 10.0 & 10.6 & 10.75 
	\\ \hline
	$\bar q'$ & 3910 & 3910 & 4070 & 5520 & 10000
	\\ \hline
	$\bar g$ & 3.36 & 4.30 & 10.0 & 10.6 & 10.74
	\\ \hline
	$\bar g'$ & 33600 & 32900 & 30000 & 44400 & 98400 
	\\ \hline \hline
	\multicolumn{6}{|c|}{$\mathbf{x = 0.3}$} 
	\\ \hline \hline
	$T' (MeV)$ & 0.0015 & 0.0321 & 0.170 & 0.315 & 1.50
	\\ \hline
	$\bar q$ & 4.015 & 4.91 & 10.3 & 10.8 & 11.0 
	\\ \hline
	$\bar q'$ & 148 & 148 & 261 & 347 & 404 
	\\ \hline
	$\bar g$ & 3.39 & 4.34 & 10.1 & 10.6 & 10.8 
	\\ \hline
	$\bar g'$ & 418 & 409 & 750 & 1080 & 1320 
	\\ \hline \hline
	\multicolumn{6}{|c|}{$\mathbf{x = 0.5}$} 
	\\ \hline \hline
	$T' (MeV)$ & 0.0025 & 0.0533 & 0.263 & 0.508 & 2.50
	\\ \hline
	$\bar q$ & 4.40 & 5.38 & 11.3 & 11.9 & 12.1 
	\\ \hline
	$\bar q'$ & 35.2 & 35.5 & 77.3 & 90.5 & 96.5
	\\ \hline
	$\bar g$ & 3.57 & 4.57 & 10.7 & 11.2 & 11.4 
	\\ \hline
	$\bar g'$ & 57.2 & 56.6 & 138 & 168 & 182 
	\\ \hline \hline
	\multicolumn{6}{|c|}{$\mathbf{x = 0.7}$} 
	\\ \hline \hline
	$T' (MeV)$ & 0.0035 & 0.0733 & 0.357 & 0.704 & 3.50
	\\ \hline
	$\bar q$ & 5.25 & 6.42 & 13.5 & 14.2 & 14.4 
	\\ \hline
	$\bar q'$ & 15.3 & 16.3 & 4.88 & 40.6 & 42.0 
	\\ \hline
	$\bar g$ & 4.17 & 5.35 & 12.4 & 13.1 & 13.3 
	\\ \hline
	$\bar g'$ & 17.4 & 18.5 & 47.8 & 53.3 & 55.5 
	\\ \hline
	\end{tabular}\\ 
	\end{center} 
	\label{tab-DOFs}
	\end{table}%

%


\def \sec-Nu-num{The number of neutrinos in the ordinary sector}
\subsubsection{\sec-Nu-num}
\label{sec-Nu-num}

\begin{figure}[htbp]
	\begin{center}
	 \includegraphics[scale=0.6]{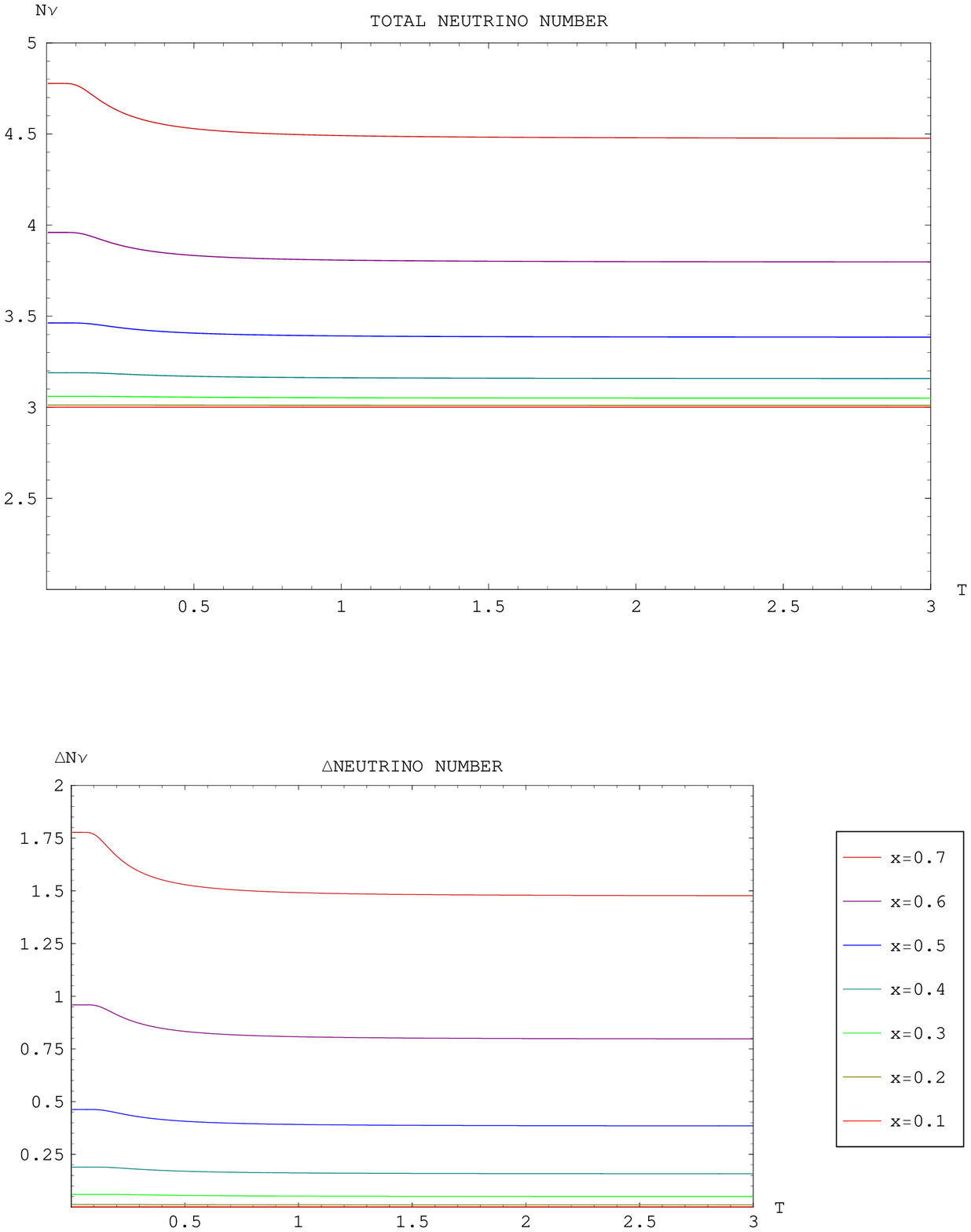}	
	\caption{{\it $N_{\nu}$ and $\Delta N_{\nu}$ in the ordinary sector for several $x$ values.}}
	\label{fig_Nnu}
	\end{center}
	\end{figure}

We know from chapter \ref{chap-The_standard_model} that the SM contains three neutrino species; the possible existence of a fourth neutrino has been longly investigated, also using BBN constraints. This is why in the literature one can often find bounds on the number of DOFs in terms of extra-neutrino number $\Delta N_{\nu}$ - see for instance  \cite{Mangano:2006ur,Lisi:1999ng}.
In general, the number of neutrinos $N_{\nu}$ is found assuming that all particles contributing to the universe energy density, to the exclusion of electrons, positrons and photons, are neutrinos; in formula that means:
	\begin{eqnarray}
	\label{Nnu_conversion}
	\bar g (T) &=& g_{e^{\pm}} (T) + g_{\gamma} + 
	\frac{7}{8}\cdot 2 N_{\nu}  \cdot \left( \frac{T_{\nu}}{T} \right)^4
	\cr\cr
	\Rightarrow N_{\nu} &=&
	\frac{\bar g - g_{e^{\pm}}(T) - g_{\gamma}}{\frac{7}{8}\cdot 2} 
	\cdot \left( \frac{T}{T_{\nu}} \right)^4
	\end{eqnarray}
We can also work out the number of extra-neutrinos $\Delta N_{\nu}$, that is the number of non standard DOFs in term of addictional neutrinos:
	\begin{eqnarray}
	\label{Delta_Nnu_conversion}
	\Delta N_{\nu} = \frac{\bar g - g_{std}}{\frac{7}{8}\cdot 2} 
	\cdot \left( \frac{T}{T_{\nu}} \right)^4
	\end{eqnarray}
Both $N_{\nu}$ and $\Delta N_{\nu}$ have been worked out with Mathematica; some data are reported in Table \ref{tab-Nnu}, while plots for several $x$ values and any temperatures from $0$ to $3\:MeV$ are shown in Figure \ref{fig_Nnu}.
	\begin{table}[htdp]
	\caption{The number of neutrinos in the ordinary sector for some special cases.}
	\begin{center}
	\begin{tabular}{|c|c|c|c|c|c|}
	\hline
	$T(MeV)$ & standard & $x=0.1$ & $x=0.3$ & $x=0.5$ & $x=0.7$
	\\ \hline\hline
	\multicolumn{6}{|c|}{{\bf ordinary sector}} 
	\\ \hline \hline
	0.005 & 3.00000 & 3.00074 & 3.05997 & 3.46270 & 4.77751
	\\ \hline
	0.5 & 3.00000 & 3.00074 & 3.05997 & 3.40706 & 4.52942
	\\ \hline
	1 & 3.00000 & 3..00071 & 3.05202 & 3.39166 & 4.49133
	\\ \hline
	5 & 3.00000 & 3.00063 & 3.04989 & 3.38430 & 4.47563
	\\ \hline
	\end{tabular}
	\end{center}
	\label{tab-Nnu}
	\end{table}
We stress that the standard values $N_{\nu} = 3$ and $\Delta N_{\nu} = 0$ are the same at any temperatures, while a distinctive feature of the mirror scenario is that the number of neutrinos raises with the temperature. 
Anyway, this effect is not a problem; on the contrary it may be useful since some recent data fits give indications for a number of neutrinos at recent times higher than at BBN \cite{Mangano:2006ur}.

\paragraph{Predictions for special temperatures}

It is possible to work out $N_{\nu}$ and $\Delta N_{\nu}$ at $T \gg T_{D\nu}$ and $T \ll T_{ann \, e^{\pm}}$ starting from the standard values - see (\ref{std_DOF_tot_eq}) and (\ref{std_DOF_tot_after_ann}):
	\begin{eqnarray}
	\label{g_extremes_std}
	g(T \gg T_{D\nu}) = 10.75  \hspace{1.5cm}
	g(T \ll T_{ann \, e^{\pm}}) \simeq 3.36
	\end{eqnarray}
Without the mirror universe, we have, as expected:
	\begin{eqnarray}
	\label{Nnu_Delta_Nnu_standard}
	N_{\nu} (T \gg T_{D\nu})  &=& \frac{10.75 - 2 - \frac{7}{8} \cdot 4}{{\frac{7}{8}\cdot 2}} = 3
	\cr\cr
	N_{\nu} (T \ll T_{ann \, e^{\pm}}) &=& \frac{3.36 - 2}{{\frac{7}{8}\cdot 2}} 
	\cdot \left( \frac{11}{4} \right)^\frac{4}{3} = 3
	\end{eqnarray}
While, when the mirror universe is present, we have $\bar g = g (1+x^4)$ at the special temperatures we are considering; hence
	\begin{eqnarray}
	\label{Nnu_Delta_Nnu_total}
	N_{\nu} (T \gg T_{D\nu})  &=& \frac{10.75(1+x^4) - 2 - \frac{7}{8} \cdot 4}{{\frac{7}{8}\cdot 2}} 
	\cr\cr\cr
	N_{\nu} (T \ll T_{ann \, e^{\pm}}) &=& \frac{3.36 (1+x^4)- 2}{{\frac{7}{8}\cdot 2}} 
	\cdot \left( \frac{11}{4} \right)^\frac{4}{3} 
	\end{eqnarray}
From the equations written above we can see that the rise in $N_{\nu}$ is:
	\begin{eqnarray}
	\label{N_nu_mir_approx}
	N_{\nu} (T \ll T_{ann \, e^{\pm}}) - N_{\nu} (T \gg T_{D\nu})  = \cr\cr
	=x^4 \cdot \frac{1}{\frac{7}{8}\cdot 2} \left[ 10.75 - 3.36 \left( \frac{11}{4} \right)^\frac{4}{3}\right]
	\simeq 1.25 \cdot x^4
	\end{eqnarray}
This leads to $N_{\nu}$ higher than the standard in presence of the mirror world and to a further rise of this parameter at low temperatures. 
For instance, if $x=0.7$ equations (\ref{Nnu_Delta_Nnu_total}) give:
	\begin{eqnarray}
	\label{}
	N_{\nu} (T \gg T_{D\nu}) \simeq 4.5
	\hspace{1cm} \mathrm{and} \hspace{1cm} 
	N_{\nu} (T \ll T_{ann \, e^{\pm}}) \simeq 4.8
	\end{eqnarray}
%

\def \sec-Nu-num-mir{The number of neutrinos in the mirror sector}
\subsubsection{\sec-Nu-num-mir}
\label{sec-Nu-num-mir}

Similarly the number of neutrinos in the mirror sector can be worked out as:
	\begin{eqnarray}
	\label{Nnu_conversion_mir}
	N'_{\nu} =
	\frac{\bar g' - g'_{e^{\pm}}(T') - g'_{\gamma}}{\frac{7}{8}\cdot 2} 
	\cdot \left( \frac{T'}{T'_{\nu}} \right)^4
	\end{eqnarray}
Once again these values are higher than the ordinary ones by a factor $x^4$; they have been numerically computed using Mathematica and some special values are reported in Table \ref{tab-Nnu-mir}.
	\begin{table}[htdp]
	\caption{The number of neutrinos in the mirror sector for some special cases.}
	\begin{center}
	\begin{tabular}{|c|c|c|c|c|}
	\hline
	$T'\,(MeV)$ & $x=0.1$ & $x=0.3$ & $x=0.5$ & $x=0.7$
	\\ \hline\hline
	\multicolumn{5}{|c|}{{\bf mirror sector}} 
	\\ \hline \hline
	0.005 & 74011 & 917.0 & 121.4 & 33.83
	\\ \hline
	0.1 & 62007 & 805.0 & 111.6 & 32.21
	\\ \hline
	0.5  & 61447 & 763.1 & 101.9 & 28.86
	\\ \hline
	1  & 61435 & 761.8 & 101.4 & 28.66
	\\ \hline
	5  & 61432 & 761.4 & 101.3 & 28.59
	\\ \hline
	\end{tabular}
	\end{center}
	\label{tab-Nnu-mir}
	\end{table}

\def \sec-elem-prod-ord{The light elements production in the ordinary sector}
\section{\sec-elem-prod-ord}
\label{sec-elem-prod-ord}
\markboth{Chapter \ref{chap-mir-BBN}. ~ \chap-mir-BBN}
                    {\S \ref{sec-elem-prod-ord} ~ \sec-elem-prod-ord}

As we have seen in \S \ref{sec-DOFs-num}, the presence of the mirror sector can be parametrized in terms of extra DOFs number or extra neutrino families; therefore, since the physical processes involved in BBN are not affected by the mirror sector, we can use and modify a pre-existing BBN code to work out the light elements production. 

We chosed the \it Kawano code for BBN \rm \cite{Kawano:1992ua} since it is a well-tested and fast program and its accuracy is large enough for our purposes.
The Kawano code is an updated and modified version of the code of R.V. Wagoner \cite{Wagoner:1972jh}. It solves for elemental abundances arising from the epoch of primordial nucleosynthesis in the early universe by mean of a second-order Runge-Kutta driver.
The parameter values we used are 
	\begin{eqnarray}
	\label{Kaw_param}
	\eta = 6.140 \cdot 10^{-10}
	\hspace{2cm}
	\tau = 885.7 
	 \, s
	\end{eqnarray}
where $\eta$ is the final baryon to photon ratio.
The number of DOFs enters the program in terms of neutrino species number; this quantity is a free parameter and can set by the user. But the program uses the same number during the whole BBN process.

Instead, the mirror sector leads to variable $N_{\nu}$! 
We therefore used before a first order approximation where $\Delta N_{\nu} = const \propto x^4$ (the so called \it zero approximation\rm) to get a feeling of the nuclides production.
Then, in order to get more accurate results, we modified the program to read 
the neutrino number for any temperatures and $x$ values as an external input. 
This input comes from a file, which has been written with a Mathematica program which solves the equations which describe the thermodynamical evolution of both sectors - see \S \ref{sec-Nu-dec-mir} and \S \ref{sec-DOFs-num}.
	\begin{figure}[htbp]
	\begin{center}
	\includegraphics[scale=1.]{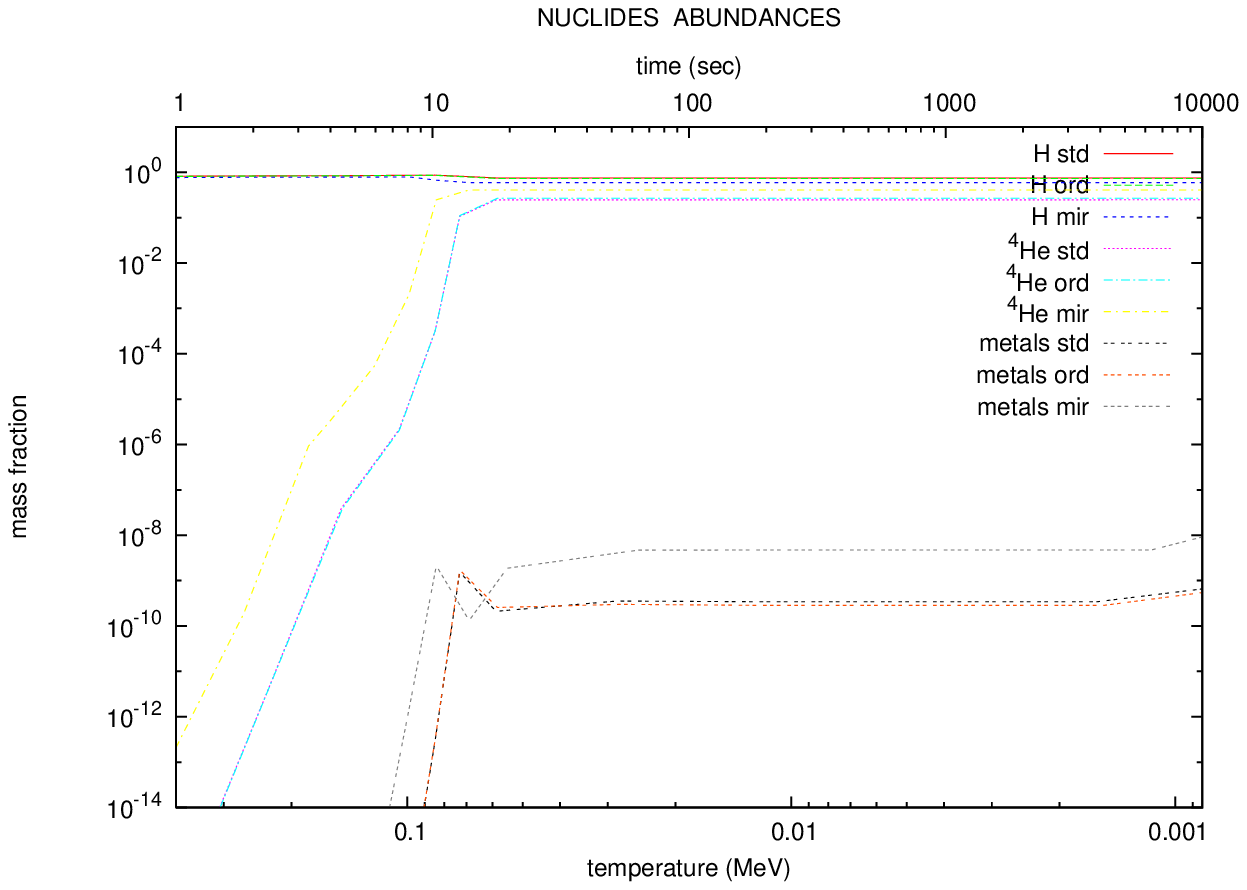}
	\includegraphics[scale=1.]{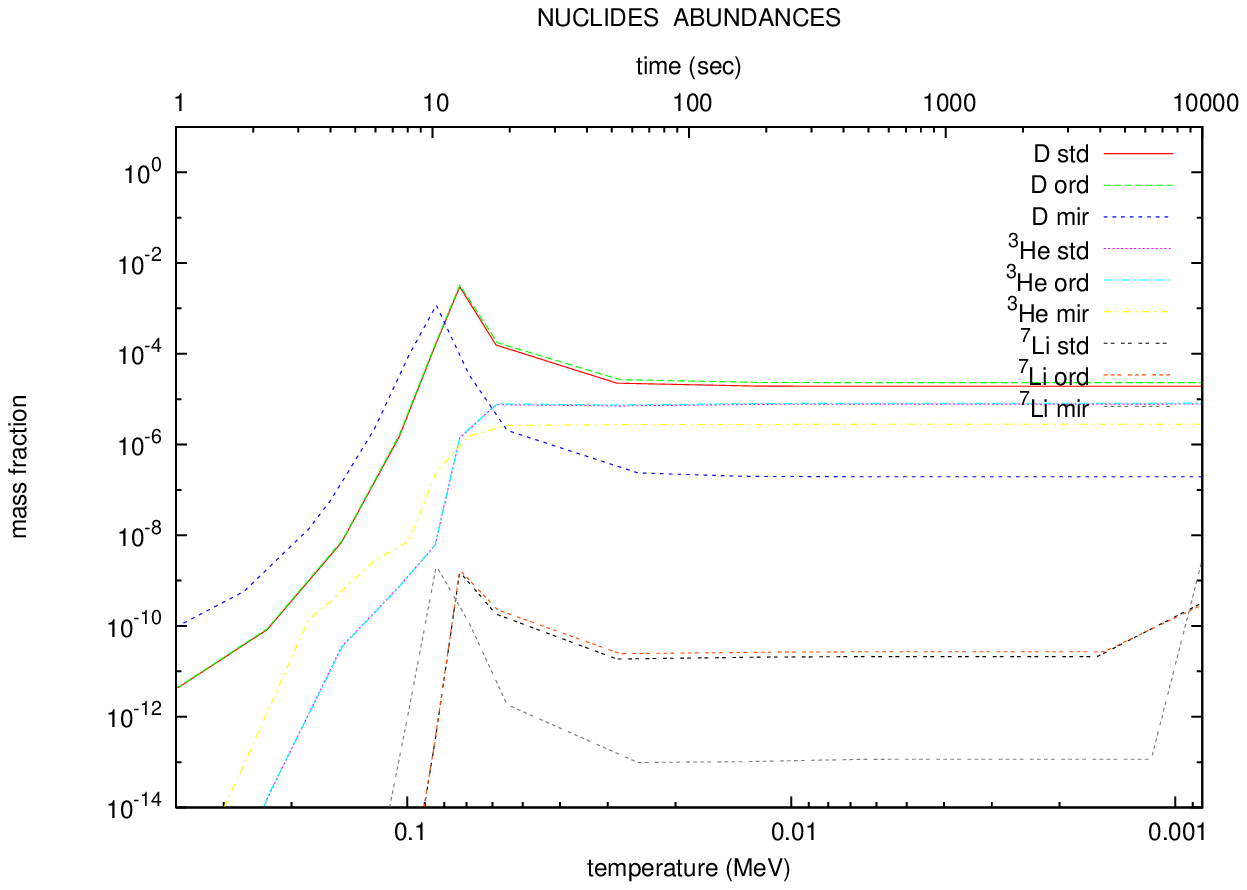}
	\caption{{\it The development of primordial nucleosynthesis for the standard
	universe (that is without the mirror sector) and in the two sectors when the mirror
	is present. To work out these data we used the parameters $x=0.7$, $\beta = 5$
	.}}
	\label{std_ord_nucl}
	\end{center}
	\end{figure}
The only parameter of the mirror sector which affects ordinary BBN is $x$; the baryon ratio $\beta$ - defined in equation (\ref{beta_DM_def}) - does not induce any changes on the production of ordinary nuclides. We will see in the following section that $\beta$ plays a crucial role for the mirror nuclides production.

%
	\begin{table}[htdp]
	\caption{Light elements produced in the ordinary sector.}
	\begin{center}
	\begin{tabular}{|c|c|c|c|c|c|c|c|}
	\hline \hline
	\bf elements & \bf std & $\mathbf{x=0.1}$ & $\mathbf{x=0.3}$ & 
	$\mathbf{x=0.5}$ & $\mathbf{x=0.7}$
	\\ \hline \hline
	$n/H \: (10^{-16})$ & 1.161
	& 1.161 & 1.159 & 1.505 & 2.044
	\\ \hline
	$p$ & 0.7518
	& 0.7518 & 0.7511 & 0.7463 & 0.7326
	\\ \hline
	$D/H \: (10^{-5})$ & 2.554
	& 2.555 & 2.575 & 2.709 & 3.144
	\\ \hline
	$T/H \: (10^{-8})$ & 8.064
	& 8.065 & 8.132 & 8.588 & 10.07
	\\ \hline
	$^3He/H \: (10^{-5})$ & 1.038
	& 1.038 & 1.041 & 1.058 & 1.113
	\\ \hline
	$^4He$ & 0.2483
	& 0.2483 & 0.2491 & 0.2538 & 0.2675
	\\ \hline
	$^6Li/H \: (10^{-14})$ & 1.111
	& 1.111 & 1.124 & 1.210 & 1.499
	\\ \hline
	$^7Li/H \: (10^{-10})$ & 4.549
	& 4.548 & 4.523 & 4.356 & 3.871
	\\ \hline	
	$^7 Be/H \: (10^{-10})$ & 4.266
	& 4.266 & 4.238 & 4.051 & 3.502
	\\ \hline	
	$^8Li + /H \: (10^{-15})$ & 1.242 
	& 1.242 & 1.251 & 1.306 & 1.464
	\\ \hline
	\end{tabular}
	\end{center}
	\label{tab-ord-BBN}
	\end{table}

In Table \ref{tab-ord-BBN} the program output, that is the light elements produced in the ordinary sector at the end of BBN process (at $T \sim 8\cdot 10^{-4}$ MeV), is reported for several $x$ values and compared with the standard. We can see that the difference between the standard and $x=0.1$ is of order $10^{-4}$ or less, but it becomes even more important while $x$ increases.

In Figure \ref{std_ord_nucl} we plot the time and temperature evolution of the mass fraction of several light elements plus metals (by metals we mean the sum of elements heavier than $^4 He$, that is $^6Li$, $^7Li$ and so on) in the ordinary case and in both sectors in the mirror model; we used for these calculations the parameters $x=0.7$ and $\beta=5$. We can see that BBN in the mirror sector is much more different from the standard than the ordinary sector one. This is a consequence of the higher number of DOFs.

	\begin{figure} [htbp]
	\begin{center}
	\includegraphics[scale=0.8]{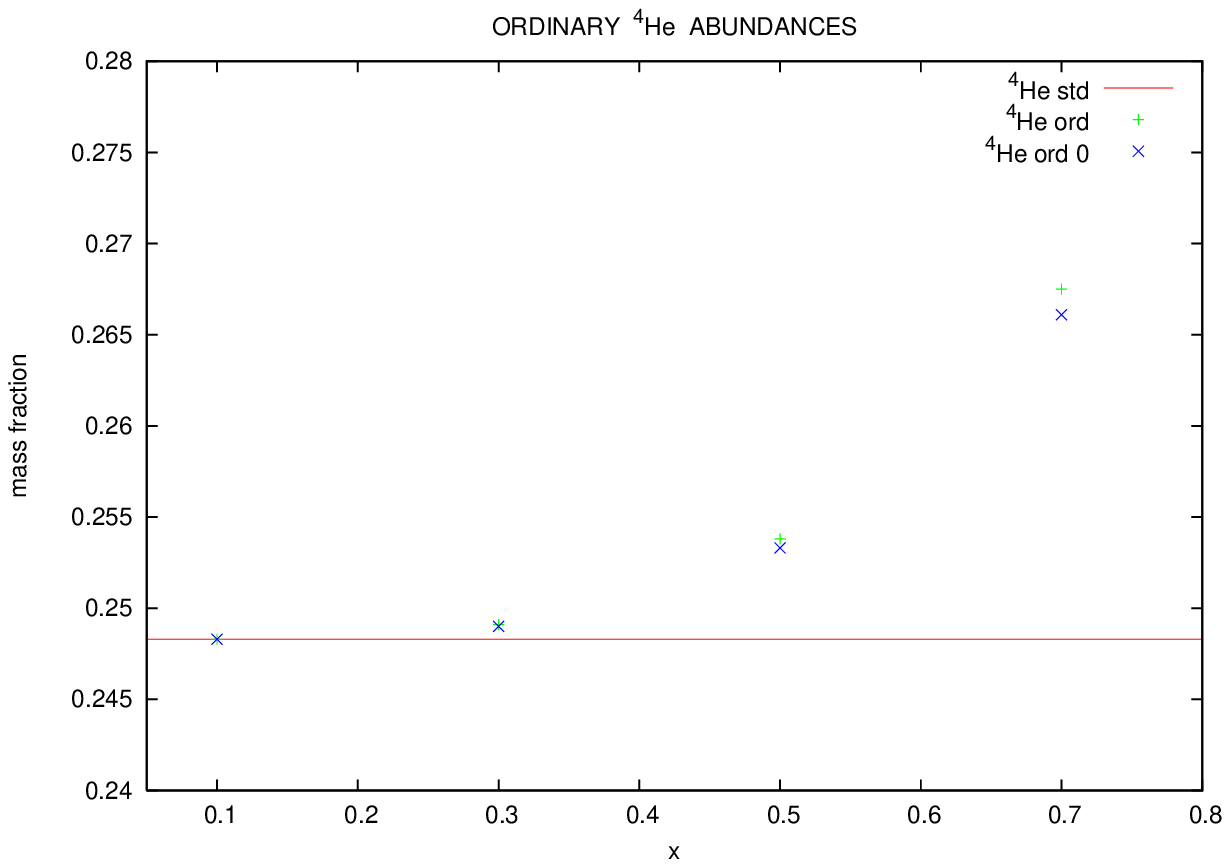} \\
	\vspace{0.5cm}
	\includegraphics[scale=0.5]{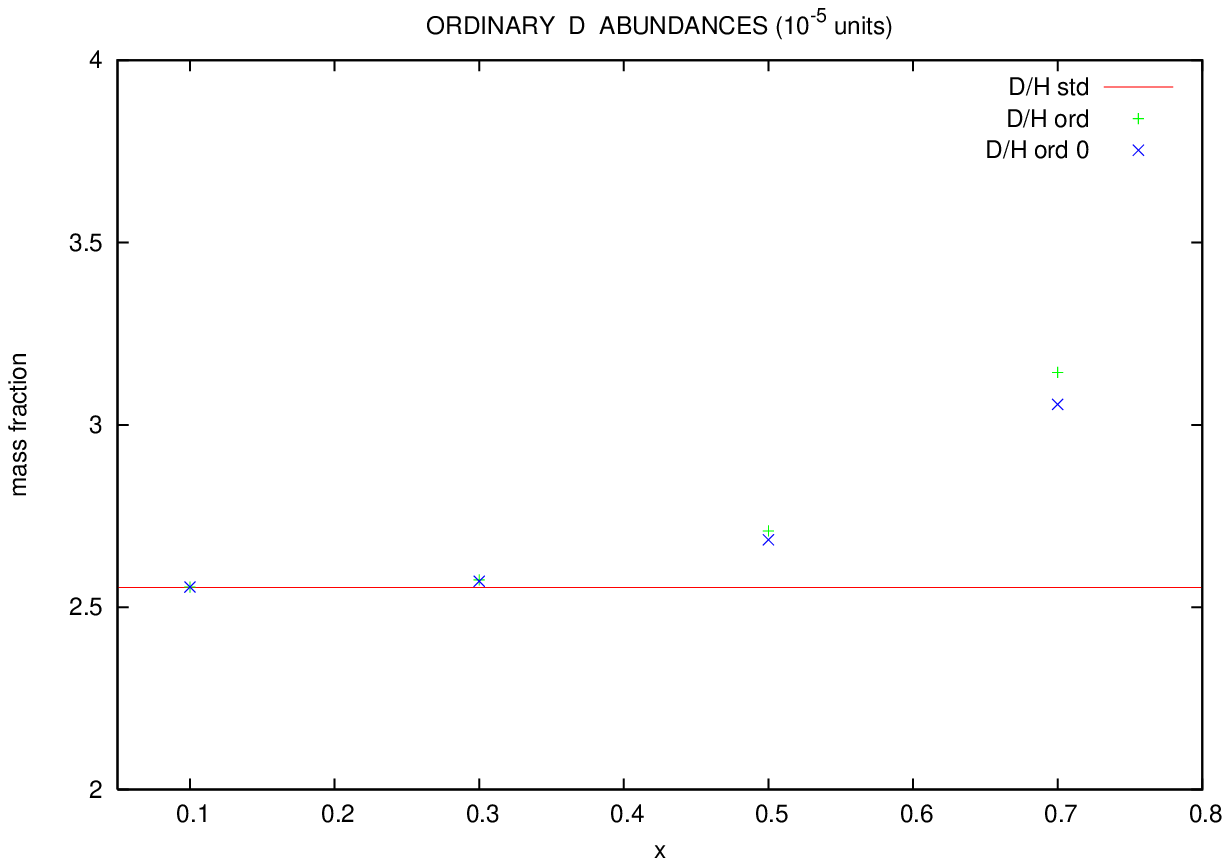} 
	\hspace{0.5cm}
	\includegraphics[scale=0.5]{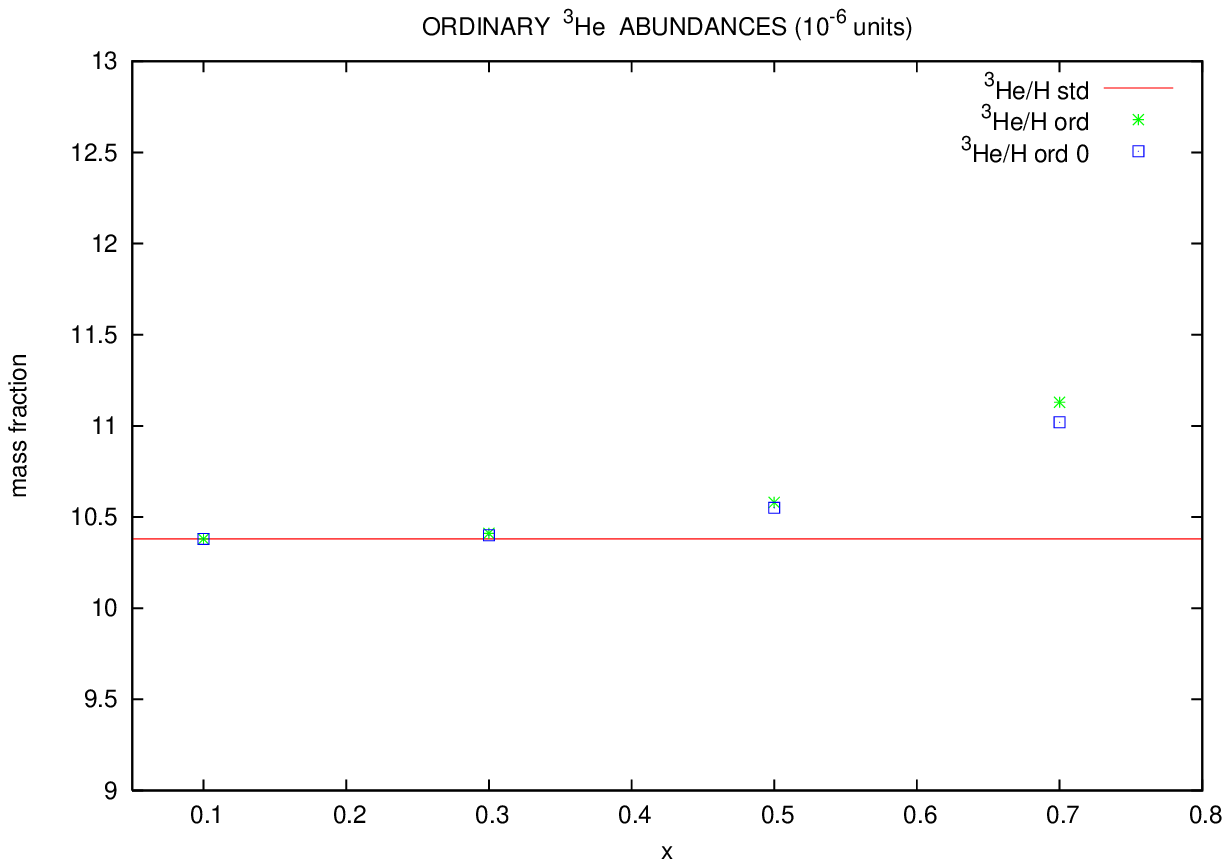}
	\hspace{0.5cm}
	\includegraphics[scale=0.5]{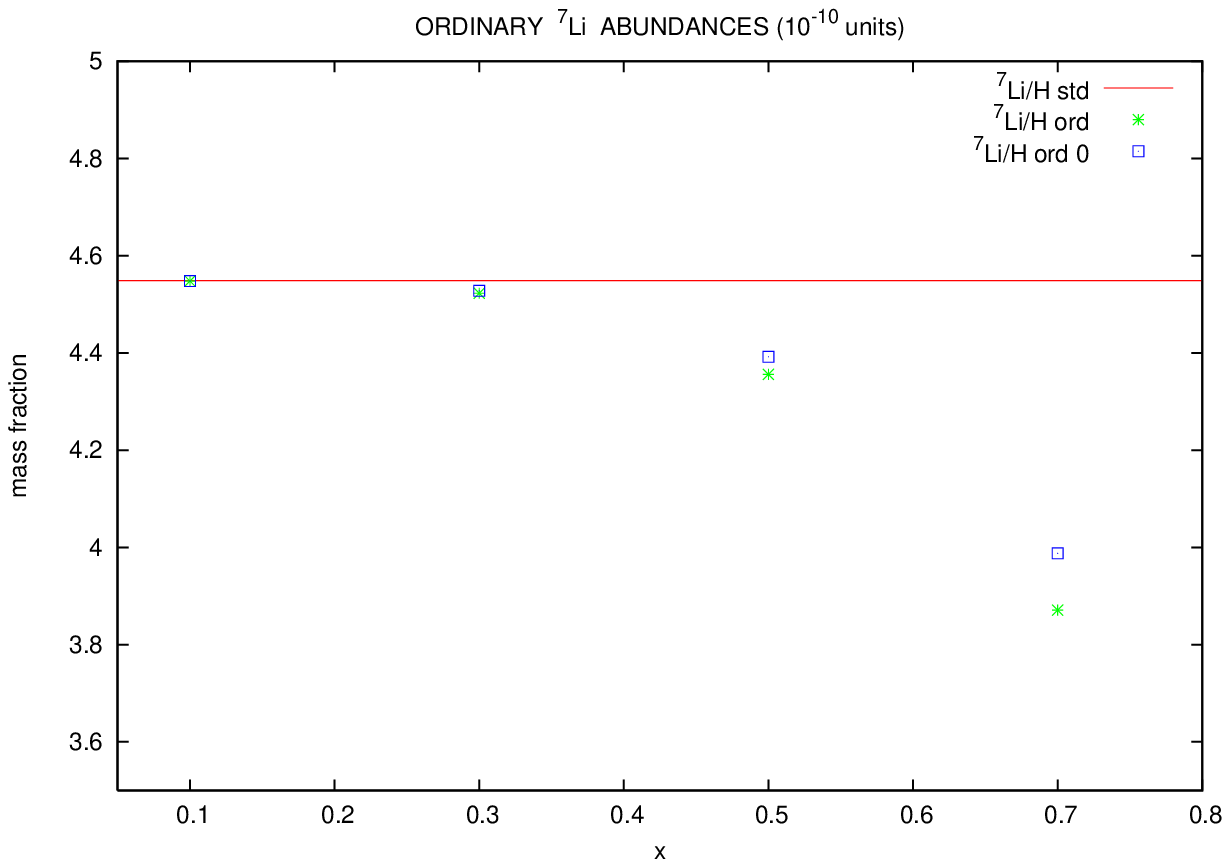}
	\hspace{0.5cm}
	\includegraphics[scale=0.5]{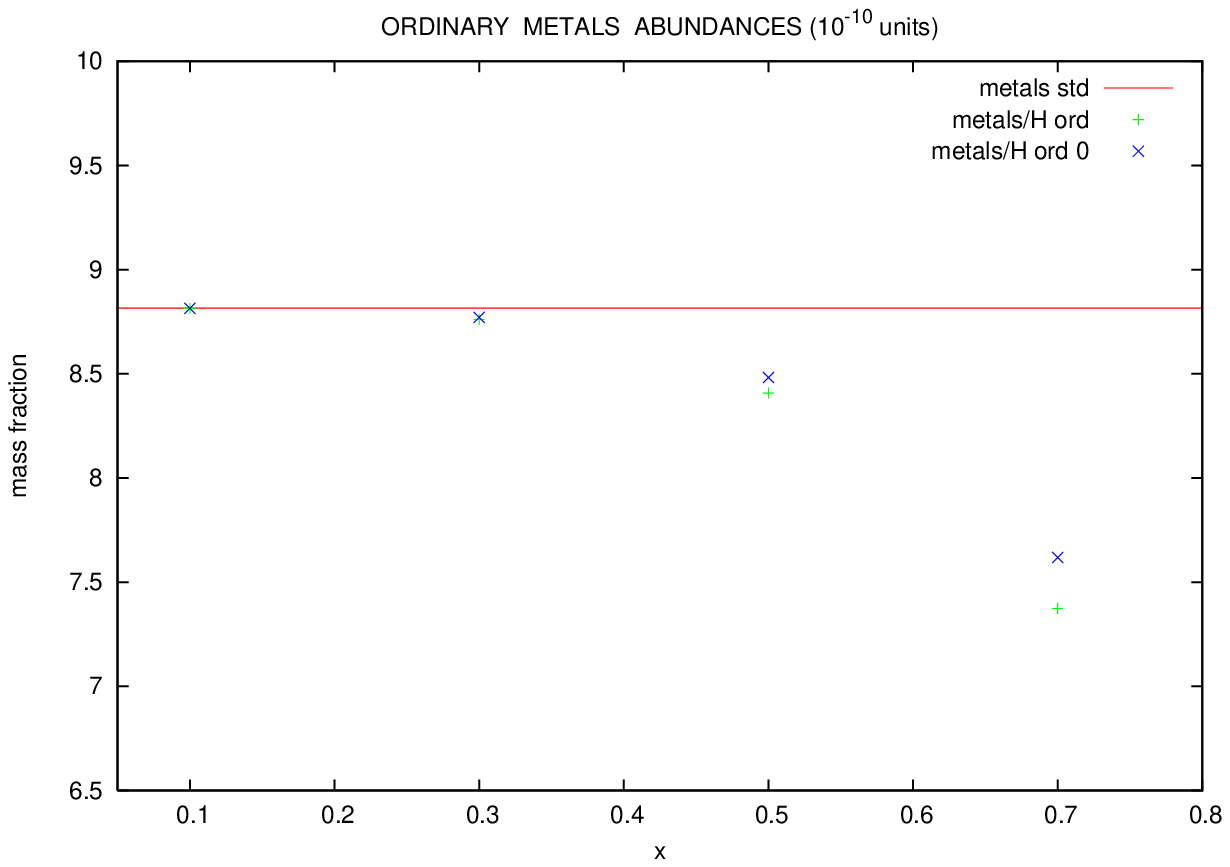}
	\bf \caption{\rm Light elements produced in the ordinary sector during BBN. 
	On every plot we report the elements ($^4He$, $D$, $^3He$, $^7Li$ and metals)
	for several $x$ values, compared with the standard and the
	"order 0" quadratic approximation of $\Delta N_{\nu}$ in equation (\ref{N_nu_mir_approx}).} \rm
	\label{fig-ord-BBN}
	\end{center}
	\end{figure}

In Figure \ref{fig-ord-BBN} we plot the final abundances of $^4He$, $D$, $^3He$, $^7Li$ and metals in the ordinary sector as a function of $x$. We can easily infer that for $x \lsim 0.3$ the light element abundances do not change more than a few percent.

\def \sec-elem-prod-mir{The light elements production in the mirror sector}
\section{\sec-elem-prod-mir}
\label{sec-elem-prod-mir}
\markboth{Chapter \ref{chap-mir-BBN}. ~ \chap-mir-BBN}
                    {\S \ref{sec-elem-prod-mir} ~ \sec-elem-prod-mir}


Even mirror baryons undergo nucleosynthesis via the same physical processe than the ordinary ones. 
But the impact of the ordinary world on the mirror BBN is dramatic since the ordinary contribution on the number of total mirror DOFs scales as $\sim x^{-4}$ - see equation (\ref{g_bar_def}).
The knowledge of the element abundances in the mirror world can be important for the study of its evoultion in later epochs as the galaxy formation, and in particular, for the formation and evolution of the mirror stars \cite{Cassisi}.

We can use the Kawano code also for the mirror nucleosynthesis; once again there are two different ways to input the neutrino number $N_{\nu}$: the order zero $\Delta N_{\nu} = const \propto x^{-4}$ and the numerical values obtained solving the equations with Mathematica. The physical parameters $\eta$, $\tau$ and $h$ have the values in equation (\ref{Kaw_param}).

An other parameter which affects mirror BBN is the mirror baryon density ($\Omega'_B \sim 1 \div 5 \Omega_B$), which raises the baryon to photon ratio $\eta'$ - see \S \ref{sec-std-BBN};
in our program $\Omega'_B$ is introduced in terms of the ratio $\beta = \Omega'_B / \Omega_B$ - see equation (\ref{beta_DM_def}). 

	\begin{table}[htdp]
	\caption{Light elements produced in the mirror sector.}
	\begin{center}
	\begin{tabular}{|c|c|c|c|}
	\hline \hline
	\bf elements & $\mathbf{x=0.1\: (\beta = 5)}$ & $\mathbf{x=0.1\: (\beta = 1)}$ & 
	$\mathbf{x=0.3} \: (\beta = 5)$ 
	\\ \hline \hline
	$n/H $ &  5.762 $\cdot 10^{-25}$ & 8.888 $\cdot 10^{-17}$ & 2.590 $\cdot 10^{-22}$ 
	\\ \hline
	$p $ & 0.1735 & 0.1772 & 0.3646
	\\ \hline
	$D/H $ & 1.003 $\cdot 10^{-12}$ & 1.331 $\cdot 10^{-6}$ & 4.838 $\cdot 10^{-9}$ 
	\\ \hline
	$T/H$ & 9.679 $\cdot 10^{-21}$ & 3.068 $\cdot 10^{-9}$ & 1.238 $\cdot 10^{-13}$ 
	\\ \hline
	$^3He/H$ & 3.282 $\cdot 10^{-6}$ & 5.228 $\cdot 10^{-6}$ & 3.740 $\cdot 10^{-6}$ 
	\\ \hline
	$^4He $ & 0.8051 & 0.8226 & 0.6351
	\\ \hline
	$^6Li/H$ & 7.478 $\cdot 10^{-21}$ & 8.638 $\cdot 10^{-15}$ & 1.309 $\cdot 10^{-17}$
	\\ \hline
	$^7Li/H $ & 1.996 $\cdot 10^{-7}$ & 5.712 $\cdot 10^{-8}$ & 3.720 $\cdot 10^{-8}$  
	\\ \hline	
	$^7 Be/H$ & 1.996 $\cdot 10^{-7}$ & 5.711 $\cdot 10^{-8}$ & 3.720 $\cdot 10^{-8}$ 
	\\ \hline	
	$^8Li +/H$ & 4.354 $\cdot 10^{-9}$ & 2.036 $\cdot 10^{-10 }$ & 5.926 $\cdot 10^{-11}$ 
	\\ \hline
	\hline \hline
	\bf elements & $\mathbf{x=0.3\: (\beta = 1)}$ & $\mathbf{x=0.7\: (\beta = 5)}$ & 
	$\mathbf{x=0.7} \: (\beta = 1)$ 
	\\ \hline \hline
	$n/H $ &  1.915 $\cdot 10^{-16}$ & 1.726 $\cdot 10^{-19}$ & 2.076 $\cdot 10^{-16}$ 
	\\ \hline
	$p $ & 0.3675 & 0.5924 & 0.6017
	\\ \hline
	$D/H $ & 7.094 $\cdot 10^{-6}$ & 3.279 $\cdot 10^{-7}$ & 2.235 $\cdot 10^{-5}$ 
	\\ \hline
	$T/H$ & 2.190 $\cdot 10^{-8}$ & 3.722 $\cdot 10^{-10}$ & 7.328 $\cdot 10^{-8}$ 
	\\ \hline
	$^3He/H$ & 6.880 $\cdot 10^{-6}$ & 4.691 $\cdot 10^{-6}$ & 9.719 $\cdot 10^{-6}$ 
	\\ \hline
	$^4He $ & 0.6326 & 0.4077 & 0.3984
	\\ \hline
	$^6Li/H$ & 1.660 $\cdot 10^{-14}$ & 3.361 $\cdot 10^{-16}$ & 1.951 $\cdot 10^{-14}$
	\\ \hline
	$^7Li/H $ & 8.930 $\cdot 10^{-9}$ & 7.962 $\cdot 10^{-9}$ & 1.120 $\cdot 10^{-9}$  
	\\ \hline	
	$^7 Be/H$ & 8.878 $\cdot 10^{-9}$ & 7.962 $\cdot 10^{-9}$ & 1.064 $\cdot 10^{-9}$ 
	\\ \hline	
	$^8Li +/H$ & 2.514 $\cdot 10^{-12}$ & 3.949 $\cdot 10^{-13}$ & 1.814 $\cdot 10^{-14}$ 
	\\ \hline
	\hline \hline
	\bf elements & $\mathbf{x=0.5\: (\beta = 5)}$ & $\mathbf{x=0.5\: (\beta = 1)}$ & 
	\\ \hline \hline
	$n/H $ &  1.840 $\cdot 10^{-20}$ & 2.058 $\cdot 10^{-16}$ & 
	\\ \hline
	$p $ & 0.4966 & 0.5028 &
	\\ \hline
	$D/H $ & 6.587 $\cdot 10^{-8}$ & 1.352 $\cdot 10^{-5}$ &  
	\\ \hline
	$T/H$ & 2.108 $\cdot 10^{-11}$ & 4.358 $\cdot 10^{-8}$ &  
	\\ \hline
	$^3He/H$ & 4.172 $\cdot 10^{-6}$ & 8.232 $\cdot 10^{-6}$ &  
	\\ \hline
	$^4He $ & 0.5035 & 0.4974 & 
	\\ \hline
	$^6Li/H$ & 1.016 $\cdot 10^{-16}$ & 1.790 $\cdot 10^{-14}$ & 
	\\ \hline
	$^7Li/H $ & 1.535 $\cdot 10^{-8}$ & 2.948 $\cdot 10^{-9}$ & 
	\\ \hline	
	$^7 Be/H$ & 1.535 $\cdot 10^{-8}$ & 2.891 $\cdot 10^{-9}$ & 
	\\ \hline	
	$^8Li +/H$ & 3.827 $\cdot 10^{-12}$ & 1.657 $\cdot 10^{-13}$ &
	\\ \hline
	\end{tabular}
	\end{center}
	\label{tab-mir-BBN}
	\end{table}
In Table \ref{tab-mir-BBN} the program output, that is the light elements produced in the mirror sector at the end of BBN process (at $T \sim 8 \cdot 10^{-4}$ MeV), is reported for several $x$ and $\beta$ values and compared with the standard. 

%
	\begin{figure} [htbp]
	\begin{center}
	\includegraphics[scale=0.8]{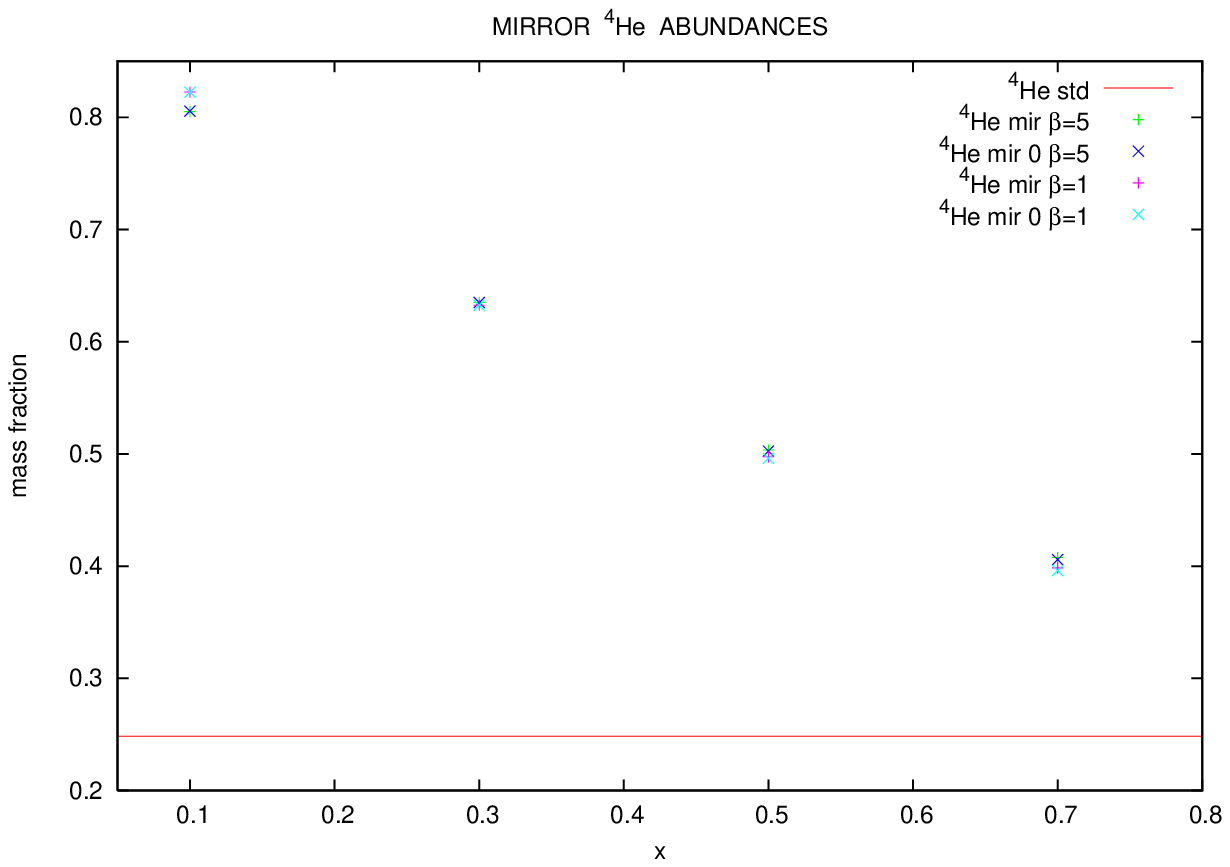} \\
	\vspace{0.5cm}
	\includegraphics[scale=0.5]{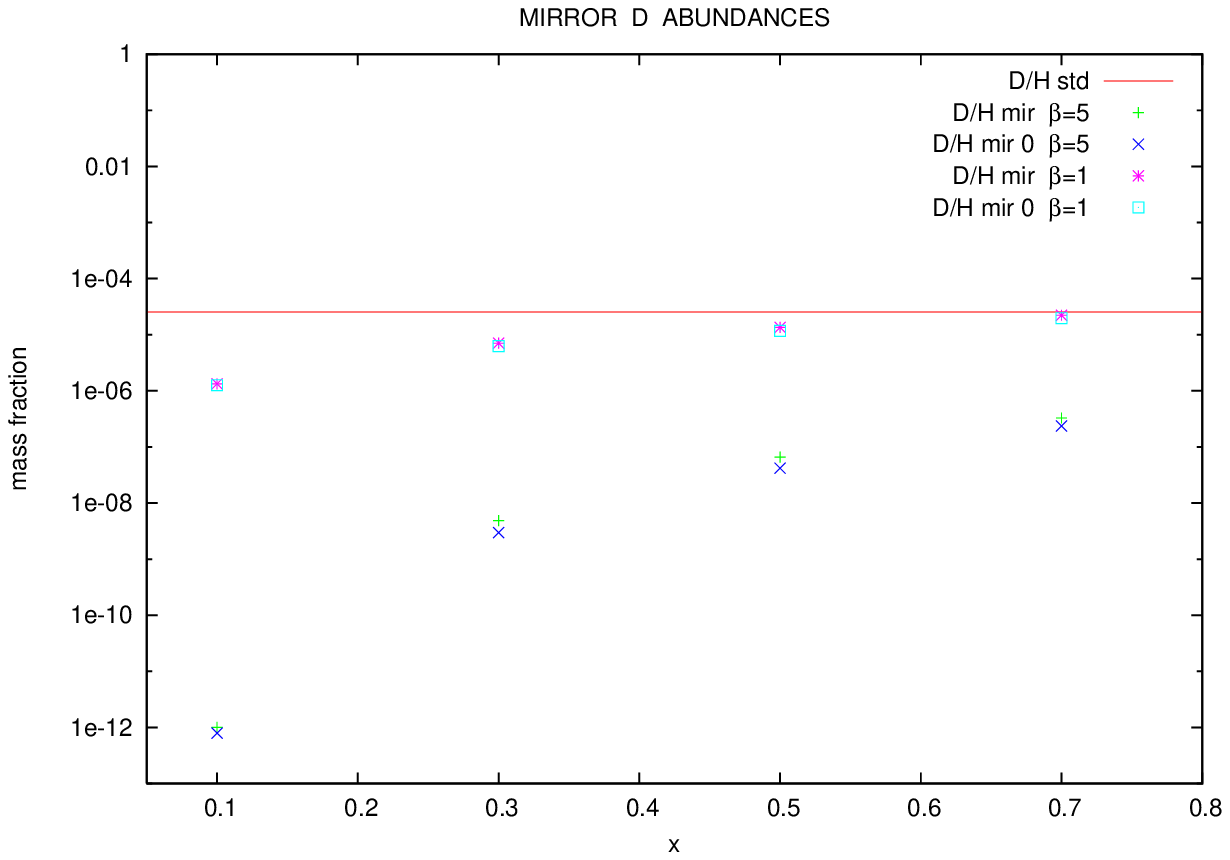} 
	\hspace{0.5cm}
	\includegraphics[scale=0.5]{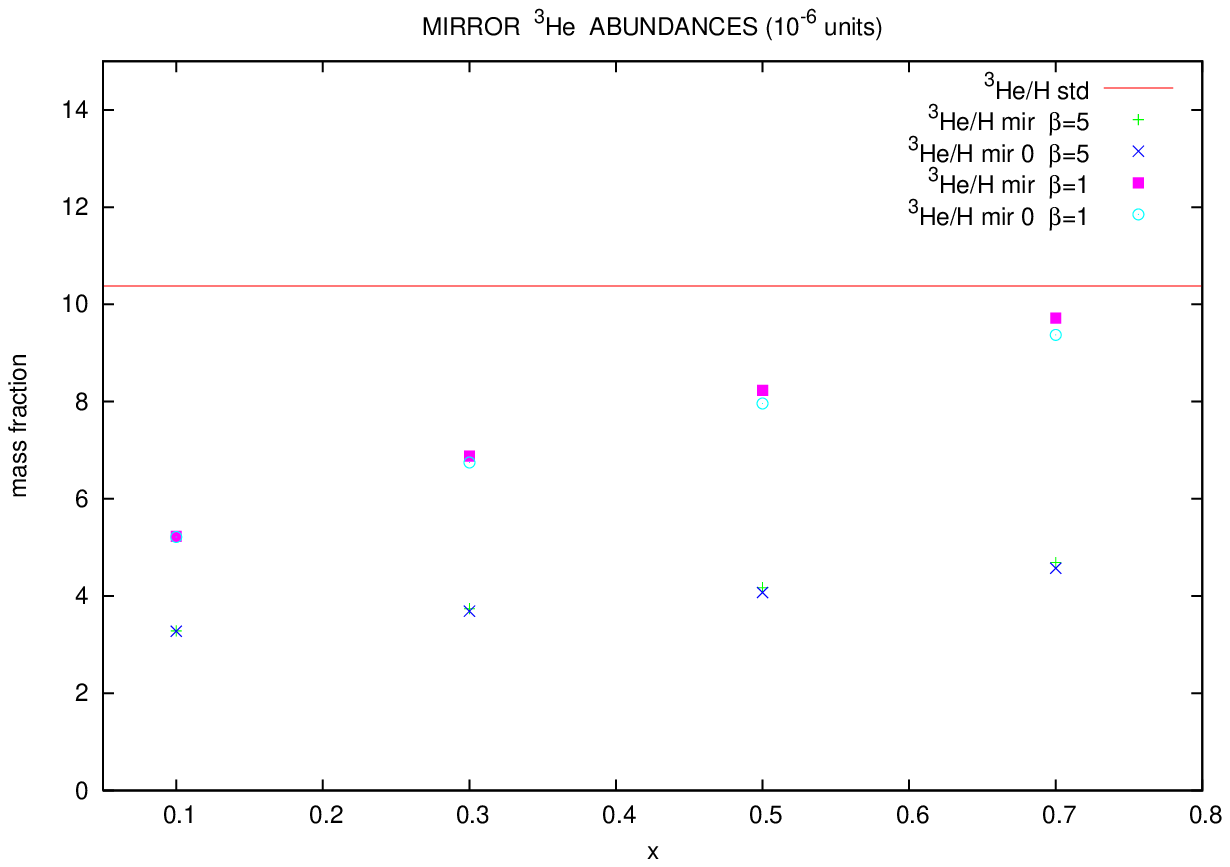}
	\hspace{0.5cm}
	\includegraphics[scale=0.5]{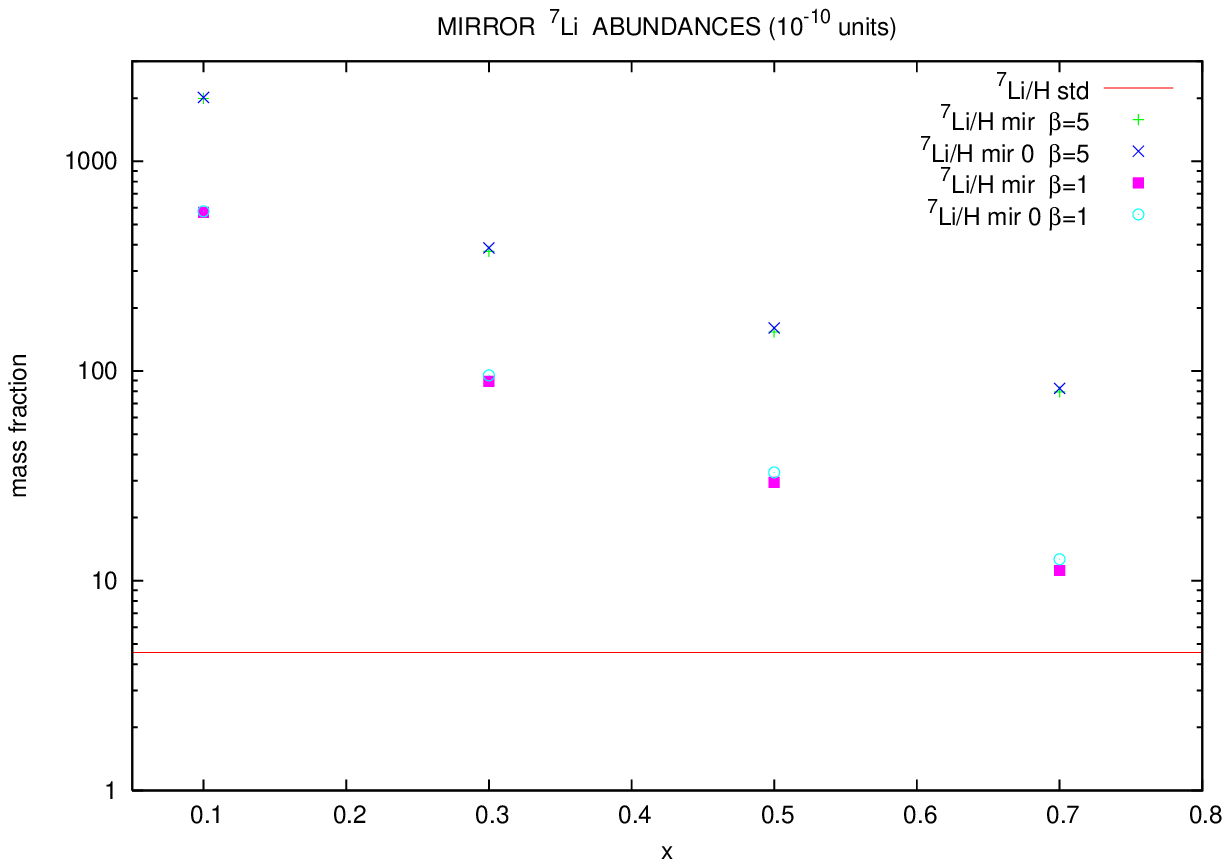}
	\hspace{0.5cm}
	\includegraphics[scale=0.5]{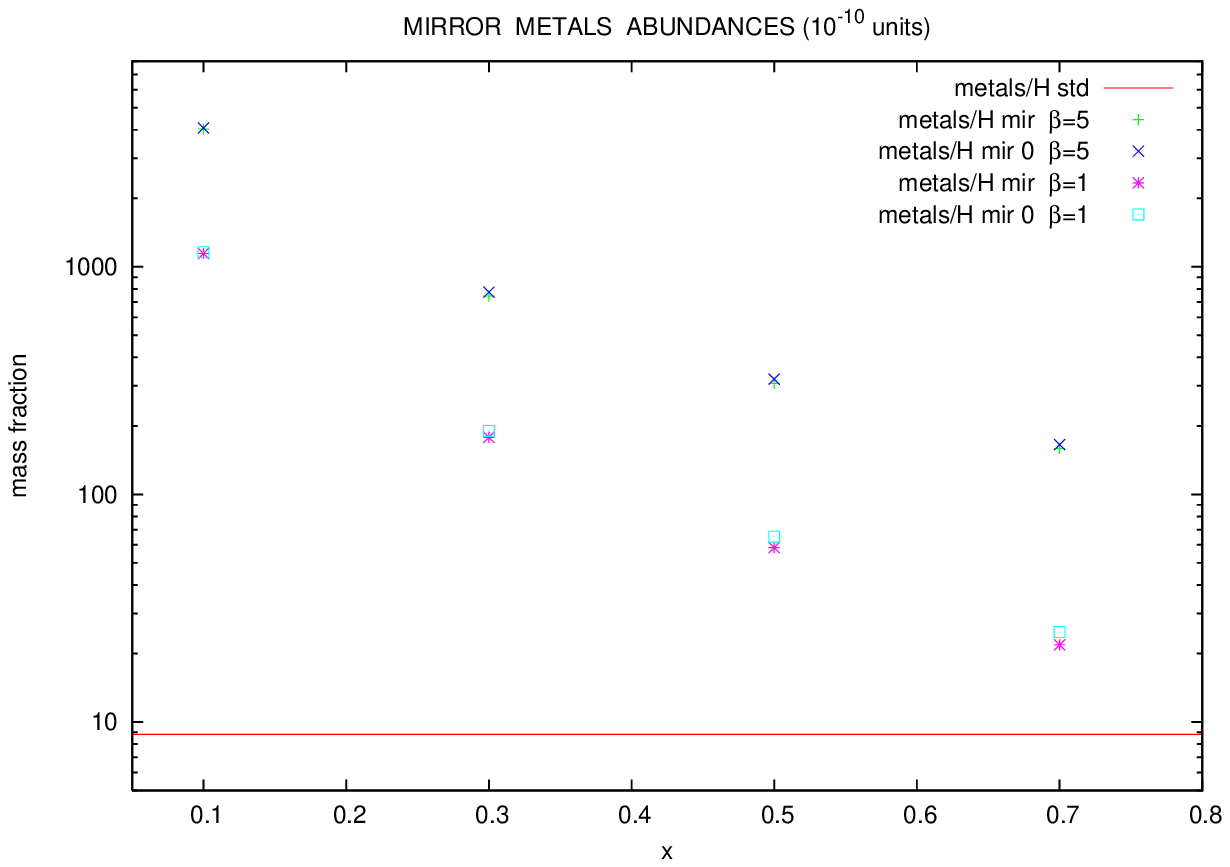}
	\bf \caption{\rm Light elements produced in the mirror sector during BBN. 
	On every plot we report the elements ($^4He$, $D$, $^3He$, $^7Li$ and metals)
	for several $x$ values, compared with the ordinary sector standard and the
	"order 0" quadratic approximation of $\Delta N_{\nu}$ in equation (\ref{N_nu_mir_approx}).
	For every $x$ we used two different values for the barion ratio 
	$\beta= \Omega'_B / \Omega_B$.} \rm
	\label{fig-mir-BBN}
	\end{center}
	\end{figure}
In Figure \ref{fig-mir-BBN} we plot the final abundances of $^4He$, $D$, $^3He$, $^7Li$ and metals in the mirror sector as a function of $x$. 
Since $T' \sim xT$, at higher $x$ the mirror sector becomes hotter; hence, we expect to have mirror abundances closer to the ordinary ones at higher $x$. 
Moreover, we note that in general lower $\beta$ implies final mass fractions closer to the standard.



A very rough approximation for the mirror $^4He$ production can be achieved from the following argument: for any given temperature $T'$, using equations (\ref{g_bar_def}) and (\ref{g_tot_with_mirror}) and assuming $\alpha \simeq 1$ we have 
	\begin{eqnarray}
	\label{hub_mir}
	H(T') \simeq 5.5 \, \sqrt{1 + \frac{1}{x^4}} \:
	\frac{T'^2}{M_P}
	\end{eqnarray}
for the Hubble expansion rate. Therefore, comparing $H(T')$ with the reaction rate $\Gamma (T') \propto {T'}^5$ - see equation (\ref{T_W}) - we find a freeze-out temperature 
	\begin{eqnarray}
	\label{}
	T'_W=(1+x^{-4})^{1/6} T_W
	\end{eqnarray}
which is larger than $T_{W}$, whereas the time scales as 
	\begin{eqnarray}
	\label{}
	t'_W = t_W/(1+x^{-4})^{5/6} < t_W
	\end{eqnarray}
(obtained using 
the relation $t \propto H^{-1}$). In addition, $\eta'$ is different from $\eta\simeq 5 \times 10^{-10}$. However, since $T_N$ depends on baryon density only logarithmically (see eq.~(\ref{T_A})), the temperature $T'_N$ remains essentially the same as $T_N$, while the time $t'_N$ scales as 
	\begin{eqnarray}
	\label{}
	t'_N = t_N/(1+x^{-4})^{1/2}
	\end{eqnarray}
Thus, for the mirror $^4$He mass fraction we obtain - see equation (\ref{helium}):  
	\begin{eqnarray}
	\label{m_helium}
	Y'_{4}\simeq 2X'_n(t'_N)= 
	{{ 2\exp^{-t_N/\tau(1+x^{-4})^{1/2}} } \over
	{1+\exp^{Q/T_W(1+x^{-4})^{1/6} } }} 
	\end{eqnarray}
We see that $Y'_{4}$ is an increasing function of $x^{-1}$.  In particular, for $x\rightarrow 0$ one has $Y'_{4}\rightarrow 1$.   
Hence, in this case $Y'_4$ is always bigger than $Y_4$.   In other words, if dark matter of the Universe is represented by the baryons of the mirror sector, it should contain considerably bigger fraction of primordial $^4$He than the ordinary world. 
In particular, the helium fraction of mirror matter is comprised between 20\% and 80\%, depending on the values of $x$ and $\eta '$. This is a very interesting feature, because it means that mirror sector can be a helium dominated world, with important consequences on star formation and evolution, and other related astrophysical aspects.


%% file: files/mirror_kin_mix.tex

\def \chap-mir-MCP{BBN when the mirror sector is millicharged}
\chapter{\chap-mir-MCP}
\label{chap-mir-MCP}
\markboth{Chapter \ref{chap-mir-MCP}. ~ \chap-mir-MCP}
                    {Chapter \ref{chap-mir-MCP}. ~ \chap-mir-MCP}

If the mirror particles are milli-charged with respect to the ordinary photon, see chapter \ref{The_mirror_universe}, there can be processes, like the pair annihilation and production, which lead to entropy and energy exchanges between ordinary and mirror sectors. 

A consequence of this is that the entropy densities ratio $x$ we defined in \S \ref{mirror_cosmology} is not a constant of time anymore, but changes during the universe evolution.

In this chapter we analyze these processes and work out rough bounds on the photon mixing parameter $\epsilon$; finally we will critically discuss them in order to propose further developments of the work began in this thesis.



\def \sec-O-M-processes{Physical processes involving mirror and ordinary matter}
\section{\sec-O-M-processes}
\label{sec-O-M-processes}
\markboth{Chapter \ref{chap-mir-MCP}. ~ \chap-mir-MCP}
                    {\S \ref{sec-O-M-processes} ~ \sec-O-M-processes}

The processes of lower order involving mirror and ordinary matter are proportional to $\epsilon^2$. 
There are three: the \it pair annihilation-production \rm $e^+e^-\leftrightarrow e'^+e'^- $, the \it scattering \rm $e e' \leftrightarrow e e'$ and the \it plasmon decay \rm $\gamma \rightarrow e'^+ e'^-$.
It can be shown that the plasma process is negligible: the plasma frequency in a relativistic plasma indeed is \cite{Raffelt:1996wa}:
	\begin{eqnarray}
	\label{}
	\omega_P = \sqrt{ \left(\frac{4\pi}{9} \alpha \right)} \: T \simeq 0.1 T < 1 MeV 
	\hspace{0.2cm} \mathrm{at \; BBN}
	\end{eqnarray}
Since we are interested in particles having the same or higher mass than the electron, the plasma process is below its threshold $(2m_e)$ and can be ignored.
The other two processes are analyzed below.

\def \sec-ee->ff{Pair annihilation and production}
\subsection{\sec-ee->ff}
\label{sec-ee->ff}
\markboth{Chapter \ref{chap-mir-MCP}. ~ \chap-mir-MCP}
                    {\S \ref{sec-ee->ff} ~ \sec-ee->ff}

The cross section for the process $e^+e^-\leftrightarrow e'^+e'^-$ is worked out in detail in appendix \ref{e_e_ebar->f_f_bar}. The total cross section for the special case $m_e=m_{e'}=m$  is:
	\begin{eqnarray}
	\label{}
	\sigma =  \frac{4}{3} \pi \alpha^2 \; \frac{\epsilon^2}{s^3} \left( s + 2 m^2\right)^2
	\end{eqnarray}
where the calculation has been performed in the center of mass frame of reference and $s$ stands for the total four-impulse squared: $s = |p_1 + p_2|^2$ - see \ref{s_average}.

Once the cross section is known, $\Gamma$ can be worked out; at low temperature we can follow Gondolo and Gelmini \cite{Gondolo:1990dk}, who demonstrated that the thermal average can be achieved solving a single integral:
	\begin{eqnarray}
	\label{Gondolo_Gelmini_Moller}
	\langle \sigma v_{Mo\!\!\!/l} \rangle = 
	\frac{1}{2} \cdot
	\frac{1}{8m^4TK_2^2(x)} 
	\int_{4m^2}^{\infty} \sigma \cdot (s-4m^2) \sqrt{s} K_1\left( \frac{\sqrt{s}}{T} \right) \;ds
	\end{eqnarray}
where $x=\frac{m}{T}$, $K_i$ are the modified Bessel functions of order $i$ and the factor $\frac{1}{2}$ arises because the initial particles are not identical. \it This formula is valid for particles with Maxwell-Boltzmann statistics or having any other statistics provided that $T\leq 3m$\rm . The corresponding quantity in the center of mass (CM) frame is:
	\begin{eqnarray}
	\label{}
	\langle \sigma v \rangle^{CM} = 
	2 \left[ 1+ \frac{K_1^2(x)}{K_2^2(x)} \right]^{-1} \langle \sigma v_{Mo\!\!\!/l} \rangle
	\end{eqnarray}
Recalling $n$ from equation (\ref{number_density}), $\Gamma$ is finally obtained as:
	\begin{eqnarray}
	\label{Gamma_Gondolo_ee_ff}
	\Gamma = \langle \sigma v \rangle^{CM} n
	\end{eqnarray}
At high temperature we can use an ultra-relativistic approximation, which is described below.
	\begin{figure}[htbp]
	\begin{center}
	 \includegraphics[scale=0.5]{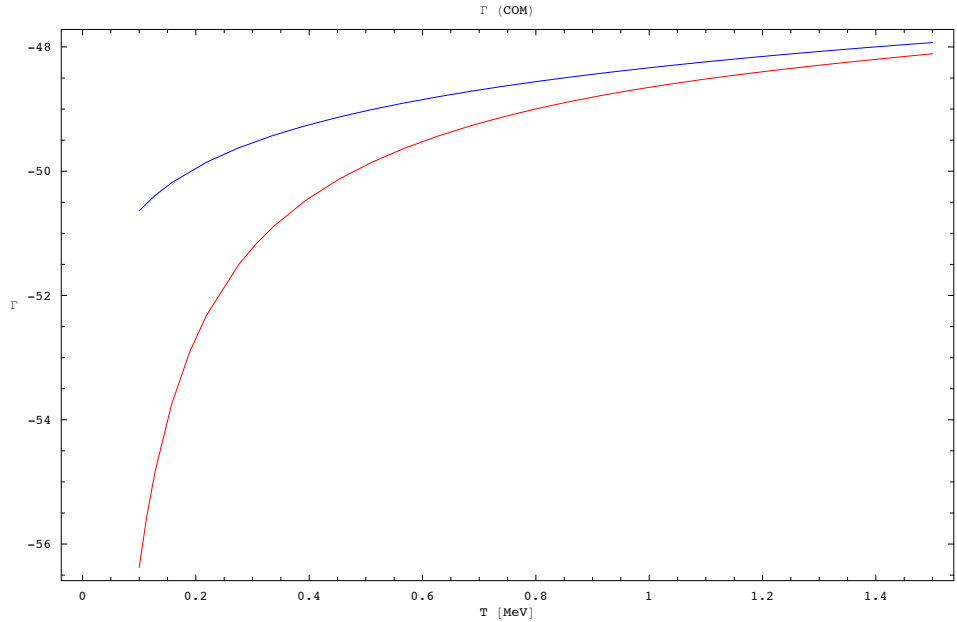}	
	 \hspace{0.1cm}
	\caption{{\it $\Gamma$ for $\epsilon=10^{-8}$ calculated from Gondolo and Gelmini 
	formula (\ref{Gamma_Gondolo_ee_ff}) 
	(red) and in the ultrarelativistic approximation in (\ref{Gamma_Davidson_ee_ff}) (blue). 
	$\Gamma$ is in units $\log(\Gamma/\mathrm{MeV})$}}
	\label{Gamma_e_ebar_f_fbar}
	\end{center}
	\end{figure}

\subsubsection{Ultrarelativistic approximation}

For ultrarelativistic (UR) fermions - see equation (\ref{td_fcts_ultra_rev}) 
- we have:
	\begin{eqnarray}
	\label{}
	\sigma_{UR} &=&  \frac{4}{3} \pi \alpha^2 \; \frac{\epsilon^2}{s}  
	\hspace{1.5cm} s \propto T^2
	\cr\cr
	n(T)_{UR} &=& \frac{3}{4} \frac{\zeta_3}{\pi^2} T^3
	\hspace{1.5cm} v=c  =1
	\end{eqnarray}
This way we get the ultrarelativistic interaction rate:
	\begin{eqnarray}
	\label{Gamma_Davidson_ee_ff}
	\Gamma_{UR} = n \langle c \sigma \rangle= \frac{\zeta_3}{2\pi} \epsilon^2 \alpha^2 T 
	\simeq 0.2 \epsilon^2 \alpha^2 T
	\end{eqnarray}
In Figure \ref{Gamma_e_ebar_f_fbar} $\Gamma$ is plotted for $\epsilon=10^{-8}$ in comparison with the ultrarelativistic approximation. We can see that the relative difference between the two formulae is of order $10\%$ or less in the range $1 \lsim T \lsim 5 $MeV, but it becomes much higher for any other temperatures.

Comparing the interaction rate (\ref{Gamma_Davidson_ee_ff}) with the Hubble parameter $H$ - see equations (\ref{eq_criterion}) and (\ref{H_Mp}) - we can work out a rough bound on the millicharge $\epsilon$. Indeed, to preserve mirror matter out of equilibrium when the neutron-proton reactions (\ref{reaz_eq_p_n}) freeze-out, that is at $T = T_W \simeq 0.8 MeV$ - see equation (\ref{T_W}) - we must have
	\begin{eqnarray}
	\frac{\Gamma}{H} &=& \frac{\zeta_3}{2\pi}\frac{1}{1.66 \sqrt{g}} (\epsilon \alpha)^2
	\frac{M_P}{T_W}<1
	\end{eqnarray}
from which we obtain the bound on the photon kinetic mixing parameter
	\begin{eqnarray}
	\label{eps_bound_exp}
	\epsilon \leq 571 \, \sqrt{\frac{T_W}{M_P}} \simeq 4.7 \cdot 10^{-9}
	\end{eqnarray}
%

%

\def \sec-ef->ef{Scattering}
\subsection{\sec-ef->ef}
\label{sec-ef->ef}
\markboth{Chapter \ref{chap-mir-MCP}. ~ \chap-mir-MCP}
                    {\S \ref{sec-ef->ef} ~ \sec-ef->ef}


The calculations we made to calculate the cross section for the process $e e' \leftrightarrow e e'$ are reported in appendix \ref{e_f->e_f}. The differential cross section is:
	\begin{eqnarray}
	\label{}
	\frac{d\sigma}{d\Omega}& =& 	\left[ 2E^4 + p^4 [1+\cos^2(\theta)] 
	+ 4 E^2 p^2 \cos^2\left( \frac{\theta}{2}\right) - 
	2m^2 (E^2 - p^2 cos\theta) + 2m^4 \right]  \cdot \cr\cr
	&\cdot& 
	\frac{\epsilon^2 \alpha^2}{32 \sin^4\left( \frac{\theta}{2}\right)} \cdot 
	\frac{1}{E^2 \, p^4} 
	\end{eqnarray}
Clearly, the differential cross section and its integral diverge at small angles; anyway this process can be \it ignored \rm at first order in $x$, as we can see from the following argument. 

Let us consider for simplicity relativistic particles, so that $n \propto T^3$, $\Gamma \propto T$ and $\langle E \rangle \propto T$; to compare the scattering process with the pair annihilation, we can use the order-zero approximation for the energy loss - see equation (\ref{ener_loss_appr}):
	\begin{eqnarray}
	\label{}
	\frac{d \rho_{ann}}{dt} \simeq - \Gamma \, n_{e^{\pm}} \langle E_{e^{\pm}} \rangle
	\propto T^5
	\end{eqnarray}
For the process $e e' \leftrightarrow e e'$ instead we have $s \propto T T'$ from which follows $\Gamma \propto n_{e^{\pm}}s \propto \frac{T^2}{T'}$ and therefore
	\begin{eqnarray}
	\label{}
	\frac{d \rho_{scat}}{dt} \simeq 
	- \Gamma \, n_{e' \bar e'} \langle E_{e' \bar e'} \rangle
	\propto \frac{T^2}{T'} T'^3 T'
	\simeq \frac{d \rho_{ann}}{dt} x^3
	\end{eqnarray}
BBN bounds impose $x \leq 0.64$ and therefore we can ignore $\frac{d \rho_{scat}}{dt}$ at first order since
	\begin{eqnarray}
	\label{}
	\frac{d \rho_{scat}}{dt} \leq \frac{1}{4} \, \frac{d \rho_{ann}}{dt} 
	\end{eqnarray}

\def \sec-E-exch{Energy transfer between the two sectors}
\section{\sec-E-exch}
\label{sec-E-exch}
\markboth{Chapter \ref{chap-mir-MCP}. ~ \chap-mir-MCP}
		{\S \ref{sec-E-exch} ~ \sec-E-exch}

In this section we will work out the energy exchange between the mirror and the ordinary sectors using various approximations.

For any processes the number of interacting particles $N$ changes in agreement with
	\begin{eqnarray}
	\label{ptc-num-loss}
	\frac{dN}{dt} = - \Gamma N
	\end{eqnarray}
where $\Gamma \equiv \langle \sigma v \rangle n_{target}$ as usual and 
%
	$N \equiv n V$,  
	$V \propto a^3$.
%

\def \sec-rho-no-exp{Neglecting the universe expansion}
\subsection{\sec-rho-no-exp}
\label{sec-rho-no-exp}
\markboth{Chapter \ref{chap-mir-MCP}. ~ \chap-mir-MCP}
		{\S \ref{sec-rho-no-exp} ~ \sec-rho-no-exp}

The roughest approximation we can make is to neglect the universe expansion, that is to assume $\dot{a} = 0$; in this case we will have:
	\begin{eqnarray}
	\label{}
	\frac{d n}{dt} = - \Gamma n 
	\hspace{1.5cm} 
	\frac{d \rho}{dt} \simeq \frac{d n}{dt}  \cdot \langle E (T) \rangle
	\end{eqnarray}
If ordinary particles, which are hotter that the corresponding mirror ones, annihilate by this process creating mirror matter and viceversa, the change in the ordinary world energy density will roughly be:
	\begin{eqnarray}
	\label{ener_loss_appr}
	\frac{d \rho_{o}}{dt} &\simeq& -\left[ \frac{d n_{o\rightarrow m} (T) }{dt} 
	 \langle E_o (T) \rangle - 
	\frac{d n_{m \rightarrow o}(T')}{dt}  \langle E_m (T') \rangle \right] \cr\cr
	&=& -\Gamma (T) n_o(T) \langle E_o (T) \rangle + \Gamma (T') n_m(T') 
	\langle E_m (T') \rangle  \cr\cr
	&=& - -\Gamma (T) \rho_o(T) + -\Gamma (T') \rho_m(T') \cr\cr
	&=& - \frac{\pi^2}{30} \left[ g(T) \Gamma(T) - g'(T') \Gamma(T') \left(\frac{T'}{T}\right)^4
	\right]  \cdot  T^4 \cr\cr
	&\simeq&  - \frac{\pi^2}{30} \left[ g(T) \Gamma(T) - g'(T') \Gamma(T') \left(x\right)^4
	\right]  \cdot  T^4
	\end{eqnarray}
where the index $o$ stands for "ordinary" and the index $m$ stands for "mirror".
In the previous calculation the thermodynamic functions - $ \rho, \, \langle E \rangle, \, g$ - are not of the whole universe but just of the involved particle species.

We stress that the term describing the energy transfer from the mirror to the ordinary sector is of order $\sim x^4 < 0.15$; thus, considering the present accuracy level, we can neglect it. In the following we will consider the only energy transfer from the ordinary to the mirror sector.


\subsubsection{Order of magnitude of the energy loss}

Using the ultrarelativistic interaction rate for the pair annihilation in equation (\ref{Gamma_Davidson_ee_ff}) in the order-zero approximation for the energy density loss (\ref{ener_loss_appr}) we get 
	\begin{eqnarray}
	\label{}
	\frac{d\rho}{dt} &\simeq& - \frac{\pi}{60} \zeta_3 (\epsilon \alpha)^2 g(T) T^5
	\left[ 1- x^5 \, \frac{g'(T')}{g(T)} \right] \cr\cr
	&\simeq& - 3\cdot 10^{-6} \epsilon^2 g(T) T^5
	\left[ 1- x^5 \, \frac{g'(T')}{g(T)} \right] \cr\cr
	&\simeq& - 3\cdot 10^{-6} \epsilon^2 g(T) T^5
	\end{eqnarray}
The substitutions $1MeV \simeq 1.6 \cdot 10^{21} s^{-1}$ and $ g(T) \simeq 4$ will finally lead us to
	\begin{eqnarray}
	\label{Glashow_energy_loss}
	\left| \frac{d\rho}{dt} \right| \simeq 2 \cdot 10^{16} \epsilon^2 T^5 \; (MeV \; s)^{-1} 
	\end{eqnarray}
which is the same energy loss used by Carlson and Glashow \cite{Carlson:1987si}. We can use this formula to work out a bound on $\epsilon$ more accurated than the one in equation (\ref{eps_bound_exp}). Indeed, the energy transferred to the mirror sector is about:
%
	\begin{eqnarray}
	\label{}
	\Delta \rho'(T) = \int
	\frac{d\rho}{dt'} dt' = 0.602 \: g_{e^{\pm}} \cdot 10^{-6} \epsilon^2 \bar g^{- 1/2} \: M_P T^3
	\end{eqnarray}
where we used the substitution in equation (\ref{t_vs_T}) and ignored the $g$ dependence on $T$, that is
	\begin{eqnarray}
	\label{}
	t \; \simeq \; 0.301 \; \bar g^{- 1/2} \; {M_P \over T^2} \hspace{1.5cm}
	dt \; \simeq \; - 0.602 \; \bar g^{- 1/2} \; {M_P \over T^3} \; dT
	\end{eqnarray}
We recall also that $\Delta g = 2 \cdot \frac{7}{8} \cdot \Delta N_{\nu} = 1.75 \cdot \Delta N_{\nu} $ and thus 
	\begin{eqnarray}
	\label{}
	\frac{\rho'}{\rho} \leq \frac{1.75 \: \Delta N_{\nu}}{10.75} \simeq 0.163 \: \Delta N_{\nu}
	\end{eqnarray}
But, since $\rho' \geq \Delta \rho'$, we have 
	\begin{eqnarray}
	\label{}
	\frac{\Delta \rho'}{\rho} =
	\epsilon^2 \, \frac{g_{ e^{\pm}} }{g \sqrt{\bar g} } \:
	\frac{30}{\pi^2} \: 0.6 \cdot 10^{-6} \frac{M_P}{T} 
	\leq \frac{\rho'}{\rho} \leq 0.163 \: \Delta N_{\nu}
	\end{eqnarray}
%
%
%
In conclusion, $\epsilon$ is bounded to
	\begin{eqnarray}
	\label{}
	\epsilon \leq \sqrt{ \frac{1}{6} \: \frac{\pi^2}{30} \frac{g \sqrt{\bar g}}{g_{ e^{\pm}}} \: \frac{1}{0.602} 
	\, 10^6 \frac{T_W}{M_P}} \cdot \sqrt{\Delta N_{\nu}}
	\end{eqnarray}
where as usual $T_W \simeq 0.8$ MeV, $M_P \simeq 1.2 \cdot 10^{22}$ MeV; at $T_W$ we have - see \S \ref{sec-Num-calc}: 
	\begin{eqnarray}
	\label{DOFs_num_T_W}
	g \simeq 10.45 \hspace{1.5cm} 10.45 \leq \bar g \leq 13.0
	\hspace{1.5cm} g_{ e^{\pm}} \simeq 3.39
	\end{eqnarray}
Using these values and assuming the conservative bound $\Delta N_{\nu} = 0.5$ we finally get the bound on the kinetic mixing parameter:
	\begin{eqnarray}
	\label{eps_bound_2}
	\epsilon \leq 0.246 \cdot 10^{-8} 
	\sqrt{\frac{g \sqrt{\bar g}}{g_{ e^{\pm}}}}
	\sqrt{\Delta N_{\nu}} \leq 
	0.80 \cdot 10^{-8}  \sqrt{\Delta N_{\nu}} \leq
	5.7 \cdot 10^{-9}
	\end{eqnarray}

\def \sec-n-exp{The number density in the expanding universe}
\subsection{\sec-n-exp}
\label{sec-n-exp}
\markboth{Chapter \ref{chap-mir-MCP}. ~ \chap-mir-MCP}
		{\S \ref{sec-n-exp} ~ \sec-n-exp}

Let us now take in account of the well-known fact that the universe expands and consider the ordinary particles annihilation followed by the mirror particles creation, that is the process $e^+e^-\leftrightarrow e'^+ \,  e'^- $; we inferred in \S \ref{sec-O-M-processes} that this is the most important process involving particles from both sectors.

We can use equation (\ref{ptc-num-loss}), which still holds and implies:
	\begin{eqnarray}
	\label{}
	\frac{dN_e}{dt} = - \Gamma N_e 
	\hspace{1cm} \mathrm{and} \hspace{1cm} 
	\frac{dN_{e'}}{dt} = \Gamma N_e
	\end{eqnarray}
Making the substitutions 
	\begin{eqnarray}
	\label{}
	N_{e'} \propto n_{e'} a^3  \hspace{1.2cm}  N_e \propto n_e a^3
	\hspace{1.2cm} H = \frac{\dot{a}}{a}
	\end{eqnarray}
we get the Boltzmann equation
	\begin{eqnarray}
	\label{Boltz_eq}
	\dot{n}_{e'}+ 3 n_{e'} H = \Gamma n_e
	\end{eqnarray}
Let us now rewrite equations (\ref{H_Mp}) and (\ref{t_vs_T}) as:
	\begin{eqnarray}
	\label{}
	t \simeq a \frac{M_P}{T^2}		\hspace{2cm}
	H \simeq b \frac{T^2}{M_P} 
	\end{eqnarray}
where $a$ and $b$ are constants and we neglet the $g$ dependence on $T$. Equation (\ref{Boltz_eq}) becomes in terms of the temperature:
	\begin{eqnarray}
	\label{Boltz_eq_T}
	\frac{dn_{e'}}{dT} - 3 \, \frac{n_{e'}}{T} = f(T)
	\end{eqnarray}
where
	\begin{eqnarray}
	\label{}
	f(T) = \frac{-2a \Gamma(T) n_e(T) \, M_P}{T^3}
	\end{eqnarray}
The first step to resolve the differential equation (\ref{Boltz_eq_T}) is to find a solution to the homogeneous associated equation, which gives:
	\begin{eqnarray}
	\label{n_homog}
	n_{e'} ^ {h}= c\, T^3
	\end{eqnarray}
where $h$ stands for homogeneous.
Then we make the constant $c$ vary in order to find the general solution. Assuming $c = c(T)$ in (\ref{n_homog}) and substituting it in (\ref{Boltz_eq_T}) we finally get
	\begin{eqnarray}
	\label{}
	c(T) = c_1 + \int_{\infty}^T \frac{f(y)}{y^3} dy
	\end{eqnarray}
where $c_1$ is a constant. Therefore the general solution of (\ref{Boltz_eq_T}) is
	\begin{eqnarray}
	\label{n_f_gen_sol}
	n_{e'} (T) = \left[ c_1 + \int_{\infty}^T \frac{f(y)}{y^3} dy \right] T^3
	\end{eqnarray}

\paragraph{Initial condition: empty mirror sector}

At high temperatures - such that $T \gg m_e$ - electrons are ultrarelativistic and we have $n_e \propto T^3$, $\Gamma \propto T$ - see \S \ref{sec-ee->ff}; it follows that $f(T) = c_2 T$ where 
	\begin{eqnarray}
	\label{}
	c_2 =  \frac{3 \cdot 0.301 \, \zeta_3^2}{\pi^3 \, \sqrt{\bar g}} \, \frac{g_{e^{\pm}}}{4} \, 
	M_P \, \alpha^2 \epsilon^2 \simeq
	5.6 \cdot 10^{-7} \cdot 
	\frac{g_{e^{\pm}}}{\sqrt{\bar g}} \,
	M_P \, \epsilon^2 
	\end{eqnarray}
%
Solving the integral in (\ref{n_f_gen_sol}) we finally get
	\begin{eqnarray}
	\label{}
	n_{e'}(T) = c_1 T^3 + c_2 T^2
	\end{eqnarray}
Let us impose now the initial condition requiring that the number of mirror partilces can be neglected respect to the number of ordinary particles, that is, the mirror sector is empty: 
	\begin{eqnarray}
	\label{}
	\lim_{T \rightarrow \infty} \frac{n_{e'}(T)}{n_e(T)} \propto 
	\lim_{T \rightarrow \infty} \frac{n_{e'}(T)}{T^3}= 0
	\end{eqnarray}
This implies $ c_1 = 0 $ and therefore the general solution  of equation (\ref{Boltz_eq_T}) becomes
	\begin{eqnarray}
	\label{n_f_gen_sol_bc}
	n_{e'} (T) = T^3 \int_{\infty}^T \frac{f(y)}{y^3} dy 
	\end{eqnarray}
which becomes, in the ultrarelativistic limit
	\begin{eqnarray}
	\label{}
	n_{e'} = c_2 T^2
	\end{eqnarray}

\paragraph{Initial condition: $x$ dependence}

An other possible initial condition is that at high temperatures the mirror fermions $f$ are at equilibrium with the other mirror species, that is:
	\begin{eqnarray}
	\label{}
	\lim_{T \rightarrow \infty} n_{e'} = \frac{3}{4} \, \frac{\zeta_3}{\pi^2} \, 4 \, T'^3
	\hspace{1cm} \mathrm{that \: is} \hspace{1cm} 
	\lim_{T \rightarrow \infty}  \frac{n_{e'}}{n_e} = \frac{T'^3}{T^3}
	\end{eqnarray}
This implies that 
	\begin{eqnarray}
	\label{}
	\lim_{T \rightarrow \infty} \frac{n_{e'}}{n_e} = 
	\frac{c_1 T^3}{\frac{3}{4} \, \frac{\zeta_3}{\pi^2} \, 4 \, T^3} =
	\frac{c_1 T'^3}{ x^3 \frac{3}{4} \, \frac{\zeta_3}{\pi^2} \, 4 \, T^3} =
	\frac{T'^3}{T^3}
	\end{eqnarray}
from which follows
	\begin{eqnarray}
	\label{}
	c_1 = \frac{3}{4} \, \frac{\zeta_3}{\pi^2} \, 4 x^3
	\end{eqnarray}

\def \sec-en-density-kin-mix{The energy density}
\subsection{\sec-en-density-kin-mix}
\label{sec-en-density-kin-mix}
\markboth{Chapter \ref{chap-mir-MCP}. ~ \chap-mir-MCP}
		{\S \ref{sec-en-density-kin-mix} ~ \sec-en-density-kin-mix}

The energy density of mirror particles can be written approximately as
	\begin{eqnarray}
	\label{}
	\rho_{e'} \simeq n_{e'} \langle E_{e'} \rangle
	\end{eqnarray}
while the energy transfer from the ordinary to the mirror sector is 
	\begin{eqnarray}
	\label{}
	\Delta \rho_{e'} = \Delta n_{e'} \langle E_{e'} \rangle 
	\end{eqnarray}
We worked out $n_{e'}$ as a function of the ordinary sector temperature $T$ in \S \ref{sec-n-exp}. 
%
%
In the ultrarelativistic limit we have $\Delta n_{e'} = c_2 T^2$, $\langle E_{e'} \rangle \simeq 3.15 \, T$ and we can use these formulae to work out the energy transfer from the ordinary to the mirror sector:
%
	\begin{eqnarray}
	\label{delta-rho-f}
	\Delta \rho_{e'} (T) \simeq
	3.15 \, c_2 T^3
	\end{eqnarray}
From which follows, applying the same procedure of \S \ref{sec-rho-no-exp}:
	\begin{eqnarray}
	\label{}
	\frac{\Delta \rho'}{\rho} \simeq
	\frac{1.8 \cdot 10^{-6} \cdot 
	\frac{g_{e^{\pm}}}{\sqrt{\bar g}} \,
	M_P \, \epsilon^2 T^3}{\frac{\pi^2}{30} \,g\, T^4} \leq 
	\frac{\rho'}{\rho} \leq \frac{1}{6} \Delta N_{\nu}
	\end{eqnarray}
Using again (\ref{DOFs_num_T_W}), we get the bound on the kinetic mixing parameter:
	\begin{eqnarray}
	\label{eps_bound_3}
	\epsilon \leq 
	\sqrt{ \frac{1}{6} \: \frac{\pi^2}{30} \frac{g \sqrt{\bar g}}{g_{ e^{\pm}}} \: \frac{1}{1.8} 
	\, 10^6 \frac{T_W}{M_P}} \cdot \sqrt{\Delta N_{\nu}}
	\simeq 3.3 \cdot 10^{-9}
	\end{eqnarray}

\def \sec-res-prosp{Prospects for future researches}
\section{\sec-res-prosp}
\label{sec-res-prosp}
\markboth{Chapter \ref{chap-mir-MCP}. ~ \chap-mir-MCP}
		{\S \ref{sec-res-prosp} ~ \sec-res-prosp}

In the previous chapters of this thesis we analyzed the features of the mirror model and worked out in detail its primordial nucleosynthesis if millicharged interactions are not present. 

Future researches should focus on BBN in presence of these interactions to work out more precise bounds on $\epsilon$ and $x$. 
In particular, the processes we analyzed in \S \ref{sec-O-M-processes} lead to energy and entropy exchanges between the sectors and therefore to a not trivial evolution of the entropy ratio $s'/s$ with time. In general we expect millicharged interactions to raise the mirror sector temperature, energy and entropy and lower the corresponding quantities in the ordinary sector, making $s'/s$ grow with time till the temperature becomes smaller than the threshold of the involved processes, that is $T \sim m_e /2$.

It is also important to stress that the presence of the mirror universe and its being a candidate for dark matter does not affect only BBN, but also other cosmological and astrophysical processes. 
The parameters $x$ and $\beta$, which are enough to unambigously fix the symmetric model, must be therefore chosen such that all these processes take place and are compatible with the experimental observations.

For instance, the structure formation in presence of mirror dark matter has been analyzed by Berezhiani et al. and the previsions of this model on the cosmic microwave background (CMB) and the large scale structure (LSS) have been compared with those of the cold dark matter (CDM) model \cite{Berezhiani:2003wj,Ciarcelluti:2003wm}.

What came out from these comparisons is that, if $x \leq 0.2$, CMB and LSS power spectra are equivalent for the CMB and CDM models and it is possible that all DM is made of mirror baryons, while for higher $x$ these spectra strongly depend on the amount of mirror baryons, which must, to save the compatibility with data, constitute not more than $20 \%$ of DM.
This happen because decoupling of matter and radiation must take place before in the mirror sector and be complete before the ordinary one begins.

Let us now recall the three rough bounds on $\epsilon$ we worked out in equations (\ref{eps_bound_exp}), (\ref{eps_bound_2}) and (\ref{eps_bound_3}). All them have the same order of magnitude, that is $\epsilon \leq (3.3 \div 5.7) \times 10^{-9}$ and are proportional to $\sqrt{\Delta N_{\nu}}$ \footnote{Cosmological limit on $\epsilon$ will be discussed in more details elsewhere \cite{paper_prep}}.
We recall also from equation (\ref{x_N_nu}) that $\Delta N_{\nu} \propto x^4$ and hence
	\begin{eqnarray}
	\label{}
	\epsilon_{max} \propto \sqrt{\Delta N_{\nu}}
	\propto 
	x^2
	\end{eqnarray}
Therefore we can see that imposing $x \leq 0.2$ 
implies $\Delta N_{\nu} \leq 10^{-2}$ and hence the bounds on $\epsilon$ must be scaled by a factor 
	\begin{eqnarray}
	\label{}
	 \sqrt{2 \cdot 10^{-2}} \sim 0.14 \sim \frac{1}{7}
	\end{eqnarray}
and hence they become about 7 times smaller, so that
%
	$\epsilon \leq (4.6 \div 8.1) \cdot 10^{-10}$
%
These new bounds are not compatible with $\epsilon \sim 5 \cdot 10^{-9}$ required to explain the DAMA signal in terms of mirror baryons scattering - see \S \ref{sec-MCP-bounds}.

Nevertheless, there are at least two ways out: the first is found assuming that DM is not only done by mirror baryons but also by some other kind of cold matter; this way $x$ is only bounded to be less than 0.64 by BBN and we have the bounds in equations (\ref{eps_bound_exp}), (\ref{eps_bound_2}) and (\ref{eps_bound_3}).

The second is based on the asymmetric mirror model introduced in \S \ref{sec-mirror_mat_DM}, where $m_B' \sim 5 \, m_B$, $m_{e'} \sim 100 \, m_e$. In this scenario mirror matter and radiation decouple before, thus the only bound on $x$ arises from BBN, that is $x<0.64$.
Moreover, since the 
threshold for the process $e^+e^- \leftrightarrow e'^+ e'^-$ is at $T_{thr} \sim \frac{m_{e'}}{2} \sim 25$ MeV, the energy and entropy exchanges between the sectors lose their efficacy before.

In particular, 
at $T<T_{thr}$ exchanges between the sectors are not efficient anymore and hence the universe evolution goes on just as in chapter \ref{chap-mir-BBN}, the ratio $s'/s$ is a constant, both $x$ and $\Delta N_{\nu}$ are approximately constant (if we neglet the small variation due to the $\alpha$ factor in equation (\ref{beta_factor})) and so on. 
Therefore we can replace $T_W$ by $T_{thr}$ in the three bounds on $\epsilon$ we worked out, that is equations (\ref{eps_bound_exp}), (\ref{eps_bound_2}) and (\ref{eps_bound_3}), and this way they become higher by a factor
	\begin{eqnarray}
	\label{}
	\sqrt{\frac{T_{thr}}{T_W}} \sim 6
	\end{eqnarray}
that is, we have
	\begin{eqnarray}
	\label{}
	\epsilon \leq 2 \div 3.5 \cdot 10^{-8}
	\end{eqnarray}
%
In such an asymmetric mirror scenario we may have enough baryons to constitute all the dark matter required by experimental observations of the universe.
At the same time, this scenario is compatible with CMB and LSS and its BBN impose bounds on the kinetic mixing parameter $\epsilon$ of order a few unity $10^{-8}$. 
This asymmetric model may also explain the DAMA/NaI annual modulation in terms of scattering of millicharged mirror baryons on ordinary matter, since $\epsilon$ is high enough.
But all these intuitive hints should be developed in a more rigorous way to work out the parameter evolution with time and the $\epsilon$ value required to explain the DAMA/NaI modulation and finally verify the compatibility of the model with all the experiments.


%% file: files/Conclusions.tex

\noindent {\huge \noindent  {\bf Conclusion}}
\noindent  \addcontentsline{toc}{chapter}{{Conclusions}}
\vspace{1.5cm}
\label{chap-concl}
\markboth{Conclusions}{}

\noindent
It is nowadays well known that particle physics and astrophysics are tighly bound; indeed, a good knowledge of the interactions between particles can help in understanding or building models for cosmology and astrophysics, as well as the present and primordial universe provide high energy environments which are unreachable in laboratory experiments but are crucial for studies concerning elementary particle interactions.
This is why in the recent years a new branch - called \it astroparticle physics \rm - has became even more popular.

This thesis is devoted to some astroparticle implications of a mirror sector, which is a parallel hidden sector of particles and interactions, completely identical to the ordinary (observable) particle sector, i.e. standard model and its possible extensions, and coupled to the latter via gravity.
In other words, the mirror sector has identical particle physics to that of the ordinary one.
The mirror sector presence restores the apparently unexplicable lack of the left-right symmetry in nature;
moreover, mirror baryons are a viable dark matter candidate, since they couple to the ordinary matter mainly via gravity and their density can be as high as the dark matter one should be (about $5$ times bigger then the ordinary one).
Therefore experiments searching for dark matter (such as DAMA) are natural candidates for detecting mirror matter.

The presence of the mirror sectore does not introduce any new parameters in particle physics (as soon as mirror sector particles and interactions are identical to the ordinary ones). However in cosmology they should have different realizations; in particular the temperature of the mirror sector $T'$ should be less than the ordinary one $T$, otherwise it would spoil standard BBN predictions, which are one of the key stones of the standard cosmology.
Therefore the value $x \simeq T' /T$ is a free parameter which value is restricted by BBN conditions to $x < 0.64$.
 

An interesting possibility is that any ordinary particles may be mixed with its mirror twin. 
In particular ordinary photons can have kinetic mixings with mirror photons, as well as ordinary neutrinos, $\pi^0$, neutron and so on can be kinetically mixed with their mirror partners.

This thesis is devoted mainly to the kinetic mixing of ordinary and mirror photons and its cosmological consequences regarding BBN epoch of the primordial production of light elements and is constructed in the following way.

The first two sections are devoted to basic concepts and models of particle physics and cosmology. 
We start with a brief review of the standard model of elementary particles and their interactions and the standard cosmological paradigm, followed by a discussion about the origin of the electric charge quantization in particle physics (third chapter), which is induced in the quantum standard model by the anomaly cancellation conditions and naturally arises in grand unified theories or in presence of magnetic monopoles. 


In particular, it is a challenging question whether "milli"charged particles, having very tiny electric charges could exist in the universe. This possibility is present in models where two photons kinetically mixed are present, what can be naturally achieved in models containing a mirror sector, in which case mirror electron, positron, proton and so on would emerge as millicharged with respect to the ordinary photon.
Of course, if mirror particles are millicharged, this would open new possibilities for their detection.

In chapter \ref{chap-mir-BBN} we worked out a detailed analysis of BBN in the mirror model as well as in the mirror sector itself. 
To perform these calculations we first found a system of equations which numerical solution gives the mirror temperature $T'$ corresponding to any ordinary temperatures $T$; once these temperatures were known, we were able to work out the number of degrees of freedom and the number of extra-neutrinos in both sectors. 
A special feature of models containing the mirror sector is that the number of equivalent neutrinos (that is the total number of degrees of freedom expressed in terms of neutrinos) changes during the universe evolution.
Finally we modified the Kawano code for BBN to let it read the number of neutrinos as an external input changing with time. 
These calculations had never been done before with the same accuracy.

In chapter \ref{chap-mir-MCP} we insterted the kinetic mixing of ordinary and mirror photons and  analyzed the main processes leading to energy and entropy exchanges between ordinary and mirror sectors. 
We made some extimates of the maximum value for the photon mixing parameter $\epsilon$, which came out to be a few units $ \times 10^{-9}$.

An interesting possibility is that mirror baryons scattering on ordinary baryons may explain the DAMA/NaI annual modulation if $\epsilon \sim 5 \cdot 10^{-9}$, that is of order of our upper bound.

We would like now to stress some final remarks which constitute a starting point for further future researches. 
In our extimates the maximum value of $\epsilon$ (let us call it $\epsilon_{max}$) is proportional to $x^2$; the structure formation bounds $x$ to be smaller than $0.2$ and this leads $\epsilon_{max}$ to be about one order of magnitude stronger, that is, not compatible with the DAMA/NaI signal.

Anyway, the bound $x \leq 0.2$ applies only if all dark matter is made of mirror baryons; hence we can assume that at least $80\%$ of it is made of some other kind of cold matter and save compatibility with DAMA.

Otherwise we may use an asymmetric mirror model, where $m'_B \sim 5 \, m_B$, $m'_e \sim 100 \, m_e$. In this scenario the condition $\Omega'_B \sim 5 \, \Omega_B$ is naturally achieved if mirror and ordinary baryons have the same density. Moreover, having heavier mirror electrons lead to an earlier freezing of the pair annihilation-production processes and this allows $\epsilon_{max}$ about one order of magnitude bigger ($\sim 3 \cdot 10^{-8}$), that is, compatible with the DAMA interpretation.

\it In conclusion\rm , the mirror model is a challenging theory, supported by theoretical symmetry reasons as well as by its providing a natural candidate for dark matter. 
The mirror sector interacts with the ordinary one via gravity, but it may also have small interactions via photon mixing (the mirror model also provides a natural explication for this smallness). 
Observable effects of the mirror matter may be found in experiments searching for dark matter as well as in the primordial universe. 
In particular, calculations of the light elements produced during BBN should be performed with higher accuracy for both symmetric and asymmetric mirror models and the compatibility of the parameters with the DAMA/NaI annual modulation should be verified.


%% file: files/Ringraziamenti.tex

\noindent {\huge \noindent  {\bf Acknowledgments}}
\noindent  \addcontentsline{toc}{chapter}{{Acknowledgments}}
\vspace{1.5cm}
\markboth{Acknowledgments}{}

\noindent
Vorrei innanzitutto ringraziare le persone che con il loro aiuto e i loro suggerimenti hanno reso possibile questo lavoro: il mio relatore, Prof. Zurab Berezhiani, il Dr. Paolo Ciarcelluti ed il Dr. Francesco Vissani.

I am thankful to Dr. Andreas Ringwald for introducing me in the millicharged world.

Un ringraziamento speciale va a Paolo per la vicinanza e l'amore costantemente dimostratomi nel corso del tempo nonostante io l'abbia a volte messo a dura prova (anche oggi con questi ringraziamenti).

Grazie a mamma, pap\`a, Stefania \& Sergio, Federica \& Luca, Elisa, Francesco e Massimo per il sostegno e l'incoraggiamento datomi durante i lunghi anni che hanno preceduto questo risultato.

Non posso ovviamente dimenticare gli amici: spero mi passerete la menzione d'onore a Federica, gli altri sono in ordine rigorosamente alfabetico: (cat)Alessia, Andrea (sei un l. c.!!!), Chiaretta-Nemo
, Denise DoDe, Elenin, Eleonora-Dory, Enrica, Federi'o, Francesco (nonostante il Pescara calcio), la piccola Letizia, Marco (pace e bene), Maria Claudia, Michele (nonostante la tarantola), tutti gli altri che hanno contribuito a rendere il Dipartimento di fisica (quasi) vivibile e gli amici di ingegneria.


%% file: files/appendix_cosmology.tex


\chapter{Useful formulae for cosmology and thermodynamics}
\def \appen-cosm-mod{Useful formula for cosmology and thermodynamics}
\label{appen-cosm-mod}

\def \gen_rel{General Relativity}
\section{\gen_rel}
\label{gen_rel}
\markboth{Appendix \ref{appen-cosm-mod}. ~ \appen-cosm-mod}
                    {\S \ref{gen_rel} ~ \gen_rel}

The essence of Einstein's theory is to transform gravitation from being a force to being a property of space-time, which may be curved. The interval between two events 
	\begin{eqnarray}
	\label{metric}
	ds^2=g_{\alpha \beta}(x)~dx^{\alpha} dx^{\beta}
	\end{eqnarray}
is fixed by the {\sl metric tensor} $g_{\alpha \beta }$ which describes the space-time geometry\footnote{Repeated indices imply summation and $ \alpha, \beta $ run from 0 to 3; $ x^0 = t $ is the time coordinate and $x^i$ ($i = 1,2,3$) are space coordinates.} ($ g^{\alpha \mu }g_{\mu \beta } = \delta^{\alpha}_{\beta} $). 
For the Riemannian spaces, the {\sl tensor of curvature} is
	\begin{eqnarray}
	\label{riemanntensor}
	R^\mu _{\nu \alpha \beta} = 
	{\partial \Gamma^\mu _{\nu \beta} \over \partial x^\alpha } - 
	{\partial \Gamma^\mu _{\nu \alpha } \over \partial x^\beta } + 
	\Gamma^\mu _{\sigma \alpha } \Gamma^\sigma _{\upsilon \beta } - 
	\Gamma^\mu _{\sigma \beta } \Gamma^\sigma _{\nu \alpha } 
	\end{eqnarray}
where the $\Gamma$'s are {\sl Christoffel symbols}
	\begin{eqnarray}
	\label{christoffel}
	\Gamma^\mu _{\alpha \beta } =
	 {1 \over 2} \; g^{\mu \sigma }\left[{\partial g_{\sigma \alpha } \over \partial x^\beta }+
	 {\partial g_{\sigma \beta } \over \partial x^\alpha }-
	 {\partial g_{\alpha \beta } \over \partial x^\sigma }\right]
	\end{eqnarray}
The equation of motion of a free particle is determined by the space-time metric
	\begin{eqnarray}
	\label{geodesic}
	{d^2 x^\alpha  \over ds^2} + 
	\Gamma^\alpha _{\beta \gamma } {d x^\beta  \over ds}{d x^\gamma  \over ds} = 0
	\end{eqnarray}
so that free particle moves on a {\sl geodesic}.

On the other hand, the metric $g_{\alpha \beta }$ is itself determined by the distribution of matter, described by the {\sl energy-momentum tensor} $ T_{\alpha \beta } $, according to the {\em Einstein equations}
	\begin{eqnarray}
	\label{einsteinequation}
	G_{\alpha \beta } \equiv  R_{\alpha \beta } - 
	{1\over 2} R g_{\alpha \beta } = {8\pi G} T_{\alpha \beta } + \Lambda g_{\alpha \beta }
	\end{eqnarray}
where 
	\begin{eqnarray}
	\label{riccitensor}
	R_{\alpha \beta } = 
	R^\gamma _{\alpha \gamma \beta } ~,~~~~~~~~~~~ R = g^{\alpha \beta }R_{\alpha \beta }
	\end{eqnarray}
are respectively the {\sl Ricci tensor} and the {\sl Ricci scalar}, and $\Lambda$ is the {\sl cosmological constant}. For the FRW metric, the non zero components of the Ricci tensor and the value of the Ricci scalar are
	\begin{eqnarray}
	\label{ricci}
	& & R_{00} = -3 {\ddot a \over a } ~,~~~~~~~~~ 
	R_{ij} = - \left[ {\ddot a \over a } + 2{\dot a^2 \over a^2 } + { 2k \over a^2 } \right] g_{ij} \;, 
	\nonumber \\
	& & R = -6 \left[ {\ddot a \over a } + {\dot a^2 \over a^2 } + { k \over a^2 } \right] 
	\end{eqnarray}
For a perfect fluid the energy-momentum tensor has the form
	\begin{eqnarray}
	\label{energymomentum}
	T_{\alpha \beta } = (p + \rho) u_\alpha  u_\beta  - p g_{\alpha \beta }
	\end{eqnarray}
where $\rho$ and $p$ are respectively the energy density and pressure of the fluid, $u_\alpha = g_{\alpha \beta } \, dx^\beta / ds $ is the fluid four-velocity. Considering the symmetries of the FRW metric (uniformity and isotropy), which demand that $u^0 = 1$ and $u^i = 0$ in the comoving coordinate system, we obtain $T_{\alpha\beta} = diag(\rho,-p,-p,-p)$. This is valid also in presence of the cosmological constant, if we substitute $ p $ and $ \rho $ as indicated in the following
	\begin{eqnarray}
	\label{p_rho_wl}
	p\, \to p - p_\Lambda ~,~~~~~~~~~~ 
	\rho\, \to \rho + \rho_\Lambda ~;~~~~~~~~~~ \rho_\Lambda = - p_\Lambda = {\Lambda \over 8\pi G} 
	\end{eqnarray}
Therefore, the Einstein equations for a Universe described by the FRW metric are reduced to equations (\ref{friedmann1}) and (\ref{friedmann3}), where the relations between the energy and pressure densities are in general related as $ p = w \rho $. In particular, for the dominance of relativistic and non relativistic matter we have respectively $ w = 1/3 $ and $ w = 0 $, while for the vacuum energy dominance one has $ w = -1 $.

\def \app-flat{Flat models}
\section{\app-flat}
\label{app-flat}
\markboth{Appendix \ref{appen-cosm-mod}. ~ \appen-cosm-mod}
                    {\S \ref{app-flat} ~ \app-flat}

Here we report two special cases of the relationships (\ref{aflat})-(\ref{rhoflat}): {\sl dust} or {\sl matter dominated Universe} ($w = 0$)
	\begin{eqnarray}
	\label{aflatMD}
	a(t) & = & a_0 {\left( t \over t_0 \right)}^{2/3} ~ \\
	\label{tflatMD}
	t & = & t_0 {\left( 1+z \right)}^{-3/2} ~ \\
	\label{HflatMD}
	H & = & {2 \over 3t} = H_0 {\left( 1+z \right)}^{3/2} ~ \\
	\label{qflatMD}
	q_0 & = & {1 \over 2} ~ \\
	\label{t0flatMD}
	t_0 & = & {2 \over {3 H_0}} ~ \\
	\label{rhoflatMD}
	\rho_m & = & { 1 \over {6 \pi G t^2}} ~
	\end{eqnarray}
%
and {\sl radiation dominated Universe} ($w = 1/3$)
	\begin{eqnarray}
	\label{aflatRD}
	a(t) & = & a_0 {\left( t \over t_0 \right)}^{1/2} ~ \\
	\label{tflatRD}
	t & = & t_0 {\left( 1+z \right)}^{-2} ~ \\
	\label{HflatRD}
	H & = & {1 \over 2t} = H_0 {\left( 1+z \right)}^2 ~ \\
	\label{qflatRD}
	q_0 & = & 1 ~ \\
	\label{t0flatRD}
	t_0 & = & {1 \over 2 H_0} ~ \\
	\label{rhoflatRD}
	\rho_r & = & { 3 \over {32 \pi G t^2}} ~
	\end{eqnarray}

\def \thr_fun{Thermodynamic functions in special cases}
\section{\thr_fun}
\label{Td_ultrarev_ptc}
\markboth{Appendix \ref{appen-cosm-mod}. ~ \appen-cosm-mod}
                    {\S \ref{Td_ultrarev_ptc} ~ \thr_fun}

In this section the number density $n$, the energy density $\rho$ and the pressure $p$ functions in some special simpler limits are reported. 
\paragraph{Ultrarelativistic limit: $T \gg m \; , \; T \gg \mu$}
	\begin{eqnarray}
	\label{td_fcts_ultra_rev}
	\rho = 3p &=& \Biggl\{ 
	\begin{array}{c} \pi^2/30 \, g_{spin_A} T^4 \hspace{1.3cm} \rm{(BOSE)}
	\\ \frac{7}{8} \cdot \pi^2/30 \, g_{spin_A} T^4 \hspace{1cm} \rm{(FERMI)}
	\end{array}\  
%
%
	\cr\cr\cr
	n &=& \Biggl\{ 
	\begin{array}{c} \zeta(3)/\pi^2 \, g_{spin_A} T^3 \hspace{1.85cm} \rm{(BOSE)}
	\\\ \frac{3}{4} \cdot \zeta(3) / \pi^2 \, g_{spin_A} T^3 \hspace{1.4cm} \rm{(FERMI)}
	\end{array}\
	\end{eqnarray}
\paragraph{Non-relativistic limit: $T \ll m$}
	\begin{eqnarray}
	\label{td_fcts_non_rev}
	n &=&  g_{spin_A} \left( \frac{mT}{2\pi}\right)^{\frac{3}{2}}
	\exp^{\frac{-(m-\mu)}{T}} \cr\cr
	\rho &=& mn \propto \exp^{\frac{-(m-\mu)}{T}}  \cr\cr
	p &=& nT = \frac{\rho \, T}{m} \ll \rho
	\end{eqnarray}

\def \entr_cons{Entropy: definition and conservation}
\section{\entr_cons}
\label{entropy_derivation}
\markboth{Appendix \ref{appen-cosm-mod}. ~ \appen-cosm-mod}
                    {\S \ref{entropy_derivation} ~ \entr_cons}

The second law of thermodynamics states that
	\begin{eqnarray}
	\label{II_law_td_entropy}
	TdS = d(\rho V) + pdV - \mu \, d(nV)= d[(p+\rho)V] - Vdp - \mu \: d(nV)
	\end{eqnarray}
where $p$ and $\rho$ are the equilibrium energy density and pressure; 
%
%
differentiating equation (\ref{pressure}) with respect to $T$ we get
	\begin{eqnarray}
	\label{der_P_T}
	\frac{dp}{dT}&=& \frac{g}{6 \pi^2} \int_0^{\infty} \frac{k^4}{E} \cdot
	\frac{1}{\left( \exp^{\frac{E-\mu}{T}} \pm 1 \right)^2} 
	\cdot \exp^{\frac{E-\mu}{T}} \cdot
	\left[ \frac{E}{T^2} + \frac{d}{dT} \left( \frac{\mu}{T} \right) \right]
	\:dk \cr\cr\cr
	&=& - \frac{g}{6 \pi^2}  \int_0^{\infty} k^3T \cdot
	\frac{df}{dk} \cdot \left[ \frac{E}{T^2} + \frac{d}{dT} \left( \frac{\mu}{T} \right) \right]
	\:dk 
	\end{eqnarray}
where we used 
	\begin{eqnarray}
	\label{}
	f &=&  \frac{1}{ \exp^{\frac{E-\mu}{T}} \pm 1 } \cr\cr\cr
	\frac{df}{dk} &=& - \frac{k}{ET}  \cdot
	f^2 \cdot
	\frac{E-\mu}{T}
	\end{eqnarray}
	\begin{eqnarray}
	\label{}
	E^2=k^2+m^2 \;\;\;\;\;\;\;\;
	\Longrightarrow \;\;\;\;\;\;\;\;
	\frac{dE}{dk}= \frac{k}{E}
	\end{eqnarray}
Integrating equation (\ref{der_P_T}) by parts and using equations (\ref{energy_density}) and (\ref{pressure}) we get:
	\begin{eqnarray}
	\label{der_p_T_final}
	\frac{dp}{dT}&=&
	- \frac{g}{6 \pi^2} \left[ k^3T \cdot
	 f\cdot \left( \frac{E}{T^2} + \frac{d}{dT} \left( \frac{\mu}{T} \right) \right) \right]_{k=0}^{k=\infty}
	\cr\cr\cr
	&+& \frac{g}{6 \pi^2} \int_0^{\infty} f k^2 \cdot \left[ 3 T 
	\left( \frac{E}{T} + \frac{d}{dT} \left( \frac{\mu}{T} \right) \right) +
	k T  \left(  \frac{k}{ET} \right) \right] 
	\cr\cr\cr
	&=& \frac{\rho + p}{T} + nT \cdot \frac{d}{dT} \left( \frac{\mu}{T}\right)
	\end{eqnarray}
Using this equation in (\ref{II_law_td_entropy}) we get:
	\begin{eqnarray}
	\label{entropy_generic_bis}
	dS &=& \frac{1}{T} d \left[ (\rho + p) V \right] -
	\frac{1}{T} V \left[ \frac{p + \rho}{T} + nT \cdot \frac{d}{dT} \left( \frac{\mu}{T}\right) \right] \: dT 
	- \frac{ \mu d(nV)}{T} \cr\cr
	&=& d \left[ V \cdot \frac{\rho + p -n\mu}{T}\right] 
	\end{eqnarray}
that is equation (\ref{entropy_generic}) if we assume as usual $V \propto R^3$.  In the following we will always assume $\mu \ll T$. To demonstrate that $S=constant$ we recall the first law of thermodynamics (\ref{FirstLaw2}):
%
%
	\begin{eqnarray}
	\label{}
	d(\rho V) = - pdV - \mu d(nV)
	\end{eqnarray}
Sobstituting this identity in (\ref{entropy_generic_bis}) and using again (\ref{der_p_T_final}) we finally get:
	\begin{eqnarray}
	\label{}
	dS=0
	\end{eqnarray}

\def \s_average{The four-impulse average for massless particles}
\section{\s_average}
\label{s_average}
\markboth{Appendix \ref{appen-cosm-mod}. ~ \appen-cosm-mod}
                    {\S \ref{s_average} ~ \s_average}

Let us consider a process where two particle $1$ and $2$ with four impulses $p_i = (E_i, \mathbf{p_i})$, $i=1,2$ interact.
When they are ultrarelativistic the interaction rate for processes having $\sigma \propto s$ - see \S \ref{sec-nu-scat} - is 
	\begin{eqnarray}
	\label{}
	\Gamma \equiv n \langle v\sigma\rangle \simeq  n \langle \sigma \rangle
	\propto  n \langle s \rangle
	\end{eqnarray}
It is therefore useful to work out the four-impulse squared average;
the total four-impulse squared is defined as $s = |p_1 + p_2|^2$ and its average on the impulses is
	\begin{eqnarray}
	\label{s_aver_m_0}
	\langle s \rangle \equiv
	\frac{\int |p_1 + p_2|^2 f_1 (\mathbf{p_1},t) f_2(\mathbf{p_2},t) \; d^3\mathbf{p_1} \; d^3\mathbf{p_2}}
	{\int f_1(\mathbf{p_1},t) f_2 (\mathbf{p_2},t) \; d^3\mathbf{p_1} \; d^3\mathbf{p_2}}
	\end{eqnarray}
where $f_i (\mathbf{p_i},t) \; d^3\mathbf{p_i}$ has been defined in equation (\ref{eq_distr_funct}).
For massless particles we have:
	\begin{eqnarray}
	\label{s_ulrtarev}
	p_i = (|\mathbf{p_i}|, \mathbf{p_i}) 
	\hspace{0.8cm} \Longrightarrow \hspace{0.8cm} 
	|p_1 + p_2|^2 = 2 |\mathbf{p_1}| |\mathbf{p_2}| (1-\cos \theta) 
	\end{eqnarray}
	\begin{eqnarray}
	\label{}
	f_i \propto \frac{1}{\exp^{\frac{|\mathbf{p_i}|}{T}} \pm 1} 
	\end{eqnarray}
	\begin{eqnarray}
	\label{}
	\int d^3\mathbf{p_1} \; d^3\mathbf{p_2}
	= 4 \pi  \int |\mathbf{p_1}|^2 d |\mathbf{p_1}| \cdot
	2\pi |\mathbf{p_2}|^2 \int d |\mathbf{p_2}|
	\int_{-1}^{+1} d\cos\theta
	\end{eqnarray}
The denominator of (\ref{s_aver_m_0}) is:
	\begin{eqnarray}
	\label{s_aver_denom}
	\int f_1(\mathbf{p_1},t) f_2 (\mathbf{p_2},t) \; d^3\mathbf{p_1} \; d^3\mathbf{p_2} = 
	(4\pi)^2 \int \frac{p_1^2 d |\mathbf{p_1}|}{\exp^{\frac{|\mathbf{p_1}|}{T}} \pm 1}
	 \int \frac{p_2^2 d |\mathbf{p_2}|}{\exp^{\frac{|\mathbf{p_2}|}{T}} \pm 1} 
	 \end{eqnarray}
and can be easily worked out recalling that 
	\begin{eqnarray}
	\label{}
	\int \frac{p_i^2 d |\mathbf{p_i}|}{\exp^{\frac{|\mathbf{p_i}|}{T}} \pm 1}
	\propto n_{i\; \mathrm{ultrarelativistic}}
	\end{eqnarray}
where $ n_{i\; \mathrm{ultrarelativistic}}$ is given in (\ref{td_fcts_ultra_rev}). Therefore equation (\ref{s_aver_denom}) reduces to
	\begin{eqnarray}
	\label{}
	 64\pi^2 \zeta_3^2 T^6 \times
	 \Biggl\{ 
	\begin{array}{c} 1 \hspace{1.5cm} \rm{(for \; any \; present \; BOSONS)}
	\\ \frac{3}{4}  \hspace{1.cm} \rm{(for \; any \; present \; FERMIONS)}
	\end{array}
	\end{eqnarray}
Similarly the numerator of (\ref{s_aver_m_0}) turns out to be proportional to the energy density $\rho$ for ultrarelativistic particles. Using (\ref{s_ulrtarev}) we have:
	\begin{eqnarray}
	\label{s_aver_num}
	16 \pi^2 \int_{-1}^{+1} (1 - \cos\theta)  d\cos\theta
	\int \frac{p_1^3}{\exp^{\frac{|\mathbf{p_1}|}{T}} \pm 1} d |\mathbf{p_1}|
	\int \frac{p_1^3}{\exp^{\frac{|\mathbf{p_2}|}{T}} \pm 2} d |\mathbf{p_2}|
	\end{eqnarray}
We can observe that
	\begin{eqnarray}
	\label{}
	\int \frac{p_i^3}{\exp^{\frac{|\mathbf{p_i}|}{T}} \pm 1} d |\mathbf{p_i}|
	\propto \rho_{i\; \mathrm{ultrarelativistic}}
	\end{eqnarray}
where $\rho_{i\; \mathrm{ultrarelativistic}}$ is given in (\ref{td_fcts_ultra_rev}). Finally equation (\ref{s_aver_num}) becomes
	\begin{eqnarray}
	\label{}
	\frac{32}{225} \pi^{10} \, T^8 \times
	 \Biggl\{ 
	\begin{array}{c} 1 \hspace{1.5cm} \rm{(for \; any \; present \; BOSON)}
	\\ \frac{7}{8}  \hspace{1.cm} \rm{(for \; any \; present \; FERMION)}
	\end{array}
	\end{eqnarray}
Therefore $\langle s \rangle$ finally is:
	\begin{eqnarray}
	\label{s_UR}
	s = \frac{\pi^8 T^2}{450 \zeta_3^2} \times
	 \Biggl\{ 
	\begin{array}{c} 1 \hspace{1.5cm} \rm{(for \; any \; present \; BOSON)}
	\\ \frac{7}{6}  \hspace{1.cm} \rm{(for \; any \; present \; FERMION)}
	\end{array}
	\end{eqnarray}
Hence
	\begin{eqnarray}
	\label{}
	s &=& 14.6 \, T^2 	\hspace{3cm} 	\rm{(boson-boson \, scattering)} \cr\cr
	&=& 17.1 \, T^2 	\hspace{3cm} 	\rm{(boson-fermion \, scattering)} \cr\cr
	&=& 19.9 \, T^2 	\hspace{3cm} 	\rm{(fermion-fermion \, scattering)} 
	\end{eqnarray}
%


%% file: files/appendix_cross_sections.tex

\def \chap-cross-sect{Cross sections}
\chapter{\chap-cross-sect}
\label{chap-cross-sect}
\markboth{Appendix \ref{chap-cross-sect}. ~ \chap-cross-sect}
                    {Appendix \ref{chap-cross-sect}. ~ \chap-cross-sect}

In many books, such as \cite{Mandl:1985bg,Halzen:1984mc}, we can find expressions for the differential cross section $\frac{d\sigma}{d\Omega}$. Let us consider a scattering process in which two particles - leptons or photons - with four momenta $p_i = (E_i, \mathbf{p_i})$, $i=1,2$ collide and produce 2 final particles with moments $p'_f = (E'_f, \mathbf{p'_f})$, $f=1,2$. 
In the center of mass frame (CM) we have $\mathbf{p'_1} = - \mathbf{p'_2}$ and the differential cross section is
%
	\begin{eqnarray}
	\label{}
	\left( \frac{d\sigma}{d\Omega} \right)_{CM} = 
	\frac{\Pi_l (2m_l)}{64 \pi^2 (E_{1} + E_{2})^2}
	\frac{|\mathbf{p'_1}|}{|\mathbf{p_1}|} X 
	\end{eqnarray}
where $l$ runs over all the present lepton species, having masses $m_l$ and $X$ is the spin average:
	\begin{eqnarray}
	\label{}
	X = \frac{1}{2} \sum_{r,s=1}^2 |\mathcal{M}|^2  
	\end{eqnarray}
Below we will work out $\frac{d\sigma}{d\Omega}$ for some processes of interest for this thesis.

\def \sec-nu-scat{Neutrino scattering}
\section{\sec-nu-scat}
\label{sec-nu-scat}
\markboth{Appendix \ref{chap-cross-sect}. ~ \chap-cross-sect}
	{\S \ref{sec-nu-scat} ~ \sec-nu-scat}


In this section we will assume $k^2 \ll m_W^2$, $k^2 \ll m_Z^2$ where $k$ is the intermediate boson momentum, so that the weak boson propagators reduce to the Fermi constant.
 
\begin{figure}[htbp]
	\begin{center}
	 \includegraphics[scale=1.2]{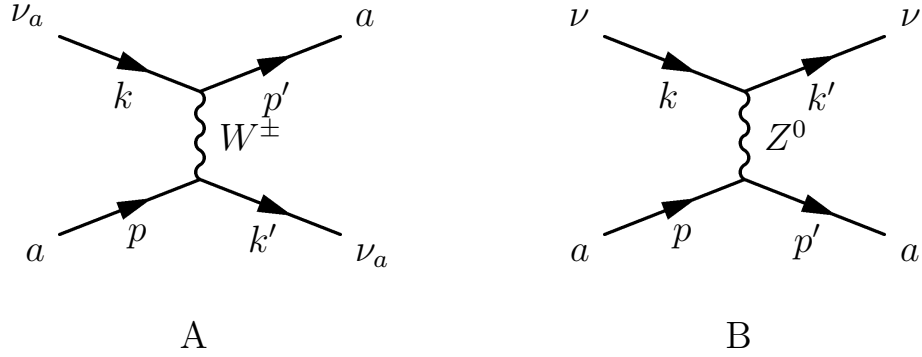}	
	\caption{{\it Neutrino scattering via charged (A) and neutral (B) currents.}}
	\label{nu_scat_fig}
	\end{center}
	\end{figure}

Let us consider the neutrino scattering $\nu_a a \leftrightarrow \nu_a a$ with $a=e,\mu,\tau$ mediated by the \it charged weak bosons $W^{\pm}$ \rm as depicted Figure \ref{nu_scat_fig} A. We will indicate below by the apex $CC$ when referring to charged currents.
The cross sections we will get are the same we would have for the scattering $\bar \nu_a \bar a \leftrightarrow \bar \nu_a \bar a$.
The Feynman amplitude for the processes we are considering is:
	\begin{eqnarray}
	\label{}
	|\mathcal{M}|^{CC} = \frac{G_F}{\sqrt{2}}
	\left[ \bar u(k') \gamma^{\mu} (1-\gamma_5) u(p) \right]
	\left[ \bar u(p') \gamma_{\mu} (1-\gamma_5) u(k) \right]
	\end{eqnarray}
from which follows 
	\begin{eqnarray}
	\label{}
	X^{CC} &=& \frac{G_F^2}{4} \mathrm{tr} 
	\left[ \gamma^{\mu} (1-\gamma_5) (p\!\!\!/ +m) \gamma^{\nu} (1-\gamma_5) k\, ' \!\!\!\!\!/ \;\,\right]
	\mathrm{tr} 
	\left[ \gamma_{\mu} (1-\gamma_5) k\!\!\!/  \gamma_{\nu} (1-\gamma_5) p\, ' \!\!\!\!\!/ \;\,\right] = \cr\cr
	&=& \frac{G_F^2}{4} \,256 \,  \left(p_{\mu} k^{\mu} \right) \left(p_{\nu}' k'^{\nu} \right)
	\end{eqnarray}
where $G_F \simeq 1.1664 \cdot 10^{-5} GeV^{-2}$ is the Fermi constant.
The differential cross section for this process is therefore
	\begin{eqnarray}
	\label{}
	\left(\frac{d\sigma}{d\Omega}\right) ^{CC} = 
	\frac{1}{64 \pi^2 s} \: 
	\frac{|p'|}{|p|} \:
	\frac{G_F^2}{2} \, 128 \, \left(p_{\mu} k^{\mu} \right)
	\left(p_{\nu}' k'^{\nu} \right)
	\end{eqnarray}
There are two extreme cases of interest in physics: the first is an interaction of a neutrino with an \it ultrarelativistic target\rm.
In this limit $E_{\nu} \gg M_a$; therefore $s = 2\,p_{\mu} k^{\mu} = 2\,p'_{\mu} k'^{\mu}$, $p' = p$ and
	\begin{eqnarray}
	\label{}
	\left( \frac{d\sigma}{d\Omega} \right) ^{CC}= 
	\frac{G_F^2}{4 \pi^2} \, s 
	\hspace{0.5cm} \Longrightarrow \hspace{0.5cm}
	\sigma^{CC} = 4 \pi \frac{d\sigma}{d\Omega} = \frac{G_F^2}{\pi} \, s
	\end{eqnarray}
The second is an interaction with an \it ultramassive target\rm, when $E_{\nu} \ll M_a$, $\mathbf{p} \simeq 0$; therefore
	\begin{eqnarray}
	\label{}
	s \simeq M_a ^2 \hspace{1.2cm} 
	p_{\mu} k^{\mu} = p'_{\mu} k'^{\mu} = M_a E_{\nu}
	\end{eqnarray}
	\begin{eqnarray}
	\label{}
	\left( \frac{d\sigma}{d\Omega} \right) ^{CC}= 
	\frac{G_F^2}{\pi^2} \, E_{\nu}^2
	\hspace{0.5cm} \Longrightarrow \hspace{0.5cm}
	\sigma^{CC} = 4 \pi \frac{d\sigma}{d\Omega} = \frac{4G_F^2}{\pi} \, E_{\nu}^2
	\end{eqnarray}
Let us now turn to the neutrino scattering $\nu_i \, a \leftrightarrow \nu_i \, a$ with $i,a=e,\mu,\tau$ mediated by the \it neutral weak bosons $Z^{0}$\rm, depicted Figure \ref{nu_scat_fig} B%
\footnote{The same process is also mediated by the Higgs boson with an amplitude $\propto \frac{1}{m_H^2}$.}.
The apex $NC$ will state for neutral currents. The Feynman amplitude for this process is:
	\begin{eqnarray}
	\label{M_NC}
	|\mathcal{M}|^{NC}=
	\frac{G_F}{\sqrt{2}}
	\left[ \bar u(k') \gamma^{\mu} (1-\gamma_5) u(k)  \right]
	\left[ \bar u(p') \gamma_{\mu} (g_V - g_A \gamma_5) u(p) \right]
	\end{eqnarray}
where 
	\begin{eqnarray}
	\label{}
	g_V \equiv 2 \sin^2 \theta_W - \frac{1}{2} \hspace{1.5cm}
	g_A \equiv - \frac{1}{2} 
	\end{eqnarray}
In the \it ultrarelativistic target \rm approximation we get the total cross section
	\begin{eqnarray}
	\label{}
	\sigma^{NC} = \frac{G_F^2 s}{3\pi}
	\left(g_V^2 + g_V g_A + g_A^2 \right)
	\end{eqnarray}
If the colliding particle is an anti-neutrino, that is if we consider the process $\bar \nu_i \, a \leftrightarrow \bar \nu_i \, a$, the amplitude will be the same than (\ref{M_NC}) with the substitution $g_A \leftrightarrow - g_A$ which leads to
	\begin{eqnarray}
	\label{}
	\sigma^{NC}_{\bar \nu} = \frac{G_F^2 s}{3\pi}
	\left(g_V^2 - g_V g_A + g_A^2 \right)
	\end{eqnarray}

\def \sec-eebar-ffbar{Pair annihilation and creation: $e^+e^-\leftrightarrow f \, \overline{f}$}
\section{\sec-eebar-ffbar}
\label{e_e_ebar->f_f_bar}
\markboth{Appendix \ref{chap-cross-sect}. ~ \chap-cross-sect}
	{\S \ref{e_e_ebar->f_f_bar} ~ \sec-eebar-ffbar}

	\begin{figure}[htbp]
	\begin{center}
	 \includegraphics[scale=0.9]{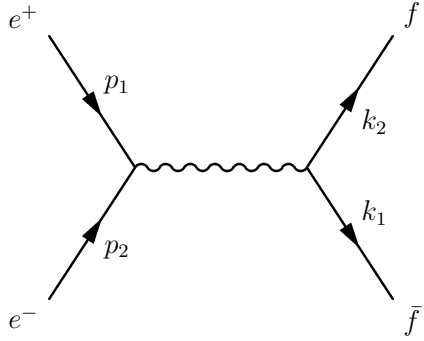}	
	\caption{{\it Pair annihilation and creation.}}
	\label{pair_ann_fig}
	\end{center}
	\end{figure}
\noindent
This process is depicted in Figure \ref{pair_ann_fig}.
The spin average factor $X$ for it can be found in textbooks and has the form:
	\begin{eqnarray}
	\label{}
	X &=&
	\frac{e^4 \epsilon^2}{2 m_e^2 m_f^2 [(p_1 + p_2)^2 ]^2} \cr\cr\cr
	&\cdot&
	\left\{ (p_1 p_1') (p_2 p_2') + (p_1 p_2') (p_2 p_1') + m_e^2 (p_1' p_2') + 
	m_f^2 (p_1 p_2) + 2 m_e^2 m_f^2 \right\} \hspace{1.1cm}
	\end{eqnarray}
Making the substitutions:
	\begin{eqnarray}
	\label{}
	p_1 p_1' &=& p_2 p_2' = E^2 - p p' \cos \theta \cr\cr
	p_1 p_2' &=& p_2 p_1' = E^2 + p p' \cos \theta \cr\cr
	p_1 p_2 &=& E^2 + p^2 \cr\cr
	p_1' p_2' &=& E^2 + p'^2 \cr\cr
	(p_1 + p_2)^2 &=& 4 E^2 \cr\cr
	E^2 &=& m_e^2 + p^2 = m_f ^2 + p'^2
	\end{eqnarray}
we finally get
	\begin{eqnarray}
	\label{}
	\left( \frac{d\sigma}{d\Omega} \right)_{CM} &=& 
	\frac{\epsilon^2 \alpha^2}{16 E^6}
	\frac{ |p'| }{ |p| }
	\left[ E^4 + p^2 p'^2 \cos^2\theta + E^2 (m_f^2 + m_e^2) \right]\cr\cr\cr
	\sigma_{CM} &=& \frac{\epsilon^2 \alpha^24\pi}{16 E^6} 
	\frac{ |p'| }{ |p| }
	\left[ E^4 + \frac{1}{3} p^2 p'^2 + E^2 (m_f^2 + m_e^2) \right]
	\end{eqnarray}
In the following we analyze some special but interesting limits which have been used in this thesis.

\subsubsection{Equal masses}

In this limit we have $m_e = m_f =m \Rightarrow p = p'$ and the differential cross section becomes
	\begin{eqnarray}
	\label{}
	\left( \frac{d\sigma}{d\Omega} \right)_{CM} = 
	\frac{\epsilon^2 \alpha^2}{16 E^6}
	[ E^4 + p^4 \cos^2\theta + 2 E^2 m^2 ]
	\end{eqnarray}
Integrating over the solid angle we get
	\begin{eqnarray}
	\label{e_ebar_f_fbar_total_sigma_equal_masses}
	\sigma_{CM} = 
	\frac{\pi}{4} 
	\frac{\epsilon^2 \alpha^2 }{3 E^6}
	(2E^2 + m^2)^2 =
	\frac{4\pi}{3} \epsilon^2 \alpha^2
	\frac{(s + 2m^2)^2}{s^3}
	\end{eqnarray}
where $\alpha = \frac{e^2}{4\pi}$ and $s = ( p_1 + p_2)^2 $ has for this special case the value $s = 4E^2$. In the ultrarelativistic limit $m \simeq 0$ eq.(\ref{e_ebar_f_fbar_total_sigma_equal_masses}) gives
	\begin{eqnarray}
	\label{}
	\sigma_{CM}^{rel} = 
	\frac{4\pi}{3} \epsilon^2 \alpha^2
	\frac{1}{s}
	\end{eqnarray}
which is the usual cross section for lepton production at high energies - see for instance the process $e^+e^-  \rightarrow \mu^+ \mu^-$.

\subsubsection{Near the energy threshold}

Let us define now the variables
	\begin{eqnarray}
	\label{}
	y = \frac{m_f}{E} \hspace{1.5cm} z =  \frac{m_e}{E}
	\end{eqnarray}
%
which can be used to recast the total cross section as:
	\begin{eqnarray}
	\label{}
	\sigma_{CM} = 
	\frac{\pi}{4} \frac{\epsilon^2 \alpha^2}{E^2}
	\sqrt{\frac{1-y^2}{1-z^2}}
	\left[ 1 + \frac{1}{3} (1-z^2) (1- y^2) + y^2 + z^2
	\right]
	\end{eqnarray}
Since we are assuming $m_f \geq m_e$, the threshold is found when $E \rightarrow m_f$; we can have the two cases
	\begin{eqnarray}
	\label{}
	\lim_{y \rightarrow 1}\sigma_{CM} = 
	\Biggl\{ 
	\begin{array}{c} 
	\frac{3\pi}{4} \frac{\epsilon^2 \alpha^2}{m^2} \hspace{1.3cm} m_f = m_e = m
	\\ 0 \hspace{2cm} m_f > m_e
	\end{array}\  
	\end{eqnarray}

\def \sec-ef-ef{Scattering of ordinary and mirror matter: $e f \leftrightarrow e f $}
\section{\sec-ef-ef}
\label{e_f->e_f}
\markboth{Appendix \ref{chap-cross-sect}. ~ \chap-cross-sect}
	{\S \ref{e_f->e_f} ~ \sec-ef-ef}

	\begin{figure}[htbp]
	\begin{center}
	 \includegraphics[scale=0.9]{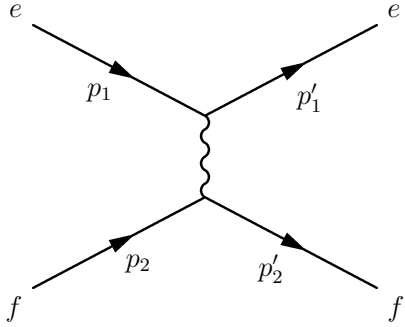}	
	\caption{{\it Mirror and ordinary matter scattering.}}
	\label{mir_scat_fig}
	\end{center}
	\end{figure}
\noindent This process is shown in Figure \ref{mir_scat_fig}.
The spin average factor $X$ has the form:
	\begin{eqnarray}
	\label{}
	X &=& 
	\frac{e^4 \epsilon^2}{2 m_e^2 m_f^2 [(p_1 - p'_1)^2 ]^2} \cr\cr\cr
	&\cdot&
	\left\{ (p_1 p_2) (p_1' p_2') + (p_1 p_2') (p_2 p_1') - m_e^2 (p_2 p_2') - 
	m_f^2 (p_1 p_1') + 2 m_e^2 m_f^2 \right\} \hspace{1.2cm}
	\end{eqnarray}
Let us assume now $m_e = m_f=m$, which implies $|\mathbf{p}| = |\mathbf{p}'|$.
Making the substitutions:
	\begin{eqnarray}
	\label{}
	p_1 p_1' &=& p_2 p_2' = E^2 - p^2 \cos \theta \cr\cr
	p_1 p_2' &=& p_2 p_1' = E^2 + p^2 \cos \theta \cr\cr
	p_1 p_2 &=& p_1' p_2' = E^2 + p^2 \cr\cr
	|p_1 - p_2|^2 &=& |(0 , \mathbf{p - p'})|^2 = -4 p^2 \sin^2\left( \frac{\theta}{2} \right)
	\end{eqnarray}
we finally get the differential cross section:
	\begin{eqnarray}
	\label{}
	\frac{d\sigma}{d\Omega} &=&
	\frac{\epsilon^2 \alpha^2}{32 \sin^4 \left(\frac{\theta}{2} \right) E^2 p^4} \cr\cr
	&\cdot& \left[ 2E^4 + p^4 (1+\cos^2\theta) +
	4E^2p^2 \cos^2 \left( \frac{\theta}{2} \right)
	- 2m^2 (E^2 - p^2\cos\theta) + 2 m^4
	\right] \hspace{1.2cm}
	\end{eqnarray}
%


%% file: tesi.bbl
\begin{thebibliography}{2}
\markboth{Bibliography}{Bibliography}
\addcontentsline{toc}{chapter}{Bibliography}




\bibitem{Glashow:1961tr}
  S.~L.~Glashow,
  Nucl.\ Phys.\  {\bf 22} (1961) 579.

\bibitem{Weinberg:1967tq}
  S.~Weinberg,
  Phys.\ Rev.\ Lett.\  {\bf 19} (1967) 1264.

\bibitem{Salam:1968}
  A.~Salam,
  ed. N. Svaratholm. Stockholm: Almquist and Forlag {\bf 19} (1968) 1264.

\bibitem{Leader:1996hk}
  E.~Leader and E.~Predazzi,
  Camb.\ Monogr.\ Part.\ Phys.\ Nucl.\ Phys.\ Cosmol.\  {\bf 3} (1996) 1.

\bibitem{Noether:1918}
  E.~Noether 
  Nachr.\ Kgl.\ Geo.\ Wiss\ Gottinger\  {\bf 235} (1918) 

\bibitem{Nambu:1960xd}
  Y.~Nambu,
  Phys.\ Rev.\ Lett.\  {\bf 4} (1960) 380.

\bibitem{Nambu:1961fr}
  Y.~Nambu and G.~Jona-Lasinio,
  Phys.\ Rev.\  {\bf 124} (1961) 246.

\bibitem{Nambu:1961tp}
  Y.~Nambu and G.~Jona-Lasinio,
  Phys.\ Rev.\  {\bf 122} (1961) 345.

\bibitem{Goldstone:1961eq}
  J.~Goldstone,
  Nuovo Cim.\  {\bf 19} (1961) 154.

\bibitem{Goldstone:1962es}
  J.~Goldstone, A.~Salam and S.~Weinberg,
  Phys.\ Rev.\  {\bf 127} (1962) 965.

\bibitem{Anderson:1963pc}
  P.~W.~Anderson,
  Phys.\ Rev.\  {\bf 130} (1963) 439.

\bibitem{Englert:1964et}
  F.~Englert and R.~Brout,
  Phys.\ Rev.\ Lett.\  {\bf 13} (1964) 321.

\bibitem{Guralnik:1964eu}
  G.~S.~Guralnik, C.~R.~Hagen and T.~W.~B.~Kibble,
  Phys.\ Rev.\ Lett.\  {\bf 13} (1964) 585.

\bibitem{Higgs:1964ia}
  P.~W.~Higgs,
  Phys.\ Lett.\  {\bf 12} (1964) 132.

\bibitem{Higgs:1966ev}
  P.~W.~Higgs,
  Phys.\ Rev.\  {\bf 145} (1966) 1156.

\bibitem{Hooft:1971rn}
  G.~'t Hooft,
  Nucl.\ Phys.\  B {\bf 35} (1971) 167.

\bibitem{Lee:1956qn}
  T.~D.~Lee and C.~N.~Yang,
  Phys.\ Rev.\  {\bf 104} (1956) 254.

\bibitem{Okun:1966}
Y.~Yu.~Kobzarev, L.~B.~Okun and I.~Ya.~Pomeranchuk,
  Yad.\ Fiz.\  {\bf 3},1154 (1966) [Sov. J. Nucl. Phys. {\bf 3},837 (1966)].
  
\bibitem{Wu:1957my}
  C.~S.~Wu, E.~Ambler, R.~W.~Hayward, D.~D.~Hoppes and R.~P.~Hudson,
  Phys.\ Rev.\  {\bf 105} (1957) 1413.

\bibitem{Politzer:1973fx}
  H.~D.~Politzer,
  Phys.\ Rev.\ Lett.\  {\bf 30} (1973) 1346.

  H.~D.~Politzer,
  Phys.\ Rept.\  {\bf 14}, 129 (1974).

\bibitem{Gross:1973ju}
  D.~J.~Gross and F.~Wilczek,
  Phys.\ Rev.\  D {\bf 8} (1973) 3633.

  D.~J.~Gross and F.~Wilczek,
  Phys.\ Rev.\ Lett.\  {\bf 30} (1973) 1343.

\bibitem{Perkins:1982xb}
  D.~h.~Perkins,
{\it  Reading, Usa: Addison-wesley ( 1982) 437p}

\bibitem{Sengupta:2000be}
  S.~Sengupta,
  Phys.\ Lett.\  B {\bf 484} (2000) 275.

\bibitem{Marinelli:1983nd}
  M.~Marinelli and G.~Morpurgo,
  Phys.\ Lett.\  B {\bf 137} (1984) 439.

\bibitem{Baumann:1988ue}
  J.~Baumann, J.~Kalus, R.~Gahler and W.~Mampe,
  Phys.\ Rev.\  D {\bf 37} (1988) 3107.

\bibitem{Meyer:1999cx}
  V.~Meyer {\it et al.},
  Phys.\ Rev.\ Lett.\  {\bf 84} (2000) 1136
  [arXiv:hep-ex/9907013].

\bibitem{Raffelt:1999gv}
  G.~G.~Raffelt,
  Phys.\ Rept.\  {\bf 320} (1999) 319.


  
\bibitem{Peskin:1995ev}
  M.~E.~Peskin and D.~V.~Schroeder,
{\it  Reading, USA: Addison-Wesley (1995) 842 p}

\bibitem{Cheng:1985bj}
  T.~P.~Cheng and L.~F.~Li,
{\it  Oxford, Uk: Clarendon ( 1984) 536 P. ( Oxford Science Publications)}

\bibitem{Kaku:1993ym}
  M.~Kaku,
{\it  New York, USA: Oxford Univ. Pr. (1993) 785 p}

\bibitem{Mandl:1985bg}
  F.~Mandl and G.~Shaw,
{\it  Chichester, Uk: Wiley ( 1984) 354 P. ( A Wiley-interscience Publication)}

\bibitem{Halzen:1984mc}
  F.~Halzen and A.~D.~Martin,
{\it  New York, Usa: Wiley ( 1984) 396p}

\bibitem{Okun:1983vw}
  L.~B.~Okun, M.~B.~Voloshin and V.~I.~Zakharov,
  Phys.\ Lett.\  B {\bf 138} (1984) 115.

\bibitem{Berezhiani:1982ww}
  Z.~G.~Berezhiani,
  Yad.\ Fiz.\  {\bf 36} (1982) 1052.

\bibitem{Babu:1989ex}
  K.~S.~Babu and R.~N.~Mohapatra,
  Phys.\ Rev.\  D {\bf 41} (1990) 271.

\bibitem{Babu:1989tq}
  K.~S.~Babu and R.~N.~Mohapatra,
  Phys.\ Rev.\ Lett.\  {\bf 63} (1989) 938.

\bibitem{Yao:2006px}
  W.~M.~Yao {\it et al.}  [Particle Data Group],
  J.\ Phys.\ G {\bf 33} (2006) 1.

\bibitem{Tegmark:2006az}
  M.~Tegmark {\it et al.},
  Phys.\ Rev.\  D {\bf 74} (2006) 123507
  [arXiv:astro-ph/0608632].

\bibitem{Georgi:1974sy}
  H.~Georgi and S.~L.~Glashow,
  Phys.\ Rev.\ Lett.\  {\bf 32} (1974) 438.

\bibitem{Adler:1969gk}
  S.~L.~Adler,
  Phys.\ Rev.\  {\bf 177} (1969) 2426.

\bibitem{Bell:1969ts}
  J.~S.~Bell and R.~Jackiw,
  Nuovo Cim.\  A {\bf 60} (1969) 47.
  
\bibitem{Adler:1969er}
  S.~L.~Adler and W.~A.~Bardeen,
  Phys.\ Rev.\  {\bf 182} (1969) 1517.
  
\bibitem{Bardeen:1969md}
  W.~A.~Bardeen,
  Phys.\ Rev.\  {\bf 184} (1969) 1848.

\bibitem{Gross:1972pv}
  D.~J.~Gross and R.~Jackiw,
  Phys.\ Rev.\  D {\bf 6} (1972) 477.
  
\bibitem{Georgi:1972bb}
H.~Georgi and S.~L.~Glashow,
Phys.\ Rev.\  D {\bf 6} (1972) 429.

\bibitem{Bouchiat:1972iq}
  C.~Bouchiat, J.~Iliopoulos and P.~Meyer,
  Phys.\ Lett.\  B {\bf 38} (1972) 519.

\bibitem{Dirac:1931kp}
  P.~A.~M.~Dirac,
  Proc.\ Roy.\ Soc.\ Lond.\  A {\bf 133} (1931) 60.

\bibitem{Hooft:1974qc}
  G.~'t Hooft,
  Nucl.\ Phys.\  B {\bf 79} (1974) 276.

  A.~M.~Polyakov,
  JETP Lett.\  {\bf 20} (1974) 194
  [Pisma Zh.\ Eksp.\ Teor.\ Fiz.\  {\bf 20} (1974) 430].

\bibitem{Georgi:1972cj}
  H.~Georgi and S.~L.~Glashow,
  Phys.\ Rev.\ Lett.\  {\bf 28} (1972) 1494.

\bibitem{Turner:1982ag}
  M.~S.~Turner, E.~N.~Parker and T.~J.~Bogdan,
  Phys.\ Rev.\  D {\bf 26} (1982) 1296.

\bibitem{Ambrosio:2002qq}
  M.~Ambrosio {\it et al.}  [MACRO Collaboration],
  Eur.\ Phys.\ J.\  C {\bf 25} (2002) 511
  [arXiv:hep-ex/0207020].

  M.~Ambrosio {\it et al.}  [MACRO Collaboration],
  Eur.\ Phys.\ J.\  C {\bf 26} (2002) 163
  [arXiv:hep-ex/0207024].

\bibitem{Holdom:1985ag}
  B.~Holdom,
  Phys.\ Lett.\  B {\bf 166} (1986) 196.

\bibitem{Foot:1991bp}
  R.~Foot, H.~Lew and R.~R.~Volkas,
  Phys.\ Lett.\  B {\bf 272} (1991) 67.

\bibitem{Foot:2000vy}
  R.~Foot, A.~Y.~Ignatiev and R.~R.~Volkas,
  Phys.\ Lett.\  B {\bf 503} (2001) 355
  [arXiv:astro-ph/0011156].

\bibitem{Carlson:1987si}
  E.~D.~Carlson and S.~L.~Glashow,
  Phys.\ Lett.\  B {\bf 193} (1987) 168.

\bibitem{Glashow:1985ud}
  S.~L.~Glashow,
  Phys.\ Lett.\  B {\bf 167} (1986) 35.

  
\bibitem{Badertscher:2006fm}
  A.~Badertscher {\it et al.},
  Phys.\ Rev.\  D {\bf 75} (2007) 032004
  [arXiv:hep-ex/0609059].

\bibitem{Foot:2003iv}
  R.~Foot,
  Phys.\ Rev.\  D {\bf 69} (2004) 036001
  [arXiv:hep-ph/0308254].

\bibitem{Foot:2004ej}
  R.~Foot,
  arXiv:astro-ph/0403043.

  R.~Foot,
  Mod.\ Phys.\ Lett.\  A {\bf 19} (2004) 1841
  [arXiv:astro-ph/0405362].

\bibitem{Bernabei:1996vj}
  R.~Bernabei {\it et al.},
  Phys.\ Lett.\  B {\bf 389} (1996) 757.

  R.~Bernabei {\it et al.},
  Phys.\ Lett.\  B {\bf 424} (1998) 195.

  R.~Bernabei {\it et al.}  [DAMA Collaboration],
  Phys.\ Lett.\  B {\bf 450} (1999) 448.

  R.~Bernabei {\it et al.}  [DAMA Collaboration],
  Phys.\ Lett.\  B {\bf 480} (2000) 23.

  R.~Bernabei {\it et al.},
  Eur.\ Phys.\ J.\  C {\bf 23} (2002) 61.

\bibitem{Bernabei:2003za}
  R.~Bernabei {\it et al.},
  Riv.\ Nuovo Cim.\  {\bf 26N1} (2003) 1
  [arXiv:astro-ph/0307403].

\bibitem{Berezhiani:2003xm}
  Z.~Berezhiani,
  Int.\ J.\ Mod.\ Phys.\  A {\bf 19} (2004) 3775
  [arXiv:hep-ph/0312335].

\bibitem{Zurab_Bento_last_minute}
  L.~Bento and Z.~Berezhiani,
    Phys.\ Rev.\ Lett.\  {\bf 96} (2006) 081801
  [arXiv:hep-ph/0507031];  
  Phys.\ Lett.\  B {\bf 365} (2006) 253
    [arXiv:hep-ph/0602227];

\bibitem{Berezhiani:2005ek}
  Z.~Berezhiani,
  ``Through the Looking-Glass: Alice's Adventures in Mirror World,''
  arXiv:hep-ph/0508233.

\bibitem{Berezhiani:1998kkk}
 Z.~Berezhiani, 
  Phys.\ Lett.\  B {\bf 417} (1998) 287.

\bibitem{Akhmedov:1992hh}
  E.~K.~Akhmedov, Z.~G.~Berezhiani and G.~Senjanovic,
  Phys.\ Rev.\ Lett.\  {\bf 69} (1992) 3013
  [arXiv:hep-ph/9205230].
  
  Z.~G.~Berezhiani and R.~N.~Mohapatra,
  Phys.\ Rev.\  D {\bf 52} (1995) 6607
  [arXiv:hep-ph/9505385].

\bibitem{AIP_Zurab}
Z.~G.~Berezhiani, AIP Conf.Proc. {\bf 878} (2006) 195
 [arXiv:hep-ph/0612371].

\bibitem{BDM} 
Z.~G.~Berezhiani, A.~D.~Dolgov and R.~N.~Mohapatra,
  Phys.\ Lett.\  B {\bf 375} (1996) 26
  [arXiv:hep-ph/9511221].

  Z.~G.~Berezhiani,
  Acta Phys.\ Polon.\  B {\bf 27} (1996) 1503
  [arXiv:hep-ph/9602326].
%
%

\bibitem{Venya}
  V.~S.~Berezinsky and A.~Vilenkin,
  Phys.\ Rev.\  D {\bf 62} (2000) 083512
  [arXiv:hep-ph/9908257].


\bibitem{KST}
  E.~W.~Kolb, D.~Seckel and M.~S.~Turner,
  Nature {\bf 314} (1985) 415.


\bibitem{Berezhiani:2000gh}
  Z.~Berezhiani, L.~Gianfagna and M.~Giannotti,
  Phys.\ Lett.\  B {\bf 500} (2001) 286
  [arXiv:hep-ph/0009290] \\
%
  Z.~Berezhiani and A.~Drago
    Phys.\ Lett.\  B {\bf 473} (2000) 281
  [arXiv:hep-ph/9911333].
  
\bibitem{Gianfagna:2004je}
  L.~Gianfagna, M.~Giannotti and F.~Nesti,
  JHEP {\bf 0410} (2004) 044
  [arXiv:hep-ph/0409185].

\bibitem{Berezhiani:1995am}
  Z.~G.~Berezhiani, A.~D.~Dolgov and R.~N.~Mohapatra,
  Phys.\ Lett.\  B {\bf 375} (1996) 26
  [arXiv:hep-ph/9511221].

\bibitem{Foot:1991py}
  R.~Foot, H.~Lew and R.~R.~Volkas,
  Mod.\ Phys.\ Lett.\  A {\bf 7} (1992) 2567.

  R.~Foot and R.~R.~Volkas,
  Phys.\ Rev.\  D {\bf 52} (1995) 6595
  [arXiv:hep-ph/9505359].

\bibitem{Kolb:1990vq}
  E.~W.~Kolb and M.~S.~Turner,
  ``The Early universe,'' (Westview Press) 
  Front.\ Phys.\  {\bf 69} (1990) 1.

\bibitem{paddybook} 
T. Padmanabhan, "Theoretical Astrophysics - Volume III: Galaxies and Cosmology" (Cambridge University Press) (2002)

\bibitem{Lisi:1999ng}
  E.~Lisi, S.~Sarkar and F.~L.~Villante,
  Phys.\ Rev.\  D {\bf 59} (1999) 123520
  [arXiv:hep-ph/9901404].

\bibitem{Lopez:1998vk}
  R.~E.~Lopez and M.~S.~Turner,
  Phys.\ Rev.\  D {\bf 59} (1999) 103502
  [arXiv:astro-ph/9807279].

\bibitem{Pagel:1991qk}
  B.~E.~J.~Pagel, E.~A.~Simonson, R.~J.~Terlevich and M.~G.~Edmunds,
  Mon.\ Not.\ Roy.\ Astron.\ Soc.\  {\bf 255} (1992) 325.

  K.~A.~Olive and G.~Steigman,
  Astrophys.\ J.\ Suppl.\  {\bf 97} (1995) 49
  [arXiv:astro-ph/9405022].

\bibitem{Izotov:1994tg}
  Y.~I.~Izotov, T.~X.~Thuan and V.~A.~Lipovetsky,
  Astrophys.\ J.\  {\bf 435} (1994) 647.

  Y.~I.~Izotov, T.~X.~Thuan and V.~A.~Lipovetsky,
  Astrophys.\ J.\ Suppl.\  {\bf 108} (1997) 1.

\bibitem{Berezhiani:2003wj}
  Z.~Berezhiani, P.~Ciarcelluti, D.~Comelli and F.~L.~Villante,
  Int.\ J.\ Mod.\ Phys.\  D {\bf 14} (2005) 107
  [arXiv:astro-ph/0312605].

\bibitem{Cassisi}
  Z.~Berezhiani, S.~Cassisi, P.~Ciarcelluti and A.~Pietrinferni,
  Astropart. \ Phys. \ {\bf 24} (2006) 495

\bibitem{Ciarcelluti:2003wm}
  P.~Ciarcelluti,
  arXiv:astro-ph/0312607.

  P.~Ciarcelluti,
  Int.\ J.\ Mod.\ Phys.\  D {\bf 14} (2005) 223
  [arXiv:astro-ph/0409633].

\bibitem{Berezhiani:2000gw}
  Z.~Berezhiani, D.~Comelli and F.~L.~Villante,
  Phys.\ Lett.\  B {\bf 503} (2001) 362
  [arXiv:hep-ph/0008105].

\bibitem{Bento:2002sj}
  Fortsch.\ Phys.\  {\bf 50} (2002) 489;
   Phys.\ Rev.\ Lett. {\bf 87} (2001) 231304 [arXiV: hep-ph/0107281];
   arXiV: hep-ph/0111116]

\bibitem{Yao:2006px}
  W.~M.~Yao {\it et al.}  [Particle Data Group],
  J.\ Phys.\ G {\bf 33} (2006) 1.

\bibitem{Kawano:1992ua}
  L.~Kawano,
FERMILAB-PUB-92-004-A (1992);

\bibitem{Wagoner:1972jh}
  R.~V.~Wagoner,
  Astrophys.\ J.\  {\bf 179} (1973) 343.

\bibitem{Mangano:2006ur}
 G.~Mangano, A.~Melchiorri, O.~Mena, G.~Miele and A.~Slosar,
 JCAP {\bf 0703} (2007) 006
 [arXiv:astro-ph/0612150].
 
\bibitem{Gondolo:1990dk}
P.~Gondolo and G.~Gelmini,
Nucl.\ Phys.\  B {\bf 360} (1991) 145.

\bibitem{Davidson:1991si}
  S.~Davidson, B.~Campbell and D.~C.~Bailey,
  Phys.\ Rev.\  D {\bf 43} (1991) 2314.

  S.~Davidson and M.~E.~Peskin,
  Phys.\ Rev.\  D {\bf 49} (1994) 2114
  [arXiv:hep-ph/9310288].

\bibitem{Davidson:2000hf}
  S.~Davidson, S.~Hannestad and G.~Raffelt,
  JHEP {\bf 0005} (2000) 003
  [arXiv:hep-ph/0001179].
 
\bibitem{Dubovsky:2003yn}
  S.~L.~Dubovsky, D.~S.~Gorbunov and G.~I.~Rubtsov,
  JETP Lett.\  {\bf 79} (2004) 1
  [Pisma Zh.\ Eksp.\ Teor.\ Fiz.\  {\bf 79} (2004) 3]
  [arXiv:hep-ph/0311189].
  
\bibitem{Raffelt:1996wa}
  G.~G.~Raffelt,
{\it  Chicago, USA: Univ. Pr. (1996) 664 p}
 
\bibitem{paper_prep}
Z. Berezhiani and A.~Lepidi 
{\it (In preparation)}





\end{thebibliography}
